# Intrinsically coupled stripes within the CuO$_2$ planes of high-T$_c$ materials

W. Winkler

*Laboratory For Materials, Wackenbergstr. 84-88, 13156 Berlin, Germany*

(Dated: September 11, 2003)

## ABSTRACT

Analysis of the electronic state of the CuO$_2$ planes of high-T$_c$ materials has been performed with special regard to the influence of the Coulomb interactions separated after moments. Using this analysis to derive the basic structure of the electronic states within the CuO$_2$ planes of the high-T$_c$ materials, different symmetry breaking effects were revealed. First of all, a commensurate charge and bonding fluctuation state (CBF) with the period $(2a,2b)$ is established which exists collinearly with the antiferromagnetic spin state. It is concluded that the CBF state and the antiferromagnetic spin state are results of the same electronic renormalizations. Furthermore, the existence of localized topological hole states under hole doping is established. As a natural consequence of this local symmetry is broken. It is proven that a quadrupolar-polarization induced attractive hole-hole interaction can exist between such topological hole states. This interaction creates an ordered topological hole structure which leads to a global symmetry breaking. This highly ordered topological hole structure which will be referred as bonded holes (b-holes) can be characterized as parallel one-dimensional electronic states (stripes) along particular ..Cu-O-Cu.. bonding directions which are intrinsically coupled to each other. The *highly ordered* topological b-hole state exists undisturbed for hole concentrations $n_h$ in the range of 0.125 holes/copper $\leq n_h \leq$ 0.25 holes/copper. The total correlation energy per b-hole $|E_{C(tot)}|$ was found to be greater than 160 meV. In addition to b-holes, holes which are not intrinsically bonded exist in the concentration range of $n_h >$ 0.125 holes/copper. These non-bonded holes are termed as free holes (f-holes). The inevitable consequence is an electronic *two fluid* behaviour (b-holes, f-holes) within the hole concentration range of 0.125 holes/copper $< n_h \leq$ 0.25 holes/copper. A comparison with experimental results is given with particular respect to peculiarities of µSR experiments, neutron scattering results (1/8 problem), Hall-effect anomalies and scanning tunneling microscopy (STM). The characteristics of the electronic state deduced by the new methodology described here enable a more in depth understanding of these experimental results.

## I. INTRODUCTION

Since the discovery of high-T$_c$ materials, a wide range of possible scenarios has been proposed to explain the mechanism responsible for superconductivity in these materials. The fundamental assumptions within the particular theoretical approaches are often completely different. In line with classical ideas, it has been proposed that charge carrier phonon mechanisms can induce an electron pairing.[1-3] Several models have also been developed which consider magnetic correlations as the mechanism responsible for superconductivity.[4-13]

Other theories take pure electronic mechanisms into account which result from couplings between the charge carriers and other electronic subsystems.[14-17] The observation of an increased oxygen isotopic effect stimulated scientists to consider the combination of antiferromagnetic correlations with phononic (breathing-, buckling) modes of the atoms within the $CuO_2$ planes.[18,19] Somewhat outside of the common treatments of the problem of high-$T_c$ superconductivity, the possibility of a time-reversal symmetry breaking has been discussed.[20,21] All of these particular theoretical models are so different that there is currently no consensus with regard to the theory of high-$T_c$ superconductivity. However, the first difficulty is not the superconductivity, but the lack of understanding of the various unusual properties within the normal state, i.e. the state above the superconducting transition temperature $T_c$.[10,22] This lack of understanding is mainly due to insufficient knowledge of the electronic state already in the hole undoped but in particular in the hole doped case.

The local density-functional (LDA) approach does not reflect the experimentally proven antiferromagnetic ground state of the undoped materials.[23] However, an antiferromagnetic ground state has been predicted by *ab initio* methods which were applied to small clusters.[24-26] Band structure calculations predict a metallic behaviour for the undoped parent compounds, which is in contrast to the experimentally observed nonmetallic behaviour.[27,28] The existence of an antiferromagnetic state is one explanation for the nonmetallic behaviour of the undoped HTC materials. This aspect has generated many attempts to describe the electronic structure using model Hamiltonians which include the antiferromagnetic correlations formally.[10] The reduced symmetry with respect to the antiferromagnetic state necessarily requires a doubling of the unit cell which leads to a gap within the half filled band. The same phenomena, however, may occur if the doubling of the unit cell is the result of a commensurate charge fluctuation state which has a period of a double lattice spacing. The possibility that such a charge fluctuation state is the result of a complex electronic reorganization leads to the question if there is a common causality between the two phenomena, the antiferromagnetism and the charge fluctuations. This is one important subject that will be discussed in this article.

The characterization of the electronic state of the $CuO_2$ planes under hole doping proves to be particularly difficult. A great uncertainty results from the unknown relation between the absolute electron onsite energy $U$ and the band width t. In a weak coupling limit $U \ll t$, extended holes with a kind of self-trapping by the interior spin frustrations may occur.[5,10] In the opposite case where $U \gg t$, an added hole leads to the reduction of the local order parameter by disordering the spins in the vicinity of the holes.[4,10] A further problem results from the sign of $U$ and a possible intrinsic intersite coupling, $W$. Negative values of $U$ tend to lead to the formation of on-site electron pairs when $W \ll 0$, an intersite attraction is induced as given in the case of bipolarons, for example.[29] In this paper the complex paths of the electronic reorganizations separated after the moments of the Coulomb interactions are analysed and lead to a deeper insight into the causality of the mechanisms of the electronic rearrangements.

In particular, the results of this paper are related to an emerging theoretical concept termed as the formation of *stripes* or *striped phases*. Already at early stages of a theoretical description of the electronic states within the cuprates, it has been suggested that a spatial and temporal disordering of spin and charge states could occur within the copper oxide plane.[30-40] In connection with growing experimental indications, it is assumed that the charges and spins are confined to separate linear regions in the crystal resembling stripes.

Experimentally, elastic and inelastic neutron diffraction measurements (NS, INS) have revealed the existence of a charge and spin order within the $CuO_2$ planes.[41-54] In particular, static and dynamic incommensurate spin fluctuations were found which support a striped phase concept. The magnetic peaks are displaced from the antiferromagnetic wave vector $\left(\frac{1}{2},\frac{1}{2},0\right)$ by $(\pm\varepsilon,0,0)$ and $(0,\pm\varepsilon,0)$. The *incommensurability* $\varepsilon$ includes the so-called 1/8 problem which means that $\varepsilon$ increases nearly linearly with increasing hole density and saturates at $\approx 1/8$, which is simultaneously the nominal hole concentration per copper atom within the $CuO_2$ planes.[49] Additionally, a strong depression of $T_c$ can be observed below this hole concentration. Consequently, it is assumed that in the metallic/superconducting state the holes are bound to separated magnetic domains where freezing into a stationary ordered state is prevented by perpetual quantum fluctuations.[11,55-58] Earlier and recent model calculations[30-40,59-63] which support the possibility of such spin and hole striped states are based on the Hubbard representation[32-40,59,60] or its related $t$-$J$ model[30,31,61-63] of the electronic states. As usual, the important question is the extent to which such model calculations reflect the real situation within the solid. It refers to the general problem of the description of the electronic state limited by in minimum two parameters (hopping and onsite interactions in a one-band model). Additionally, some of the uncertainties of these model calculations result from the fact that Coulomb interactions are neglected or are at least not sufficiently included. That refers to the long-range Coulomb monopole interaction but more decisively to short-range higher order interactions. If considered,[63] the Coulomb interactions are usually assumed as repulsive, which counteract the formation of striped phases. The latter are thought to be the consequence of the antiferromagnetic expulsion of holes leading to hole-rich and hole-free regions.[11-13] It is demonstrated here that the inclusion of higher order Coulomb interactions can cause an intrinsic attractive hole-hole interaction.

Other theoretical approaches start from a local pairing of charge carriers.[29] Consequently, narrow bands and large electron phonon couplings are obtained. The formation of bipolaronic states which undergo a Bose-Einstein condensation below $T_c$ is one of the theoretical approaches within the local pairing conceptions.[3,29] In spite of the fact that several experimental results can be understood on this basis, the problem remains that a large mass renormalization occurs if one starts from local paired fermions within a disordered state.

On the other hand, in the present model of one-dimensionally formed stripes of paired charge carriers that result due to a magnetic proximity effect, the coupling between the stripes is assumed to be negligibly small. Under these conditions, the transition to the superconducting state is thought to be caused by a Josephson pair-tunneling of electrons between stripes.[12] It is possible that this Josephson pair-tunneling, which does not occur in direction of the stripes, could be counter-productive to the formation of the stripes. If intrinsically coupled stripes have to be assumed as a fundamental property, it is very reasonable to suppose that two-dimensional long-range phase coherence (*phase stiffness*) is also related to the coupling mechanism between the stripes. It is clearly shown here that such an intrinsic coupling exists.

This work is based on *ab initio* Hartree-Fock (HF) cluster model calculations. This method avoids the difficulties which are inherent to model calculations with adjusted parameters but carries the risk that some uncertainties arise from the limited cluster. Instead of discussing absolute quantities as it is often done, the work described here concentrates on relative changes. The aim of this work is to give a description of the general structure of the electronic state of the $CuO_2$ planes. It is accomplished by a deductive chain of proof based on appropriately chosen cluster models. The HTC-materials are antiferromagnetically correlated systems. Most of the theoretical approaches concentrate on this particular interaction. However, Löwdin has previously shown that the energetically almost favoured HF state of a multi-electronic system is not necessarily given by a wave function which is a constant of motion of the presupposed Hamiltonian.[65] That means that there may be a large part of the correlation energy which depends on the symmetry constraints. Therefore, an important part of the correlation effects may be comprehended by HF studies with regard to symmetry breaking effects. This is accomplished here with the result that a general idea about the structure of the general ground state can be deduced. It is shown here that within a rigid lattice approach the electronic state is characterized by a globally broken symmetry in both the hole doped and undoped case. In the hole doped case, an ordered state of intrinsically coupled hole stripes is formed.

This paper is organized as follows. In Sec. II, the methodical and theoretical basis is described, which is based on Hartree-Fock cluster calculations and the introduction of the principle of "*detailed momentum balance*". In Sec. III, the electronic state of an undoped $CuO_2$ plane is derived with the result that a charge and bonding fluctuation state (CBF) exists which includes two different copper sites and the dividing of the Brillouin zone by half. It is concluded that the collinear coupling of CBF state and antiferromagnetic spin state is caused by a common causality. The electronic state of a hole doped $CuO_2$ plane is extensively analyzed in Sec. IV. It is shown that a localization of hole density with $q_h = 1$ occurs at particular oxygen atoms with a stabilizing self-energy $E_S$ which implicates a local symmetry breaking. The possibility that attractive couplings of such topological hole states can occur is theoretically proven and results in an additional coupling energy $E_C$. It is shown that the two effects ($E_S$, $E_C$) are caused by quadrupolar-polarizations in relation to the CBF state. Definitive coupling rules of pairs of topological hole states are derived. Taking the results of Secs. III and IV as the basis, the fundamental structure of the electronic state under hole doping is derived in Sec. V. The general influence of long-range Coulomb monopole interactions is considered by analysis of well-chosen topological hole configurations. It is shown that a fundamental structure of attractively bonded hole states (b-holes) exists, and that these b-holes create a global symmetry breaking. Additionally, non-bonded holes (f-holes) can exist depending on the hole concentration and can lead to a two fluid system (b-holes, f-holes). The b-hole structure is a sequence of one-dimensional electronic states (stripes) which are intrinsically coupled by the coupling energy $E_C$. An extensive comparison with a variety of experimental results is given in Sec. VI, which are directly related to the consequences resulting from the electronic structure derived in Sec. V. In conclusion, the experimental findings of μSR experiments, neutron scattering results (1/8 problem), Hall-effect anomalies and scanning tunneling microscopy (STM) are consistently explained from the deduced electronic structure.

## II. THEORETICAL AND METHODOLOGICAL BASIS

The common characteristic of the particular high-$T_c$ superconductors is that they all consist of charged $CuO_2$ planes. Nowadays there is agreement about the mechanism of the transition from the semiconducting to the conducting phase of these substances. This transition is caused by placing holes inside the $CuO_2$ planes. The formation of the superconducting state is also primarily connected with the existence of these holes. Therefore, I will concentrate on the electronic state of the $CuO_2$ planes. In a first approximation the influence of the out of plane constituents on the electronic state of the planes is reduced to Coulomb interactions only because the bond strengths between atoms inside the plane is much higher than to atoms outside of it. Two further presuppositions are made, which are certainly not fulfilled in the high-$T_c$ substances. However, these restrictions do not undermine the deciding conditions for the electronic state in the $CuO_2$ planes. Firstly, orthorhombic distortions will be neglected, which means that the lattice vectors $a$, $b$ are supposed to be identical. All conclusions obtained in the subsequent sections are unrestrictedly true, for the orthorhombic case as well. Merely some assumptions which are related to the condition $a = b$ have to be qualified for the orthorhombic case. Furthermore, Coulomb interactions from charges outside the $CuO_2$ planes being of higher order than zero have been neglected. This restriction is also justified, because the higher order interactions from constituents outside the planes are comparatively small which means that the general symmetry conditions for the electronic states within the planes are essentially unaffected.

For electronic structure calculations restricted Hartree-Fock SCF calculations have been employed. Apart from full *ab initio* calculations using standard basis sets,[66] in most cases effective core potentials have been used where the core potentials and basis sets from Hay and Wadt were adopted.[67] This makes effective cluster calculations possible, for larger clusters as well. The cluster considered in the given work were embedded in an adequate ionic environment that accounts for the potential of the out of plane constituents. This ionic environment has been systematically varied in order to qualify its influence on the electronic state of the $CuO_2$ planes. This work has proven that the ionic environment has no specific influence on the electronic states of the $CuO_2$ planes, i.e. the results of the subsequent sections are universal for all HTC materials. Basis set and pseudopotential dependent effects have been tested extensively in order to rule out some uncertainties connected with this topic.

The key problem in order to reach an understanding of the superconductivity in the HTC materials is the explanation of the unexpected properties of the normal state. It refers to Hall coefficients, d.c. resistivity, tunneling spectroscopy, neutron scatterings, angle resolved photoemission data and much more.[10] The experimental results obtained by these particular methods are hard to explain by a Fermi liquid behaviour of the charge carriers.[4,10,22] Hence, a deciding step towards the explanation of the unusual properties of the normal state is to give a reliable answer to the problem of the existence/non-existence of a two-dimensional Fermi surface in the $CuO_2$ planes on the basis of an ordinary quasi-particle conception. In a mean field representation of the Schrödinger equation as it is given for the HF approximation the Coulomb potential within the Hamiltonian can be represented by:

$$V(\mathbf{r}) = V_{Lat}(\mathbf{r}) + V_{SCF}(\mathbf{r}) \tag{1}$$

where $V_{Lat}(\mathbf{r})$ is the lattice part given by the atomic core charges and $V_{SCF}(\mathbf{r})$ is the self-consistent part which results from the calculated electronic densities. If the rotational and translational symmetry of the self-consistent field potential $V_{SCF}(\mathbf{r})$ is identical with the lattice part $V_{Lat}(\mathbf{r})$ where the latter is assumed to have a regular periodic structure, one can hope that the quantum numbers ($\mathbf{k}$) of the plane wave states which are originally determined by the symmetry of $V_{Lat}(\mathbf{r})$ will be good quantum numbers further on and the quasi-particle conception can be maintained. However, if the symmetry of the self-consistent part $V_{SCF}(\mathbf{r})$ is clearly different from the lattice contribution $V_{Lat}(\mathbf{r})$ by local or/and global symmetry breakings the quantum states will be completely reorganized. Such a behaviour can be caused either by the occurrence of localized states or by a strong collective behaviour of the electronic system. In both cases the Fermi liquid picture is inadmissible.

In the following the potential $V(\mathbf{r}_{Cu})$ with respect to a particular copper place is considered. It can be expanded as a power series at the origin of the copper atom leading to

$$V(\mathbf{r}_{Cu}) = \sum_l V^{(l)}(\mathbf{r}_{Cu}) \tag{2}$$

with $l$ as the order of the Coulomb potential. The total potential $V(\mathbf{r}_{Cu})$ can be further subdivided into an inner atomic part $V_{in}(\mathbf{r}_{Cu})$ which results from the charges on the copper atom itself and an outer ligand field part $V_{out}(\mathbf{r}_{Cu})$ in which all parts from charges surrounding the considered copper atom are collected. The action of the zero order potential part $V_{out}^{(o)}(\mathbf{r}_{Cu})$ on the copper orbital states is given by a uniform energy shift of all orbital states in Fig. 1(b) (monopole interaction). It is assumed that the copper valence states have primarily only $s$ and $d$ character which means that dipolar terms with $l = 1$ can be neglected. The quadrupolar interactions with $l = 2$ ($V_{out}^{(2)}(\mathbf{r}_{Cu})$) leads to a term splitting according to Fig. 1(b) but not to an absolute shift of the orbital energies. Therefore, the response of the total electronic state at the copper atom is completely different with respect to a changed outer monopole potential or variations of the higher order potentials, where the electric filed gradient (EFG) for $l = 2$ is most important. In covalently bonded systems a changed monopole potential at a given atom will lead in the first order merely to a charge transfer between the considered atom and the covalently bonded atoms. The bonding symmetry is not affected in the first order. Contrary to that a changed outer EFG can change directly the relative internal orbital splitting on the copper atom according to Fig. 1(b) and therefore the bonding symmetry. The electronic response happens in such a way that a changed outer EFG will induce a local quadrupole moment at the copper atom. The thereby changed covalent bonds between copper atom and ligands will again change the outer EFG. As a result a new balanced state between the outer EFG and the induced quadrupole moment at the copper atom occurs. In the case of the monopole interaction a balanced state between the outer monopole potential $V_{out}^{(o)}(\mathbf{r}_{Cu})$ and the absolute charge density at the copper atom occurs. In the following, this effect of a momentum separated electronic reorganization will be termed as "detailed momentum balance". A further deciding aspect is given by the spatial dependence of the Coulomb potentials being $1/r$ for the monopole case and $1/r^3$ for the quadrupolar interactions. The latter ($1/r^3$) weights the charges which surround the copper atom in a way that the quadrupolar interactions have a strong local character whereas the influence of the monopole potential is rather global. The whole discussion in the subsequent sections arises from this momentum separated way of looking.

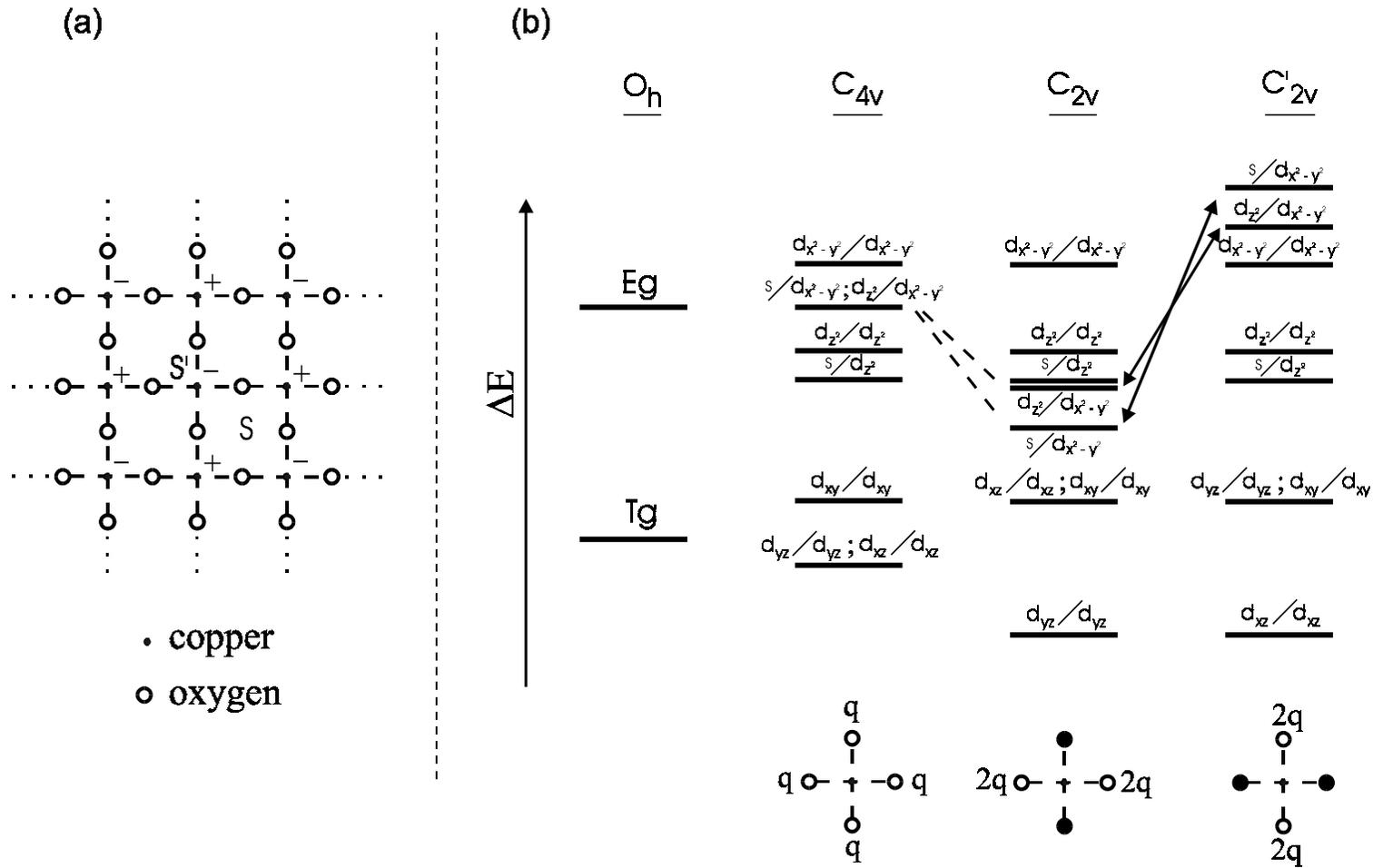

FIG. 1. (a) sector of the CuO$_2$ plane. +,− indicate different copper states/sites corresponding to the charge and bonding fluctuation state (CBF) according to FIG. 2(a). (b) term scheme of the copper orbital states in dependence on the symmetry of the ligand field originating from the surrounding negative oxygen charges $q$ located at the $\overline{Cu\text{-}O}$ bonding distance. The octahedral splitting as well as the off-diagonal terms in the case of the $C_{4v}$ symmetry are arbitrary. The filled circles correspond to hole states, i.e. they are neutral.

In discussing the orbital splitting of the copper states one usually restricts oneself to the diagonal orbital terms. This neglects essential parts which are in relation to the hybridization of the copper orbital states. However, the electronic states in the $CuO_2$ planes are decisively influenced by strong hybridizations. I will show that the strong symmetry dependent off-diagonal terms $d_{z^2}/d_{x^2-y^2}$ and $s/d_{x^2-y^2}$ in Fig. 1(b) are the deciding quantities for the electronic rearrangements under hole doping.

### III. THE UNDOPED ELECTRONIC STATE

In the following the electronic states are considered as a linear combination of atomic basis states (LCAO). A regular two-dimensional structure as given in Fig. 1(a) represents a global $C_{4v}$ point symmetry if a commensurable superstructure (designated by $+,-$) caused by electronic rearrangements is neglected for the moment. In the undoped state, i.e. if identical doubly charged $CuO_2^{2-}$ units exist, every copper atom is placed within a local $C_{4v}$ point symmetry, too. Hence, one can start from symmetry orbitals which have the following common structure with respect to the orbital representation:

$$\psi_i([X]) = \left(...,c_{si},c_{d_{z^2}i},c_{p_xi},c_{p_yi}...\right) \text{ and } \psi_j([Y]) = \left(...,c_{d_{x^2-y^2}j},c_{p_xj},c_{p_yj}...\right), \qquad (3)$$

with the orbital coefficients $c_s$ for the copper $s$ states, $c_{d_{x^2-y^2}}$ and $c_{d_{z^2}}$ for the copper $d_{x^2-y^2}$ und $d_{z^2}$ orbitals and $c_{p_i}$ for the oxygen $p$ orbitals. $X$ and $Y$ are appropriate irreducible representations in dependence on the considered cluster and the chosen symmetry center ($S,S'$ in Fig. 1(a)). The discussion of the electronic states within the framework of separated basis representations as defined above is justified for the moment from subsequent arguments. An outer $C_{4v}$ charge symmetry around the copper atoms within the $CuO_2$ plane gives no reason for asymmetrical tensor components $Q_{xx}$ und $Q_{yy}$ of the electronic quadrupole moment on the copper atoms (detailed momentum balance). Non-disappearing $c_s c_{d_{x^2-y^2}}$ and $c_{d_{z^2}} c_{d_{x^2-y^2}}$ densities, however, create basically asymmetrical tensor components $Q_{xx}$ und $Q_{yy}$. Hence, either symmetry orbitals according to Eq. (3) exist or if the two orbital representations in Eq. (3) are mixed the off-diagonal densities $c_s c_{d_{x^2-y^2}}$ und $c_{d_{z^2}} c_{d_{x^2-y^2}}$ have to be fully compensated for by adding over the particular symmetry orbitals. In addition, a similar argumentation results with respect to the different bonding symmetries of the $s, d_{z^2}$ orbitals and the $d_{x^2-y^2}$ orbitals. The $s, d_{z^2}$ orbital states are without notes within the $x,y$ plane and establish, therefore, the bonding conditions $c_{p_x} \varphi_{p_x} > 0, c_{p_y} \varphi_{p_y} > 0$ or $c_{p_x} \varphi_{p_x} < 0, c_{p_y} \varphi_{p_y} < 0$, where $\varphi_j$ is the oxygen $p$ orbital of the neighboured oxygen atoms which are directed towards the copper atom. On the other hand, the $d_{x^2-y^2}, p_i$ copper oxygen bonds implicate the relations $c_{p_x} \varphi_{p_x} > 0, c_{p_y} \varphi_{p_y} < 0$ or $c_{p_x} \varphi_{p_x} < 0, c_{p_y} \varphi_{p_y} > 0$. Therefore, under the condition of a local $C_{4v}$ point symmetry there is a basic conflict between these different bonding parts if the symmetry orbitals $\psi_i ([X])$ and $\psi_j ([Y])$ are mixed (at least for one-dimensional irreducible representations $X, Y = $ A, B).

For argumentative reasons only, a hypothetical state is assumed for the moment which is characterized by non-oxygen bonding parts. Each copper atom adds 11 valence electrons into the bonding states. The energetically low lying states $d_{xy}$, $d_{yz}$ and $d_{xz}$ (Fig. 1(b)) catch 6 electrons, but they are largely localized and do not contribute essentially to the covalent bonds. They are neglected in afterward discussions. There remain 5 valence electrons, 2 in each of the states $d_{z^2}$ and $d_{x^2-y^2}$ and 1 electron occupies the weakly bound copper $4s$ orbital. This weakly bound $4s$ electron is transferred to the non-bonding oxygen $p$ orbital, i.e. the oxygen $p$ orbital that is perpendicularly oriented with respect to the Cu-O-Cu bonding direction. It forms a lone pair together with one of the oxygen electrons. This tendency is caused by the widely non-bonding character of this oxygen orbital and the large electronegativity of the oxygen atoms. Such a behaviour was basically verified by all subsequent calculations. The remaining 4 electrons will occupy weakly bound states with bonding and antibonding character according to

$$\psi_b = N_b (c_d(1) + c_d(2)) \tag{4}$$

$$\psi_a = N_a (c_d(1) - c_d(2)), \tag{5}$$

whith $c_d$ as the corresponding coefficient with respect to the $d_{z^2}$ and $d_{x^2-y^2}$ states at neighboured copper atoms (1) and (2) as well as $N_a$ and $N_b$ representing the corresponding normalization coefficients. If an admixture of oxygen states into the states of Eqs. (4) and (5) according to Eq. (3) occurs, the bonding structure is decisively changed as a result of the occurrence of Cu-O-Cu bonding sequences. In dependence on the polarity of the oxygen $p$ orbital different copper-oxygen bonding sequences will be created. Within the state of Eq. (4) either a Cu-antibonding-O-bonding-Cu sequence (A,B), or a Cu-bonding-O-antibonding-Cu sequence (B,A) will be created. Equation (5) leads basically to a Cu-bonding-O-bonding-Cu structure, (B,B). The occupation of the orbitals is not changed because the states represented by Eqs. (4) and (5) have closed shell character.

In order to discuss the electronic states a cluster as depicted in Fig. 2(a) is considered firstly. It consists of 4 copper and 4 oxygen atoms of the $CuO_2$ plane forming a cyclic arrangement. This cluster is appropriate to verify the electronic behaviour within the $CuO_2$ plane. The general $C_{4v}$ point symmetry with respect to the symmetry center $S$ is reflected and the cyclic structure of the bonds around $S$ (Fig. 1(a)) creates cyclic symmetry orbitals with periodic boundary conditions. It permits common conclusions with regard to the properties of the electronic state within the $CuO_2$ planes. The realization of the bonding structure according to the Eqs. (4) and (5) leads to four orthonormal symmetry orbitals $\psi_1'$, $\psi_2'$, $\psi_3'$, $\psi_3''$, with the degenerate states $\psi_3^{'}$, $\psi_3^{''}$ (Table I, config. I). Only (B,B)-sequences as discussed above lead to a net Cu-O bonding part. From this point of view the formation of oxygen parts is therefore only useful if it occurs in energetically higher Cu-Cu antibonding states according to Eq. (5). The positions within the particular symmetry orbitals, where the occupation of oxygen states creates net bonding Cu-O parts, has been signed by ($\bullet$) in Table I. The terms A and B in Table I indicate antibonding and bonding states with respect to the Cu-Cu bonding, respectively. This means, the occupation of the oxygen states is determined by the Cu-Cu bonding structure as long as the general Cu-Cu bonding structure is maintained. In order to reach a rather independent occupation of the oxygen states the predetermined Cu-Cu bonding structure has to be changed with the result that the regular Cu-Cu bonding structure is resolved. That can be reached when new states are formed with only every second Cu place being occupied within the given symmetry orbital. In the simplest way that can be reached by forming new states $\psi_1$, $\psi_2$, $\psi_3$, $\psi_4$ by adding and subtracting the initial states $\psi_1^{'}$, $\psi_2^{'}$, $\psi_3^{'}$, $\psi_3^{''}$ (Table I, config. II). Now, oxygen states can be occupied within every symmetry orbital and the copper states are free atomic states, i.e. the constraints caused by the Cu-Cu bondings have been removed. If such new electronic states prove to be energetically favourable some new fundamental properties result.

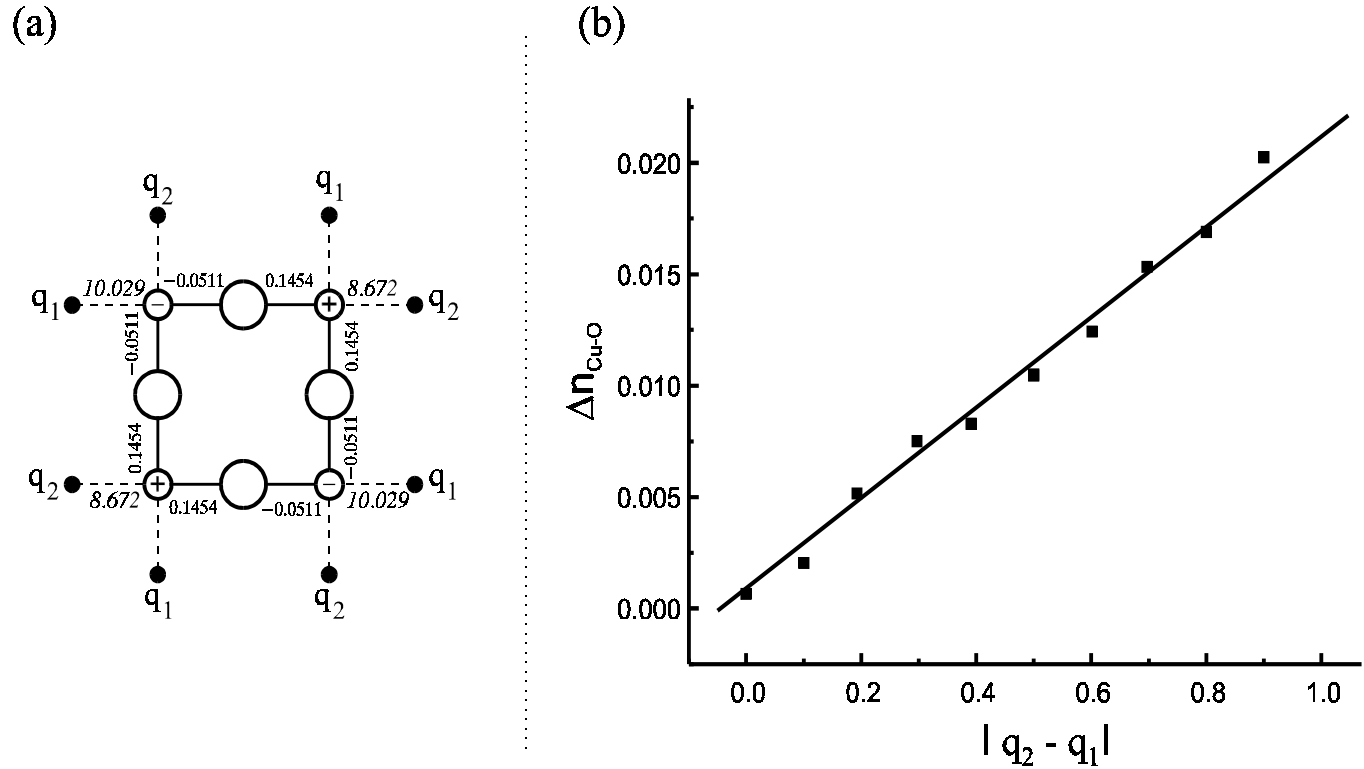

FIG. 2. (a) cluster consisting of 4 Cu and 4 O atoms and additional negative charges $q_1$, $q_2$ located at the $\overline{Cu\text{-}O}$ bonding distance with $(q_1 + q_2)/2 = -1.7 = $ const. The calculations were performed for a neutral cluster. The signs on the copper atoms indicate two different Cu sites corresponding to the electronic CBF state. The + sign corresponds to a deficiency and the − sign to an excess of the local copper electronic densities (italic numbers) in comparison to an equalized charge distribution around the copper atoms ($q_1 = q_2$). The numbers between copper and oxygen atoms stay for the overlap densities ($n_{Cu\text{-}O}$) of the Cu-O bond ($q_1 = q_2$). (b) dependence of the asymmetry of the Cu-O bond in $x$ and $y$ direction ($\Delta n_{Cu\text{-}O}$) of the + copper state from the local outer charge asymmetry corresponding to the cluster of (a).

TABLE I. Generalized representation of different configurations of symmetry orbitals of the cluster in Fig. 2a. Oxygen populations are depicted by full and open circles where full and open circles in $\psi_1$ to $\psi_4$ indicate different populations. +,− indicate the sign of the LCAO coefficients. A, B means antibonding and bonding states with respect to Cu-Cu bonds. The broken line rectangles indicate definite renormalization couplings/influences. The arrows indicate spin states assuming open shell spin orbitals. The orbitals $\psi_{1A}$ to $\psi_{4A}$ have the same electronic structure as the orbitals $\psi_1$ to $\psi_4$ but with antibonding copper oxygen states.

| Cu-orbital | config. | symmetry orbitals | | Cu$_1$ | O$_1$ | Cu$_2$ | O$_2$ | Cu$_3$ | O$_3$ | Cu$_4$ | O$_4$ | Cu$_1$ |
|---|---|---|---|---|---|---|---|---|---|---|---|---|
| $d_{x^2-y^2}$ | I | $\psi_1'$ | | + | B | + | B | + | B | + | B | + |
| | | $\psi_2'$ | | + | A ● | − | A ● | + | A ● | − | A ● | + |
| | | $\psi_3'$ | | + | A ● | − | B | − | A ● | + | B | + |
| | | $\psi_3''$ | | + | B | + | A ● | − | B | − | A ● | + |
| | II | $\psi_1 = \tfrac{1}{2}(\psi_1' + \psi_2')$ | | + ↓ | [●●] | 0 | [●●] | + ↓ | [●●] | 0 | [●●] | + ↓ |
| | | $\psi_2 = -\tfrac{1}{2}(\psi_1' - \psi_2')$ | | 0 | [●●] | − ↑ | [●●] | 0 | [●●] | − ↑ | [●●] | 0 |
| | | $\psi_3 = \tfrac{1}{2}(\psi_3' + \psi_3'')$ | | + | ● | 0 | ● | − | ● | 0 | ● | + |
| | | $\psi_4 = \tfrac{1}{2}(\psi_3' - \psi_3'')$ | | 0 | ● | + | ● | 0 | ● | − | ● | 0 |
| | III | $\psi_1$ | | + ↓ | [○●] | 0 | [○●] | + ↓ | [○●] | 0 | [○●] | + ↓ |
| | | $\psi_2$ | | 0 | [●] | − ↑ | [●] | 0 | [●] | − ↑ | [●] | 0 |
| | | $\psi_3$ | | + | ○ | 0 | ○ | − | ○ | 0 | ○ | + |
| | | $\psi_4$ | | 0 | ● | + | ● | 0 | ● | − | ● | 0 |
| | | $\psi_A^n := \{\psi_{1A}, \psi_{3A}\} \leftarrow$ $\psi_A^{ch} := \{\psi_{2A}, \psi_{4A}\} \leftarrow$ | $\psi_5'$ | | + | | + | | + | | + | |
| | | | $\psi_6'$ | | − | | + | | − | | + | |
| | | | $\psi_7'$ | | − | | − | | + | | + | |
| | | | $\psi_7''$ | | + | | − | | − | | + | |

If one starts from the common case that the oxygen states within the symmetry orbitals $\psi_1$, $\psi_2$, $\psi_3$, $\psi_4$ are differently occupied (Table I, config. III), then all these states are energetically non-degenerate. That means that the symmetry orbitals $\psi_1$, $\psi_2$, $\psi_3$, $\psi_4$ belong to the symmetry group $C_{2v}$ and not any longer to the original point group $C_{4v}$. This symmetry reduction is not a property of a particular cluster but is a general property of the electronic system caused by the renormalization of the electronic states. Therefore, it is a general property of the electronic state of the $CuO_2$ planes within the solid. The division by half of the Brillouin-zone with respect to the corresponding bands which are related to the states $\psi_1$, $\psi_2$, $\psi_3$, $\psi_4$ is a direct consequence by these renormalizations. One has to start from at least two correlated bands which are related to the states $\psi_1, \psi_3$ and $\psi_2, \psi_4$, respectively, having the lattice period $2a, 2b$. The performed HF calculations of the above cluster verify basically the formation of non-degenerate states with alternating copper occupation densities similar to the states $\psi_1$, $\psi_2$, $\psi_3$, $\psi_4$. When generalizing the results of all performed calculations one can conclude that all symmetric clusters with an even number of copper atoms reveal the formation of superstructures with twice the lattice period $(2a, 2b)$. However, the question remains to what extent a reduction of orbital and bonding symmetry has consequences for the reduction of the total point group symmetry from $C_{4v}$ to $C_{2v}$. The calculations of the cluster in Fig. 2(a) as well as all other cluster calculations performed in this work reflect distinctly the formation of a charge fluctuation state with respect to the copper atoms, which has the lattice period $2a, 2b$ corresponding to a $C_{2v}$ point group symmetry with respect to the symmetry center $S$ (Figs.1(a) and 2(a)). This means, the self-consistent rearrangements of the electronic states $\psi_1$, $\psi_2$, $\psi_3$, $\psi_4$ create commensurable charge density fluctuations with the period $2a, 2b$, i.e. a highly correlated collective electronic state is formed. This fluctuation state is decisively determined by different occupation of the oxygen states in $\psi_1, \psi_3$ in comparison to $\psi_2, \psi_4$, signed by the filled and open circles of config. III in Table I and by the exclusive occupation of only one class of antibonding orbital states ($\psi_A^n$) in the case of a neutral $Cu_4O_4$ cluster (see below). This implies that a remarkable asymmetry within the Cu-O bonds is formed. From Fig. 2(a), it is obvious that the net population of the Cu-O bonds is different for copper atoms which are placed at different diagonals. Two of the diagonally positioned copper atoms create a net bonding state while for the other two copper atoms even a net antibonding population results. Further cluster calculations have confirmed this result. It is therefore conclusive that the charge fluctuation state with the period $2a,2b$ is causally related to a bonding/antibonding fluctuation state in the above sense. This charge and bonding fluctuation state will be termed as CBF state in the following.

The formation of charge density fluctuations includes changes of the long-range Coulomb interactions (monopole interactions) which are quantitatively insufficiently comprehended in cluster calculations. However, it can be verified that the above deduced commensurable charge density fluctuation state will also lead to an energy lowering as a consequence of the changed long-range Coulomb interactions. For this reason atomic HF calculations have been performed in dependence on the copper valence state and with varying the outer Coulomb monopole potential.

The separated influence of monopole potentials could be investigated by placing additional charges far away from the considered atom. It can clearly be shown (Table II) that the dependence of the electronic energy of the copper atom on the outer monopole potential $\phi_o$ is strictly linear

$$E_{Cu} = \alpha\,\phi_o + a \;. \tag{6}$$

This behaviour was expected because the pure monopole effect causes only an alteration of the effective nuclear charge. However, further decisive facts can be outlined. The slope $\alpha$ of the straight line as well as the constant term a change with different electronic occupation numbers $n$. It is obvious that the slope is independent of the short-range higher order distortions (quadrupolar interactions) (Table II) where the influence of the latter is manifested only by differences of the constant term a which are comprehended in cluster calculations. Therefore, for the electronic energy $E_{Cu}$ the functional

$$E_{Cu}(n,\phi_o,\phi_i) = \alpha(n)\phi_o + a(n,\phi_i) \tag{7}$$

results where $n$ is the occupation number, $\phi_o$ the monopole potential and $\phi_i$ are higher multipole potentials (quadrupolar potential). In the solid, the constant term a will be additionally and decisively determined by the actual bonding situation which is comprehended by cluster calculations.

TABLE II. Relative dependence of the atomic copper electronic energies on the outer monopole potential $\phi_o$ [a.u.];$E=\alpha\cdot\phi_o+a$ [a.u.], (PA). $U$ according to Eq. (9) reflects the influence of the higher order coulomb interactions. The charges $q$ are placed at the corresponding oxygen positions according to Fig. 1(b) ($C_{4v}$-symmetry).

| coefficients | number of valence electrons | | | $q$ | $U$ [a.u.] |
| --- | --- | --- | --- | --- | --- |
| | 10 | 9.5 | 9 | | |
| $\alpha$ | -9.357 | -8.889 | -8.421 | 0 | 0.073 |
| $a$ | -45.417 | -45.100 | -44.710 | | |
| $\alpha$ | -9.357 | -8.889 | -8.421 | -1.4 | 0.043 |
| $a$ | -44.614 | -44.268 | -43.879 | | |
| $\alpha$ | -9.357 | -8.889 | -8.421 | -1.7 | 0.051 |
| $a$ | -44.446 | -44.099 | -43.701 | | |
| $\alpha$ | -9.357 | -8.889 | -8.421 | -2.0 | 0.098 |
| $a$ | -44.280 | -43.961 | -43.544 | | |

The remaining problem is the changed monopole potential around a given copper place resulting from the charge fluctuation state with the period $2a,2b$. For those copper atoms, which have an increased electronic density an increase of the outer potential $\phi_o$ will occur. It is caused by the potential of the neighbouring copper atoms having now a higher effective positive charge. As a consequence, the electronic energy is relatively decreased for the considered copper atom. In the opposite case relatively increased electronic energies can be found. Including these additional differences $\left|\Delta\phi_o\right|$ of the monopole potentials one can define an effective correlation energy according to

$$U_{eff}(\delta n,\phi_o,\phi_i) = [\alpha(\overline{n}+\delta n) - \alpha(\overline{n}-\delta n)] \cdot \left|\Delta\phi_o\right| + U(\delta n,\phi_i) . \tag{8}$$

with

$$U(\delta n,\phi_i) = E_{Cu}(\overline{n}+\delta n,\phi_i) + E_{Cu}(\overline{n}-\delta n,\phi_i) - 2 \cdot E_{Cu}(\overline{n},\phi_i) \tag{9}$$

representing the correlation energy caused by all Coulomb short-range and bonding interactions within the solid which are comprehended in cluster calculations and which favourite the CBF state ($U$ is not identical with the atomic case in Table II !). $\overline{n}(\approx 9.5)$ is the average copper occupation number. According to Table II, the first term in Eq. (8) is always negative. Hence, the changed long-range Coulomb interactions which are the consequence of commensurate charge fluctuations ($2a,2b$) around the copper atoms support the formation of the CBF state.

When calculations under symmetry restricted conditions ($C_{4v}$ symmetry) are carried out, an equalization of the electronic densities at the copper atoms and crystal symmetry adapted bonding states can be forced. In this case, a shift to higher energies lying in the range of 4.4 - 5.4 eV is obtained for the cluster in Fig. 2(a), which depends on the given configuration of the outer charges. These nominal energy shifts of 1.1 - 1.35 eV per copper atom are high enough to conclude that from the hitherto considerations a charge and bonding fluctuation state is favoured leading to a self consistently formed superstructure within the solid with a commensurate period $2a,2b$.

Another important aspect is the renormalization of the electronic states as reported above that leads to symmetry orbitals with the property that a remarkable population of only every second copper atom is observed. Therefore, there is a negligibly small covalent coupling between the populated copper atoms within every symmetry orbital, i.e. the copper states have a strong local character. Hence, one can assume that the electronic states centered at the particular copper atoms in $\psi_1,\psi_3$ and $\psi_2,\psi_4$ (i.e. at $Cu_1$ and $Cu_3$ in $\psi_1,\psi_3$ as well as $Cu_2$ and $Cu_4$ in $\psi_2,\psi_4$) are orthogonal to each other. In this case, the alignment of the spin states at $Cu_1$ and $Cu_3$ as well as $Cu_2$ and $Cu_4$ can become parallel to each other (if open-shell spin orbitals are assumed). A non-negligible antibonding Cu-Cu overlap results from the neighboured copper atoms within $\psi_1$ und $\psi_2$ in Table II. That means that the orthogonality between $\psi_1$ und $\psi_2$ is lost. In order to renormalize the states $\psi_1$ und $\psi_2$ the antibonding parts of the copper overlap have to be compensated for by additional parts of Cu-O bonding overlap. In this way the copper and oxygen states in $\psi_1$ und $\psi_2$ are coupled by these renormalizations. The short-distant Cu-O bonding interaction will energetically dominate the long distant Cu-Cu antibonding interaction. Therefore, this renormalized Cu-O bonding overlap causes an antiparallel orientation of the spin states at neighboured copper atoms which belong alternately to $\psi_1$ und $\psi_2$ (Table I). If one further assumes the spin orbit interaction within copper $d_{x^2-y^2}$ orbital to be sufficient to rotate the spins at the particular copper atoms into the $CuO_2$-plane the experimentally observed two-dimensionally antiferromagnetic order will be given in a natural way. In this case the antiferromagnetic spin state results from the same electronic renormalizations of the orbital states which are responsible for the formation of the CBF state. That means the CBF state and antiferromagnetism are two phenomena of the same causality. In other words, the CBF state and the antiferromagnetic order should be collinearly coupled. The correctness of this scenario can not be directly proven from the existing closed shell calculations, but open shell HF calculations should offer the possibility for a direct verification of the above conclusions. Corresponding model calculations have to go beyond the existing *ab initio* calculations as recently reported [68] where influences of symmetry breaking are obviously not comprehended.

The hitherto considerations were reflected on bonding states with copper $d_{x^2-y^2}$ orbitals. An equivalent argumentation can be pursued with respect to the copper $d_{z^2}$ orbitals. However, there are some peculiarities. The $d_{z^2}$ orbitals overlap only slightly with the oxygen $p_x$, $p_y$ orbitals. For this reason alone one has to include $d_{z^2} \rightarrow s$ excitations which increase the overlap population with the oxygen $p$ orbitals. Additionally, the $d_{z^2} \rightarrow s$ excitations result already from the ligand field splittings according to Fig. 1(b). All cluster calculations have proven the existence of remarkable copper orbital parts with *s*-symmetry. It has to be pointed out that these *s* contributions are not identical with atomic copper 4*s*-states which are energetically high. The problem with respect to the spin states of these orbital states is equivalent as already discussed for $d_{x^2-y^2}$ orbitals, however, the spins are not definitely polarized within a fixed plane due to strong $s,d_{z^2}$-hybridizations. Hence, a definite antiferromagnetic order in two dimensions should not be expected for these orbital states.

The electronic states discussed above were causally related to the copper orbital states. Four valence electrons remain which are supplied by the four oxygen atoms. In considering pure oxygen states these electrons would occupy symmetry orbitals which are given by the original states $\psi_5'$ to $\psi_7''$ in Table I. However, these states must again be renormalized to the already existing orbitals with broken symmetry. In consequence, orbital states $\psi_{1A}, \psi_{3A}, \psi_{2A}, \psi_{4A}$ are formed which possess the same general structure as given by the orbital states $\psi_1, \psi_3, \psi_2, \psi_4$ except with an antibonding character of the Cu-O bonds. In the case of a neutral $Cu_4O_4$ cluster, only these four oxygen electrons have to be placed within the states $\psi_{1A}$ to $\psi_{4A}$. They occupy for example two phase correlated states termed in Table I as $\psi_A^n$, without loss of generality with respect to the absolute phase. This creates a charge fluctuation state as shown in Fig. 2(a) with an accumulation of electron density at those copper atoms which are populated within the states $\psi_A^n$.

In the case of an undoped doubly negatively charged $CuO_2^{2-}$ plane the two extra electrons are consumed to form lone pair states at both oxygen atoms. The valence electrons originally occupying the copper 4*s* orbitals remain. Therefore, 4 additional valence electrons have to be placed within the cluster of Fig. 2(a). These valence electrons occupy the currently unpopulated states $\psi_A^{ch}$ in Table I. In this case, only the Cu(+) sites in Fig. 2(a) will be populated. This is generally valid, because the more positive coulomb potential at the Cu(+) sites leads basically to a capturing of electrons. In consequence, these original copper 4*s* valence electrons occupy also electronic states with only every second copper place being populated. This fact holds true for all electronic valence states as they all possess renormalized electronic structures as given in Table I. In other words, electronic states are formed which lead basically to fully occupied bands, resulting in the isolating behaviour of the undoped $CuO_2^{2-}$ plane. At this place it seems me important to remark that the explicit inclusion of antiferromagnetic couplings is not necessary in order to explain the isolating behaviour of the $CuO_2^{2-}$ planes.

The cyclic structure of the cluster in Fig. 2(a) implicates equivalent Cu-O bonds in the $x$ and $y$ directions if a symmetric outer charge distribution is given ($q_1 = q_2$). It reflects the situation for both the neutral and the undisturbed $CuO_2^{2-}$ plane, which give no reasons that the local $C_{4v}$ crystal symmetry may be broken (assuming a tetragonal structure). From this point the local orbital states are forced to preserve irreducible representations corresponding to the $C_{4v}$ point group. However, there is a strong tendency to form a CBF state on the basis of renormalized electronic states as given in Table I. Therefore, there is always the tendency to form orbital states/bands which are phase related to one copper site ($+$ or $-$ in Fig. 1(a)) and with a population of every second copper atom only. Contrary to the electronic state of a charged $(Cu_4O_4)^{4-}$ cluster as discussed above one additional electron per copper atom has to be placed into the electronic states of an undisturbed $CuO_2^{2-}$ plane. This electron is supported by the unpaired valence electron of the additional oxygen atom. These electrons occupy again phase correlated states with a general structure as given in Table I. In consequence, two additional electrons are centered at every second copper place. Therefore, a CBF state is basically maintained within the $CuO_2^{2-}$ plane. Then, at a given $Cu(-)$ site, the four assigned oxygen valence electrons have to occupy two degenerate states in order to realize identical Cu-O bonding strength in the $a$ and $b$ directions. This is possible if the local $d_{x^2-y^2}$ states are mixed with higher energetic s states. Even the existence of a local $C_{4v}$ symmetry does not preclude this, i.e. mixing the two states in Eq. (3). Adding over these two degenerate orbital states cancels the accompanied non-zero off-diagonal densities $c_s c_{d_{x^2-y^2}}$ and $c_{d_{z^2}} c_{d_{x^2-y^2}}$. This gives rise to the formation of two degenerate bands which are polarized along the coordinate axes ($a$ and $b$ respectively) and are highest in energy. However, the band states apply to the lattice periods $2a$ and $2b$, which on the other hand result from a highly correlated complex electronic behaviour.

# IV. THE HOLE DOPED ELECTRONIC STATE

## A. Breaking of the local symmetry

Now, the reorganization of the electronic state will be considered if particular electrons are removed from the $CuO_2^{2-}$ plane. This problem can not reliable be answered on the basis of limited cluster calculations, because a complex reorganization of all orbital states has to be expected. However, there are bonding theoretical arguments which allow a well-founded decision in which way the formation of holes occurs.

It seems very reasonable to assume that hole doping occurs under the premise that the general structure of the renormalized orbital states as given in Table I are maintained. Otherwise an additional increase in energy would be expected. If the holes are primarily placed on the copper orbital states, the renormalized quantum structures given in Table I could not be created. This would strongly disturb the CBF state with the periodicity $2a$, $2b$. In contrast, the general structure of the renormalized orbital states as given in Table I can be maintained if the doped holes occur at the oxygen site. Therefore, these arguments point to a doping of holes primarily at the oxygen atoms. In addition, it has to be expected from the discussions above that the holes will be doped at the oxygen sites within the energetically highest, two times degenerate and perpendicularly polarized bands with pronounced $d_{x^2-y^2}$,s hybridizations and antibonding parts of the copper-oxygen bonds.

It was experimentally proven by X-ray absorption studies [69,70], XPS [71,72] and high-energy electron energy-loss spectroscopy [73] that holes are clearly created at the oxygen site of the superconducting cuprates. Hence, in the following it will be supposed that the hole states have predominantly oxygen character, the total population on the copper atoms remains widely unchanged and the CBF state is not destroyed by hole doping.

Two further problems arise with respect to the doping of holes: a) Is the doped hole state widely spread over the $CuO_2$ plane or rather localized at particular oxygen atoms thereby leading to local symmetry breakings? b) If localized hole states are very probable the question arises if a breaking of the global symmetry can occur. At first, the problem of a possible local symmetry breaking will be considered. For this reason, the electronic state of the cluster in Fig. 2(a) has been calculated with the assumption of different charges $q_1$ and $q_2$ ($q_1 \neq q_2$) but with the restriction $q_1 + q_2 = const..$ Under these conditions the monopole interactions will basically be unchanged if $q_1$ and $q_2$ are varied. However, the higher order interactions/excitations (quadrupolar excitations) are continuously changed if $q_1$ and $q_2$ are varied, according to Fig. 1(b).

Basically, the transition from the symmetric case ($q_1 = q_2$) to the asymmetric case ($q_1 \neq q_2$) decreases the electronic energy exponentially while increasing the charge asymmetry ($|q_1 - q_2|$) (Fig. 3). That means, the electronic structure will be more stable the higher the local asymmetry of the charge distribution around the copper atoms becomes. The outer charges were chosen to be $(q_1 + q_2)/2 = -1.7$, which nearly guarantees the same Coulomb influence as it is produced by the Coulomb potentials of the oxygen atoms inside the cluster.

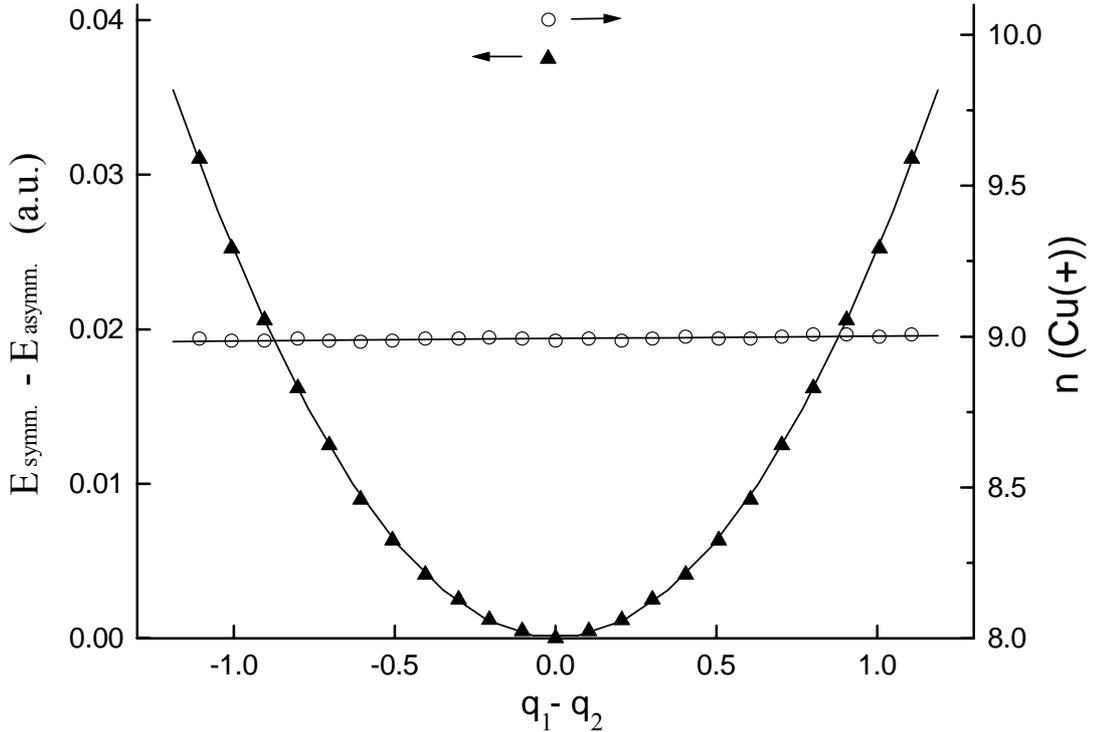

FIG. 3. Decrease in energy of the cluster from FIG. 2(a) when increasing the local asymmetry of the outer charges ($q_1$- $q_2$) with $(q_1 + q_2)/2 = -1.7 = const.$. $E_{symm.}$ corresponds to the symmetric ($q_1 = q_2$) and $E_{asymm.}$ to the asymmetric ($q_1 \neq q_2$) case, respectively. The electronic population number $n$ at the copper site corresponds to the + copper state. The mean population number is near 9.5. The calculations have been performed in pseudopotential approximation (PA).

The symmetry dependent alterations of the electronic energies can be understood on the basis of quadrupolar ligand field potentials. According to the discussions in Sec. II, the electronic distribution inside the atom will change in a way that a balanced state within the momentum representations occurs (detailed momentum balance). The monopole interaction is unchanged if $q_1$ and $q_2$ are varied. Hence, it remains to consider a balanced state between the changed EFG ($q_1 \neq q_2 \rightarrow V_{xx} \neq V_{yy}$) and the total electronic quadrupole moment at the copper atoms. In other words, an anisotropic outer charge distribution ($q_1 \neq q_2$) creates anisotropic EFG tensor copmponents at the copper atom ($V_{xx} \neq V_{yy}$) and consequently a planar electronic quadrupole moment ($Q_{xx} \neq Q_{yy}$) within the electronic states at the copper atom will be induced. Responsible for this inductive mechanism are the off-diagonal terms $d_{z^2}/d_{x^2-y^2}$ and $s/d_{x^2-y^2}$, because these are the only terms that are changed in going from the symmetric case ($q_1 = q_2$) to the asymmetric case ($q_1 \neq q_2$), Fig. 1(b). The net population of these off-diagonal terms is zero in the symmetric case. In the asymmetric case one has to suppose that these off-diagonal terms will be differently populated leading to a non-zero net population (original structures as $\psi_i$ and $\psi_j$ in Eq. (3) may be mixed). These terms may be positively or negatively populated. If the changed off-diagonal densities $\Delta\left(c_s c_{d_{x^2-y^2}}\right)$ and $\Delta\left(c_{d_{z^2}} c_{d_{x^2-y^2}}\right)$ are positive an increased density in $x$ direction and a decreased density in $y$ direction will occur. If they are negative an increased density in $y$ direction and a decreased density in $x$ direction will result. The total charge density remains unchanged. The latter behaviour is reflected by the unchanged copper populations in Fig. 3 for different values of $|q_1 - q_2|$. The different changes of the electronic densities in $x$ and $y$ direction can be visualized by the pictograms in Fig. 4(a). These pictograms will be used afterwards in the discussion of the various electronic reorganizations. Different electronic populations in $x$ and $y$ direction at the copper atoms create different Cu-O overlap populations in $x$ and $y$ direction. Hence, the difference in the total bonding strength in $x$ and $y$ direction is a measure of the above mechanism. Fig. 2(b) verifies this direct correlation between the asymmetry of the outer charges and the bonding asymmetry.

The sign of the energy shift as the result of the quadrupolar polarizations depends on the sign of $\Delta\left(c_s c_{d_{x^2-y^2}}\right)$ and $\Delta\left(c_{d_{z^2}} c_{d_{x^2-y^2}}\right)$ and the sign as well as the relative quantity of the EFG tensor components $V_{xx}$ and $V_{yy}$. If a hole is placed either at one of the four oxygen atoms surrounding a copper atom or at two oxygen atoms being in trans-position (Figs. 4(b),(3),(3') and (4),(4')) an energy shift of the off-diagonal terms $d_{z^2}/d_{x^2-y^2}$ and $s/d_{x^2-y^2}$ occurs in dependence on the position of the hole state ($x$ or $y$ directed). Taking the energy shifts according to Fig. 1(b) and the pictograms of Fig. 4(a), four possible coupling states result, as shown in Fig. 4(c). Only the two lower states in Fig. 4(c) reveal an energy lowering and will occur within the ground state.

If the off-diagonal densities $\Delta\left(c_s c_{d_{x^2-y^2}}\right)$ and $\Delta\left(c_{d_{z^2}} c_{d_{x^2-y^2}}\right)$ are not zero ($\Delta\left(c_s c_{d_{x^2-y^2}}\right) = c_s c_{d_{x^2-y^2}}$ and $\Delta\left(c_{d_{z^2}} c_{d_{x^2-y^2}}\right) = c_{d_{z^2}} c_{d_{x^2-y^2}}$ for a high symmetric tetragonal structure) the total energy shifts corresponding to a particular copper atom will be

$$\Delta\left(c_s c_{d_{x^2-y^2}}\right) \cdot \left(\Delta E_{S(s/d_{x^2-y^2})} (\pm) \Delta E_{s/d_{x^2-y^2}}\right) \tag{9}$$

and

$$\Delta\left(c_{d_{z^2}} c_{d_{x^2-y^2}}\right) \cdot \left(\Delta E_{S(d_{z^2}/d_{x^2-y^2})} (\pm) \Delta E_{d_{z^2}/d_{x^2-y^2}}\right) \tag{10}$$

with $(\pm)\Delta E_{s/d_{x^2-y^2}}$ and $(\pm)\Delta E_{d_{z^2}/d_{x^2-y^2}}$ being the real quadrupolar energy shifts according to Fig. 1(b) when hole doping leads to configurations as depicted in Figs. 4(b),(3),(3') and (4),(4'). $\Delta E_{S(s/d_{x^2-y^2})}$ and $\Delta E_{S(d_{z^2}/d_{x^2-y^2})}$ represent additionally induced energy contributions to the undisturbed symmetric case (symmetry adapted for a non-tetragonal structure) which are the consequence of non-zero densities $\Delta\left(c_s c_{d_{x^2-y^2}}\right)$ and $\Delta\left(c_{d_{z^2}} c_{d_{x^2-y^2}}\right)$, respectively. Consequently, the appearance of the terms

$$\Delta\left(c_s c_{d_{x^2-y^2}}\right) \cdot \left(\Delta E_{S(s/d_{x^2-y^2})}\right) \tag{11}$$

and

$$\Delta\left(c_{d_{z^2}} c_{d_{x^2-y^2}}\right) \cdot \left(\Delta E_{S(d_{z^2}/d_{x^2-y^2})}\right) \tag{12}$$

reflect additional contributions with respect to the symmetric case leading to an increase in energy since the symmetric case represents an energetically stable equilibrium configuration with $\Delta\left(c_s c_{d_{x^2-y^2}}\right) = 0$ and $\Delta\left(c_{d_{z^2}} c_{d_{x^2-y^2}}\right) = 0$. The emergence of non-zero off-diagonal populations $\Delta\left(c_s c_{d_{x^2-y^2}}\right)$ and $\Delta\left(c_{d_{z^2}} c_{d_{x^2-y^2}}\right)$ lead to a renormalization of all orbital populations from their stable equilibrium values with respect to the symmetric case. In order to minimize these deviations the electronic system will react in a way that the totality of the off-diagonal densities in the CuO$_2$ plane vanishes as far as possible. If for instance $\Delta\left(c_s c_{d_{x^2-y^2}}\right)$ is non-zero at a given copper atom the renormalizations to the next neighbour copper atoms will be disturbed (e.g. for the states $\psi_1$ and $\psi_2$ in Table I). This can be reversed when a density with opposite sign $-\Delta\left(c_s c_{d_{x^2-y^2}}\right)$ is induced at a next neighbour copper atom resulting in the cancellation of the corresponding terms given by Eq. (11). The same results for the influence of the local coulomb monopole potential at a given copper atom. This potential is independent of the symmetry of the surrounding charge distribution and corresponds therefore to the symmetric case. Hence, the occurrence of non-zero densities $\Delta\left(c_s c_{d_{x^2-y^2}}\right)$, $\Delta\left(c_{d_{z^2}} c_{d_{x^2-y^2}}\right)$ reflects a de-balanced state with respect to the monopole potential. Oppositely induced densities $-\Delta\left(c_s c_{d_{x^2-y^2}}\right)$ and $-\Delta\left(c_{d_{z^2}} c_{d_{x^2-y^2}}\right)$ at a neighbouring copper atom will again correct the energetic deviations. Therefore, it can be concluded that the hole induced densities $\Delta\left(c_s c_{d_{x^2-y^2}}\right)$ and $\Delta\left(c_{d_{z^2}} c_{d_{x^2-y^2}}\right)$ at a given copper site lead to additional induced densities $-\Delta\left(c_s c_{d_{x^2-y^2}}\right)$ and $-\Delta\left(c_{d_{z^2}} c_{d_{x^2-y^2}}\right)$ at a preferably next neighbour copper site. In this way a dramatic renormalization of the quantum states in relation to the symmetric case will be prevented which minimizes the corresponding terms given by Eqs. (11) and (12). That means, the electronic system will be reorganized in a way that the sum of the off-diagonal densities over all copper atoms within the CuO$_2$ plane will reach its smallest possible value, i. e. it approaches zero:

$$\sum \Delta\left(c_s c_{d_{x^2-y^2}}\right) \to 0 \tag{13}$$

and

$$\sum \Delta\left(c_{d_{z^2}} c_{d_{x^2-y^2}}\right) \to 0. \tag{14}$$

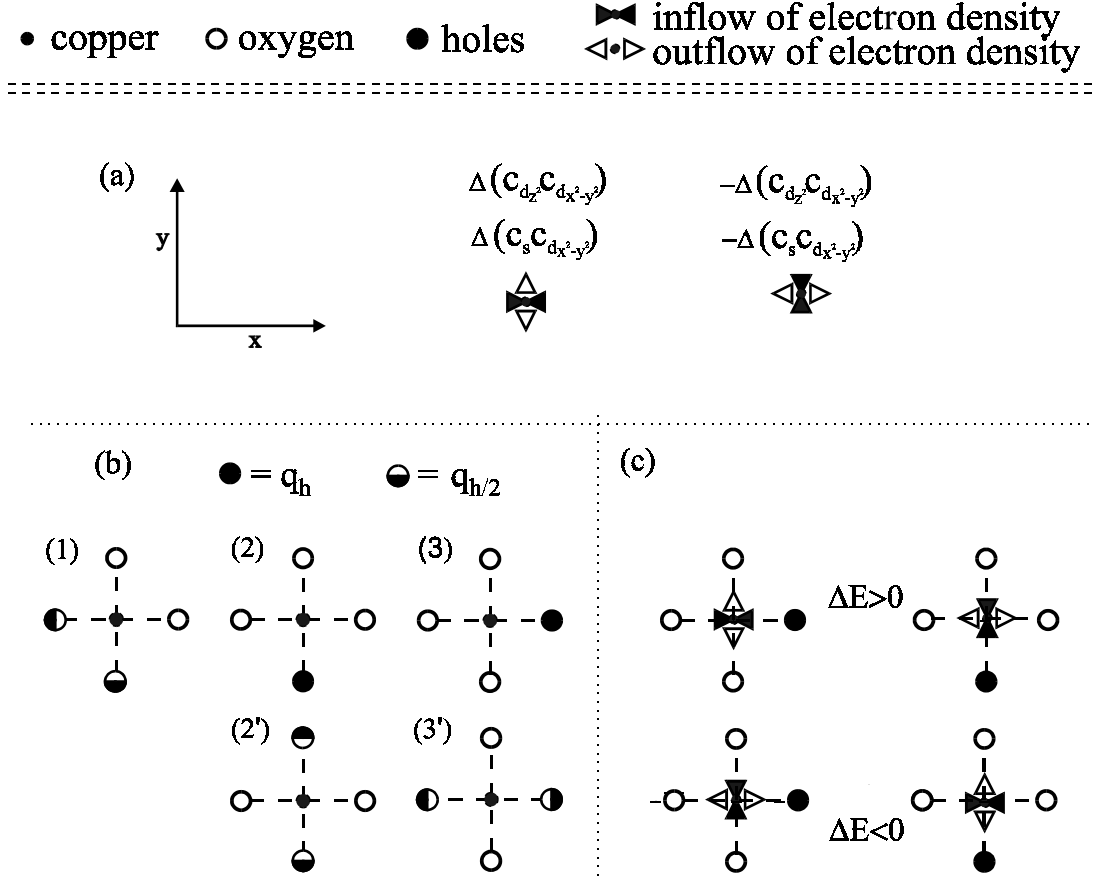

FIG. 4. (a) non-zero off-diagonal densities $\Delta\left(c_s c_{d_{x^2-y^2}}\right)$ and $\Delta\left(c_{d_{z^2}} c_{d_{x^2-y^2}}\right)$ lead to a de-balance of the electron density in $x$ and $y$ direction. If the off-diagonal densities are positive (left pictogram) the total electron density increase along the $x$ coordinate and decrease along the $y$ coordinate. In the opposite case, if the off-diagonal densities are negative a decrease along the $x$ coordinate and an increase along the $y$ coordinate occurs for the electron densities (right pictogram). (b) different configurations of the hole distribution on oxygen atoms with respect to a given copper atom ($q_h$ - hole charge): The symmetric case (1) describes an equal distribution of hole density with respect to the two bonding axes which creates symmetric EFG components, $V_{xx} = V_{yy}$. The asymmetric cases (2),(3),(2'),(3') create $V_{xx} \neq V_{yy}$. The asymmetric case with distributed hole density on the trans-positioned oxygen atoms (2'),(3') creates the same EFG at the included copper atom as in the case of the hole localization in (2),(3). (c) non-zero off diagonal densities $\Delta\left(c_s c_{d_{z^2}}\right)$ and $\Delta\left(c_{d_{z^2}} c_{d_{x^2-y^2}}\right)$ lead to increased or lowered quadrupolar-polarization induced electronic energies $\Delta E$ in dependence on the given hole position.

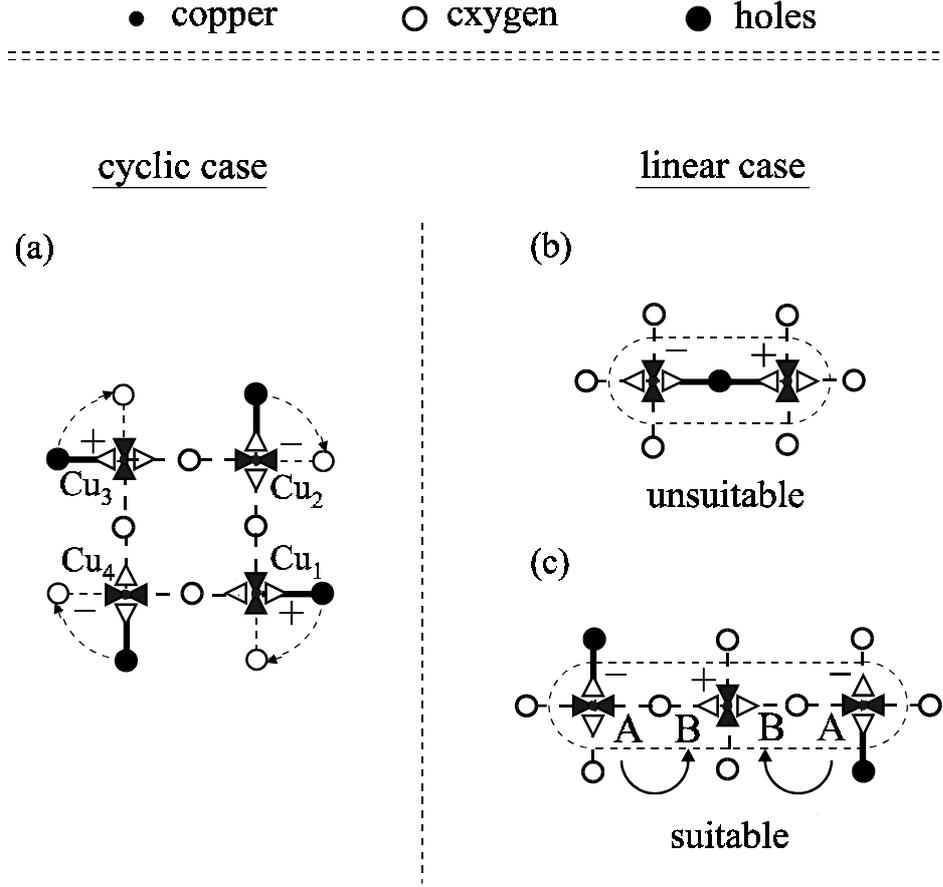

FIG. 5. (a) pictogram representation of the electronic ground state of the cluster of Fig. 2(a) according to Figs. 4(a) and (c). (b) energetic unsuitable electronic configuration within a linear hole arrangement (singular hole). (c) energetic suitable electronic configuration within a linear hole-hole arrangement. A,B means antibonding and bonding states of the Cu-O bonds, respectively. The arrows indicate a transfer of antibonding states.

If the cluster shown in Fig. 2(a) is considered the pictogram of Fig. 5(a) results for the ground state (adopting the rules of Fig. 4(c)). In this case, the pictogram representations according to Fig. 4(a) are alternating from copper atom to copper atom. It is equivalent with an alternation of the sign of $\Delta\left(c_s c_{d_{x^2-y^2}}\right)$ and $\Delta\left(c_{d_{z^2}} c_{d_{x^2-y^2}}\right)$. Therefore, the terms according to Eqs. (11),(12) can be cancelled when adding the terms around the copper atoms so that the possibility of a quenching of the off-diagonal densities according to the Eqs. 13,14 is given in a natural way. That means that in a good approximation the energy dependence according to Fig. 3 reflects only the energy lowering which is the result of the reduction of the local charge symmetry around the copper atoms. In restricting to small energy changes one can suppose that $\left|\Delta E_{s/d_{x^2-y^2}}\right| \sim |q_1 - q_2|$ and $\left|\Delta E_{d_{z^2}/d_{x^2-y^2}}\right| \sim |q_1 - q_2|$ according to the ligand field splittings given in Fig. 1(b) . The same argument can be used for the induced off-diagonal densities $\Delta\left(c_s c_{d_{x^2-y^2}}\right)$ and $\Delta\left(c_{d_{z^2}} c_{d_{x^2-y^2}}\right)$ which can also be assumed to depend linearly on the orbital energies $\left|\Delta\left(c_s c_{d_{x^2-y^2}}\right)\right| \sim \left|\Delta E_{s/d_{x^2-y^2}}\right|$ and $\left|\Delta\left(c_{d_{z^2}} c_{d_{x^2-y^2}}\right)\right| \sim \left|\Delta E_{d_{z^2}/d_{x^2-y^2}}\right|$, respectively. In

consequence , $\Delta E \sim |q_1 - q_2|^2$ results for the energy variation which is in complete accordance with Fig. 3.

Over and above, it means that the energy lowering given in Fig. 3 supports the idea that a hole is favoured to be placed along one direction ($x$ or $y$) and to be localized at a particular oxygen atom according to Figs. 4(b),(2) and (3) (bearing in mind that each hole state polarizes two neighboured copper atoms). Or in other words, the occurrence of a local symmetry breaking by the holes which surround the copper atoms results from the above considerations.

Nevertheless, there remains the question if the tendency to localize hole density at particular oxygen atoms is decisively queried by other energetic alterations within the hole doped bands, which are not yet considered. First, the orbital energies are considered. For this, the representation of the electronic states by Wannier-orbitals is taken as a basis in the subsequent discussion. If a linear relation between the occupation density of such local states and the total local energy exists then the above deduced tendency to form localized hole states is not queried. The doped holes have predominantly oxygen character, as discussed above. Hence, the Wannier-orbitals can be approximately described by the oxygen orbitals. In this case it can be shown, that a linear behaviour between hole density and oxygen atomic energy can be assumed up to about one hole charge $q_h = 1$ (see Appendix A). Hence, it seems justified to assume that a nearly linear relation between local hole density $q_h$ and the total local energy exists up to nominally one hole per oxygen site. All existing bands are related to the lattice period $2a$, $2b$ with respect to the copper sites. In relation to the continuously populated oxygen sites this means that the bands are half filled, i.e. the bands possess nominally one electron per oxygen site. However, the local electronic states are asymmetric with respect to the oxygen sites and mirror symmetric with respect to the common copper atom to which a bonding exists (see Table I). Therefore, there are also two different oxygen sites. In this case, it can supposed that a maximum of one hole per oxygen site ($q_h = 1$) can be doped.

Finally, one has to give a qualitative estimation about the changed long-ranged Coulomb interactions when hole density $q_h$ is located at particular oxygen atoms and polarized in either direction , $x$ or $y$ . The quadrupolar polarizations, as discussed above, lead to the polarization of a hole density along a particular direction ($x$ or $y$) within the CuO$_2$ plane. Or in other words, doped holes will form stripes (strings) along particular ….Cu-O-Cu… bonding lines. It is straightforward to describe the repulsive Coulomb interactions in one dimension within a rigid orbital scheme. In the following, an extended string with the length $r \cdot a$ of equidistantly ordered holes with the local density $q_h$ and a hole period of $L_o = n \cdot a$ ($a = b$ lattice constant) is considered. For comparison, the same string is considered under the premise that the entire hole density is equally distributed over the oxygen atoms along the ….Cu-O-Cu… bonding lines. Assuming a point charge approximation the mean energy difference per oxygen with respect to the long-range Coulomb interaction is given by

$$\overline{\Delta E}_{coul.} = E_{loc} - E_{distr.} \sim \frac{q_h^2}{n^2} \left( \sum_{r'=1}^{r/n} \frac{1}{r' \cdot a} - \sum_{r=1}^{r} \frac{1}{r \cdot a} \right) \tag{15}$$

with $E_{loc}$ being the Coulomb repulsion energy of the state with localized holes and $E_{distr.}$ representing the same energy if the holes are equally distributed along the string. It results $\overline{\Delta E}_{coul.} < 0$ (for $n \geq 2$) which is an evidence for the lowering of the Coulomb repulsion energy when hole density is located at particular oxygen atoms. Hence, this result as well supports the localization of hole density with $q_h = 1$ at particular oxygen atoms. I will later come back to the Coulomb repulsion problem within a generalized consideration. In summary, the considerations of this section support the idea that hole density will be localized at particular oxygen atoms. Hence, in all further discussions it will be supposed that hole density of nominally one hole ($q_h = 1$) is located at particular oxygen atoms. Even if the localization of hole density would be not complete ($q_h < 1$) the tendency to form hole stripes along particular

….Cu-O-Cu… bonding lines with strong localized hole densities will remain unrestrictedly valid. That means, that the further discussions have to be merely qualified for $q_h < 1$, but they will not be queried principally by that.

## B. Hole-hole pairing

In the previous section it has been concluded that topological hole states that are polarized along one coordinate axis ($x$ or $y$) are very probable. The driving force for this behaviour is quadrupolar polarizations which create non-zero off-diagonal densities $\Delta\left(c_s c_{d_{x^2-y^2}}\right)$ and $\Delta\left(c_{d_{z^2}} c_{d_{x^2-y^2}}\right)$. As discussed, a cyclically arranged cluster structure as given in Figs. 2(a) and 5(a) enables the quenching of the particular terms of Eqs. (11),(12) within the cluster. Therefore, the pure self-energy contribution to the topological hole state with respect to the induced local quadrupolar polarizations is comprehended in a good approximation. From Fig. 3 a nominal self-energy contribution of $E_s = 2\Delta E = $ -345 meV per topological hole ($q_h$=1) would result, corresponding to the polarization states at two next-neighbour copper atoms. This energy value (-345 meV) is only strictly valid if perpendicular polarized pairs of next-neighbour holes exist. This value has to be qualified possibly with respect to the results for singular holes and non-perpendicular polarized next-neighbour holes as given in Fig. 9. For example, a singular topological hole state creates identical polarized states (not alternating as in Fig. 2(a)) at next neighbour copper atoms which may reduce the above value (see below). On the other hand the total polarization state of each copper atom in the solid will be influenced by 4 next neighbour Cu-O bonds where only one Cu-O-Cu bond is influenced by identical polarized states if a singular topological hole is considered. Hence, the above value for $E_s$ should at least reflect the right order of magnitude for the self-energy of a singular topological hole state as long as Eqs. (13),(14) can be realized. However, one has to include an even higher polarizability of the charged $CuO_2^{2-}$ plane in comparison to a neutral cluster as used in Figs. 2(a) and 3. Therefore, the above value for $E_s$ may be even absolutely higher which, however, can not be ascertained from the existing calculations.

In the following a linear arrangement of hole states will be considered. An isolated topological hole state as depicted in Fig. 5(b) is principally unstable. The polarization state at the neighboured copper atoms is the same adopting the rules of Fig. 4(c) under the condition that $\Delta E < 0$. Hence, the particular terms according to Eqs. (11),(12) can not be quenched within this particular system but next neighbour copper atoms have to be involved. However, the accompanied polarisation of the bonds is not in ballance with the surrounding charge distribution. As a result, the energy belonging to the symmetric electronic part would be higher than the ground state energy. Hence, it is necessary that an additional hole is involved, as given in Fig. 5(c), for example. In this case the middle copper atom which has no hole in its neighbourhood is free to jump into one of the polarized states of Fig. 4(a). If it jumps into the inverted polarized state with respect to the copper atoms at the corner (Fig. 5(c)) the conditions of Eqs. (13),(14) can be realized. The inevitably reorganisation of the bonds is the same for the two hole positions. Therefore, a state as given in Fig. 5(c) is generally favoured in comparison to an isolated hole state, as given in Fig. 5(b). It is obvious that a linear hole-hole arrangement as given in Fig. 5(c) will have a different total polarization state in comparison to a cyclic arrangement as given in Fig. 5(a).

The decisive difference between the electronic state of a cyclic hole arrangement as given in Fig. 5(a) and a linear hole arrangement according to Fig. 5(c) lies in the fact that in the first case the local electronic state at all copper atoms is in equilibrium with the surrounding charge distribution but in the second case it is not. According to the concept of "detailed momentum balance" one has to state that the middle copper atom in Fig. 5(c) is not in a balanced state. The non-zero off-diagonal densities $\Delta\left(c_s c_{d_{x^2-y^2}}\right)$ and $\Delta\left(c_{d_{z^2}} c_{d_{x^2-y^2}}\right)$ induced at the middle copper atom create a planar quadrupole moment ($Q_{xx} \neq Q_{yy}$) at the middle copper atom. However, the planar EFG components at the middle copper atom caused by the surrounding charge distribution are further unchanged ($V_{xx} = V_{yy}$) if some anisotropies are neglected that are caused by different overlap densities in $x$ and $y$ direction which are already accompanied by the nonzero off-diagonal densities $\Delta\left(c_s c_{d_{x^2-y^2}}\right)$ and $\Delta\left(c_{d_{z^2}} c_{d_{x^2-y^2}}\right)$. Hence, the electronic system will react in a way that the local outer charge distribution will be changed in order to reach a new balanced state with $V_{xx} \neq V_{yy}$ and $\Delta E<0$. As discussed, the CBF state creates two different copper sites indicated in Figs. 1(a) and 2(a) by + and −, where different Cu-O bonding states are included. The bonding Cu-O character belongs predominantly to the + copper place whereas the − copper place may even possess an overall antibonding Cu-O character (Fig. 2(a)). A bonding state is principally connected with a negative Cu-O overlap charge and an antibonding state with a positive Cu-O overlap charge. Hence, one way to create an asymmetric EFG ($V_{xx} \neq V_{yy}$) consists in a transfer of antibonding states from the copper − site to free antibonding states of the copper + site along the bonding line in Fig. 5(c) (signed by the arrows). This is principally possible since within the renormalized electronic states according to Table I (config. III) only every second copper atom is populated within every symmetry orbital and two principally different orbital states exist ($\psi_{1A}, \psi_{3A}$ and $\psi_{2A}, \psi_{4A}$), which differ in phase with respect to the populated copper atoms. It enables a free exchange of the oxygen parts between these two groups of electronic states. As a result, existing antibonding parts can be transferred from the copper − site to the free antibonding state of the copper + site. Of course, a change in the location of the Cu-O antibonding parts results in a deviation from the given equilibrium configuration and, therefore, it is principally connected with an increase in energy. On the other hand, the additionally induced EFG may lead to lower energetic states according to Fig. 4(c). In which way the total energy is changed depends on the given configuration of the hole locations and on the phase of the CBF state. On this basis the following general conclusions can be drawn.

First, a hole arrangement according to Figs. 5(c) or 6(a),(1) will be examined. A quadrupolar polarized state with $\Delta E < 0$ with respect to the middle copper atom $Cu_M$ requires an induced EFG which is equivalent to place a "phantom-hole" near $Cu_M$ along the bonding line in Figs. 5(c) or 6(a),(1) when following the rules of Fig. 4(c). It can be reached if Cu-O antibonding parts belonging to the corner copper atoms (copper − states) are transferred to the middle copper atom (copper + state), as discussed above (Fig. 5(c)) . Then $\Delta E < 0$ results for the middle copper atom and simultaneously a net enhancement of the bonding states along the bonding line with respect to the corner copper atoms increases the effective negative charge along the bonding direction around the corner copper atoms. In consequence, the in-plane anisotropy of the EFG ($V_{xx} \neq V_{yy}$) at the corner copper atoms is further increased . As a result, the energy with respect to the corner copper atoms will additionally be lowered, too. Contrary to that, if the polarized states at the corner copper atoms are opposite caused by hole locations as given in Fig. 6(a),(5) the negative charge around the middle copper atom along the bonding line has to be increased in order to obtain $\Delta E < 0$. However, within the above mentioned bonding transfer this is not possible. It can only be realized when the total bonding strength is changed. In consequence, the realization of the Eqs. (13),(14) by electronic rearrangements is strongly hindered in the case of Fig. 6(a),(5). Therefore, the total electronic energies of the hole configuration of Fig. 6(a),(1) and the hole configuration according to Fig. 6(a),(5) should be distinctly different. The energy in Fig. 6(a),(5) should be generally increased with respect to states which reflect the undisturbed hole self-energy contribution as given in Fig. 5(a), for example.

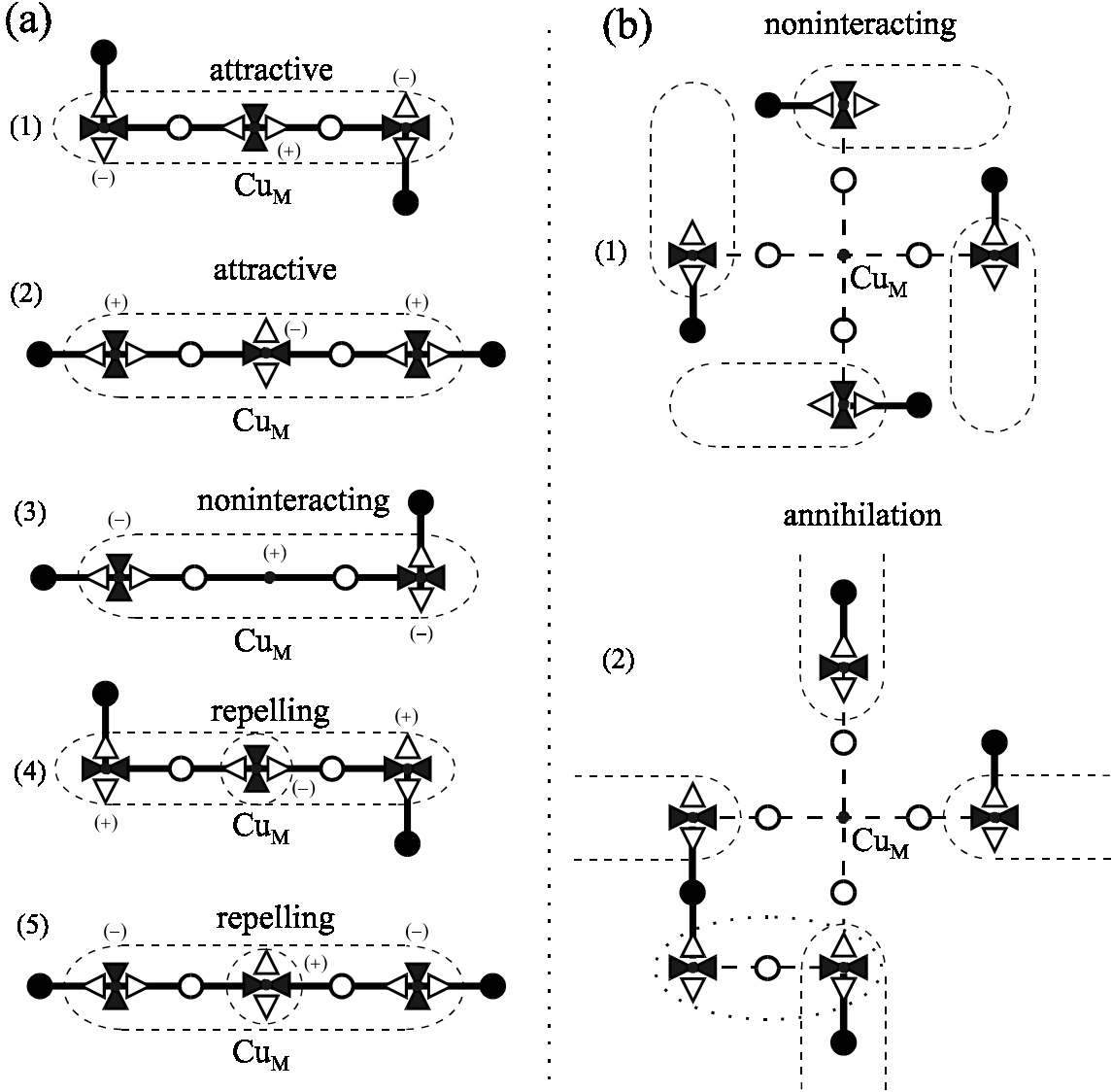

FIG. 6. Hole-hole coupling states in dependence on the hole-hole topology and the phase of the CBF state (signed by +,–). There exist two states [(a),(1)] and [(a),(2)] with an attractive hole-hole interaction. Two non-interacting hole-hole configurations [(a),(3)] and [(b),(1)] are depicted and a state [(b),(2)] where different polarization states at $Cu_M$ annihilate each other. The repelling states [(a),(4)] and [(a),(5)] are energetically unfavourable.

In order to prove the above conclusions a second cluster (Fig. 7) is considered. First, a continuous transition between the configurations of Fig. 6(a),(1) and Fig. 6(a),(5) has been simulated by changing the charges $q_1$ and $q_2$ relatively. The total charge has always been conserved, $\big((q_1 + q_2)/2 = -1.7\big)$ and $q_3$ has been fixed and chosen to be $q_3 = -1.7$ in order to secure an approximate $C_{4v}$ charge symmetry around $Cu_M$.

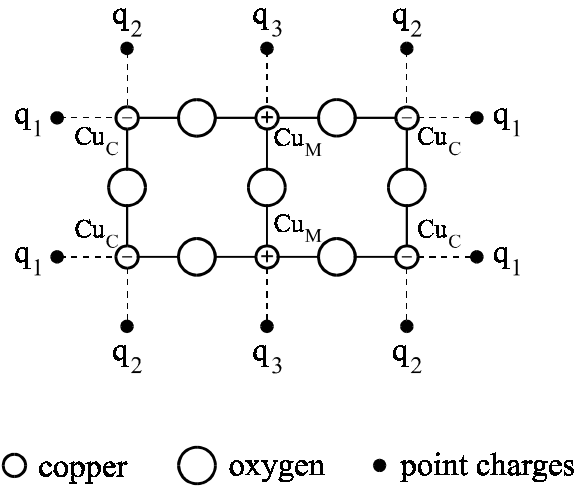

FIG. 7. Cluster consisting of 6 Cu and 7 O atoms and additional negative charges $q_1, q_2, q_3$ located at the $\overline{\text{Cu - O}}$ bond distance. The signs on the copper atoms indicate again different copper states. $Cu_M$ is the middle copper atom and $Cu_C$ is termed as corner copper atom.

A different behaviour of the electronic reorganization between the configurations (1) and (5) in Fig. 6(a) is clearly verified in Fig. 8. The sum and difference of the Cu-O bonding strengths with respect to the middle copper atom ($Cu_M$) and the corner copper atoms ($Cu_C$) is depicted. Configuration (1) is reflected by $q_2 - q_1 > 0$, and for $q_2 - q_1 < 0$ a situation corresponding to configuration (5) is given. The strong rise of the difference and the simultaneously occurring strong lowering of the sum of the bonding strengths in dependence on $|q_2 - q_1|$ when

$q_2 - q_1 < 0$ reflects especially an increased antibonding population. This inevitably leads to an additional increase of the total electronic energy. Contrary to that, the small increase of the sum and the small decrease of the difference of the bonding strengths in dependence on $\left|q_2 - q_1\right|$ for $q_2 - q_1 > 0$ reflects the above described transfer of antibonding parts from copper $-$ to copper $+$ sites without dramatic changes in the total bonding strengths. Therefore, the above conclusions regarding the different degrees of freedom for the electronic rearrangements in the cases of the configurations (1) and (5) in Fig. 6(a) have been verified. Hence, it has to be expected that in Fig. 6(a) the energy of the configuration (5) is basically higher in comparison to the configuration (1).

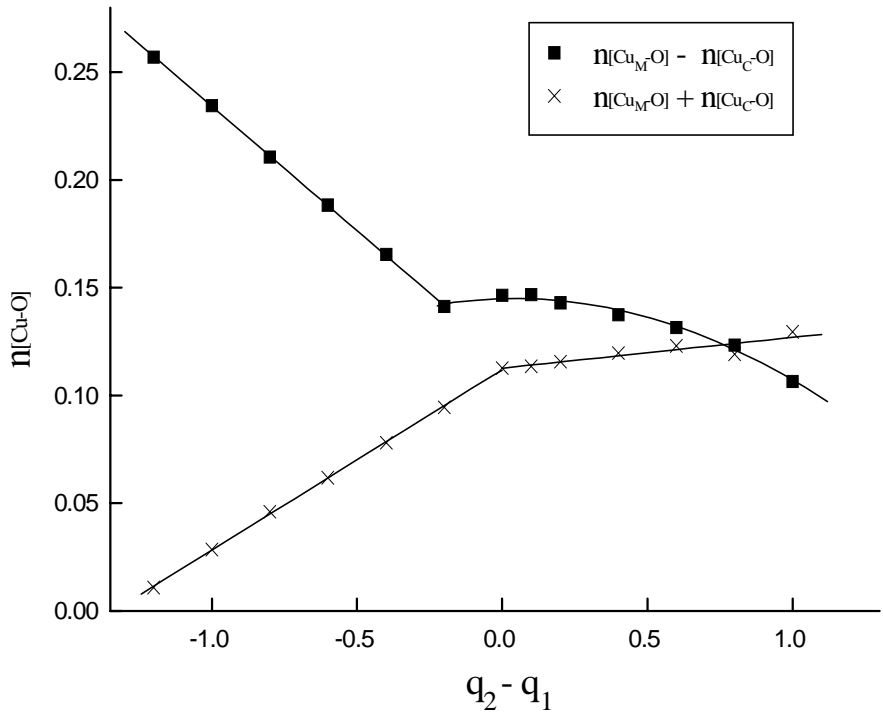

FIG. 8. Sum and difference of the overlap densities between the $Cu_M$-O bond and the $Cu_C$-O bond of the cluster in Fig. 7 in dependence on $q_2 - q_1$, with $(q_2 + q_1)/2 = -1.7$, $q_3 = -1.7$ and the CBF sequence $-,+,-$. The calculations were performed for a cluster with two negative charges.

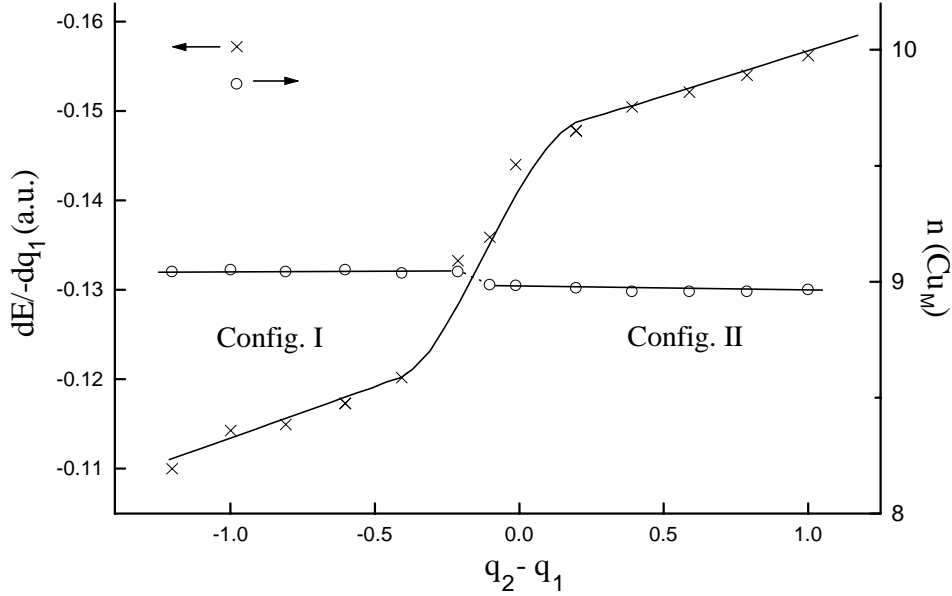

FIG. 9. Dependence of the first derivative $\mathrm{d}E/\mathrm{d}q_1$ of the electronic energy and the electronic occupation number $n$ at the central copper atom $\mathrm{Cu_M}$ (+ state) on the charge difference $q_2 - q_1$ with $(q_2 + q_1)/2 = -1.7$, $q_3 = -1.7$ and the CBF sequence $-,+,-$ (Fig. 7), (PA). A jump of the first derivative occurs at $(q_2 - q_1) \approx 0$. The calculations were performed for a cluster with two negative charges.

In Fig. 9 the first derivative of the electronic energy $E$ with respect to the outer charge state is depicted as a function of the charge asymmetry. One can see that an almost linear dependence of $-\mathrm{d}E/\mathrm{d}q_1$ on the charge asymmetry occurs, if $q_2 - q_1$ is taken away from zero. In addition, the slope is nearly the same for the two parts of the curve termed as configuration I and II. This linearity again reflects the hole self-energy caused by the charge anisotropy around the corner copper atoms ($\mathrm{Cu_C}$). This self-energy is, however, not identical with that one deduced from Fig. 3. A regular CBF state is only given in one direction (Fig. 7). Furthermore, the identical hole topology with respect to the corner copper atoms ($\mathrm{Cu_C}$) creates identically (i.e. not alternating) polarized states at the corner copper atoms contrary to the conditions in Fig. 2(a). These identical hole polarization states at the next neighbour corner copper atoms ($\mathrm{Cu_C}$) should at least partly responsible for the absolute reduction of the hole self-energy in contrast to the cluster of Fig. 2(a). However, the jump in $-\mathrm{d}E/\mathrm{d}q_1$ at $q_2 - q_1 \approx 0$ is most interesting, now. The configurations I and II belong to the same CBF state, because only small changes in the electronic densities at the inner copper atoms $\mathrm{Cu_M}$ occur when going from configuration I to II. The energetic instability at $q_2 - q_1 \approx 0$ is connected with changes of the induced EFG tensor components from $|V_{xx}| < |V_{yy}|$ to $|V_{xx}| > |V_{yy}|$ at the corner copper atoms. The additionally changed monopole potentials when varying $q_1$ and $q_2$ give merely rise to an additional offset in Fig. 9. The most important result from Fig. 9 is that an additional energy change occurs depending linearly on $|q_2 - q_1|$ which is reflected by the jump between configuration I and II . This energy jump is at least in the order of magnitude of the undisturbed hole self-energy part in Fig. 3 for high $|q_2 - q_1|$ values. Hence, this additional energy part is noticeable and has to be considered in discussing the hole-hole rearrangements. The different kinds of electronic rearrangements discussed above which are evidently confirmed in Fig. 8 give an evidential support that the configuration Fig. 6(a),(5) leads to an increase in energy. Therefore, the jump can be related to the relative energy lowering of configuration II (which corresponds to Fig. 6(a),(1)) with respect to configuration I (which corresponds to Fig. 6(a),(5)) with an absolute energy lowering of configuration II.

In order to get a further support a second simulation has been performed. Now, the inner copper atom $Cu_M$ has been changed by varying the charge $q_3$ and setting $q_1 = -2.2$ and $q_2 = -1.2$ . In studying the electronic reorganization by varying the charge $q_3$, one has to include an intense alteration of the long-range monopole potentials. However, this monopole interaction affects the entire cluster in a global manner, so that the resulting shift of energy is expected to vary linearly, again. In Fig. 10 the main results are depicted. Basically two configurations are found. Configuraion I which corresponds to configuration II in Fig. 9 and configuration II which belongs to the CBF sequence corresponding to Fig. 6(a),(4). Near $q_3 = -1.7$ a configuration interchange occurs, which means that configuration I and II have the same energy at this point. Near $q_3 = -1.7$ the transition from $|V_{xx}| < |V_{yy}|$ to $|V_{xx}| > |V_{yy}|$ appears at $Cu_M$, i.e. a change in the direction of the quadrupolar polarization at $Cu_M$ takes place. This is again an indication of different inductive effects of the quadrupolar polarizations with respect to the two configurations I and II. A substitution of hole density at the place of $q_3$ is accompanied by a polarization effect at the corner copper atoms. Hence, an increased hole density at the place of $q_3$ acts in the case of configuration II (Fig. 6(a),(4)) in the same way as an increased hole density at the corner copper atoms in Fig. 6(a),(1). In consequence, an energy lowering has to be expected. A substitution of additional hole density at the place of $q_3$ in the case of configuration I ((Fig. 6(a),(1)) acts in the same way as a substitution of hole density at corner copper atoms in Fig. 6(a),(4). Hence, no energy lowering has to be expected but an energy increase. Therefore, $dE/dq_3$ should be shifted to distinctly lower negative values in the case of configuration II in comparison to configuration I. In Fig. 10 this conclusion is evidentially confirmed by the high energy jump between the two configurations I and II. The progress in energy in dependence on $q_3$ in Fig. 10 is equivalent to Fig. 9 with respect to configuration I but different in relation to configuration II, because $dE/dq_3$ is nearly constant in dependence on $q_3$. The latter reflects probably some non-linear mechanisms between the polarized states at $Cu_M$ and at the corner copper atoms ($Cu_C$), because, the CBF configuration of Fig. 6(a),(4) is unstable with respect to the quadrupolar polarizations. In conclusion, the jump in energy between Configuration I and II in Fig. 10 can be consistently explained from the general ideas about the mechanisms of the quadrupolar polarizations.

All in all one can state that the cluster calculations support the idea of an induced hole-hole coupling by quadrupolar polarizations. In a simplified picture, one can say that an attractive hole - hole interaction is given if the electronic reorganizations include a stream of electron density in the direction of the charge fluctuation state as given in Figs. 6(a),(1) and (2). If the electronic flux is directed opposite to the charge fluctuation state (Figs. 6(a),(4) and (5)), repelling interactions have to be assumed. Hence, it is assumed in the following discussion that in addition to a self-energy part ($E_s$) of the topological holes, an attractive hole-hole coupling energy ($E_c$) exists corresponding to the configurations of Figs. 6(a),(1) and (2) which leads to the greatest total lowering of the energy.

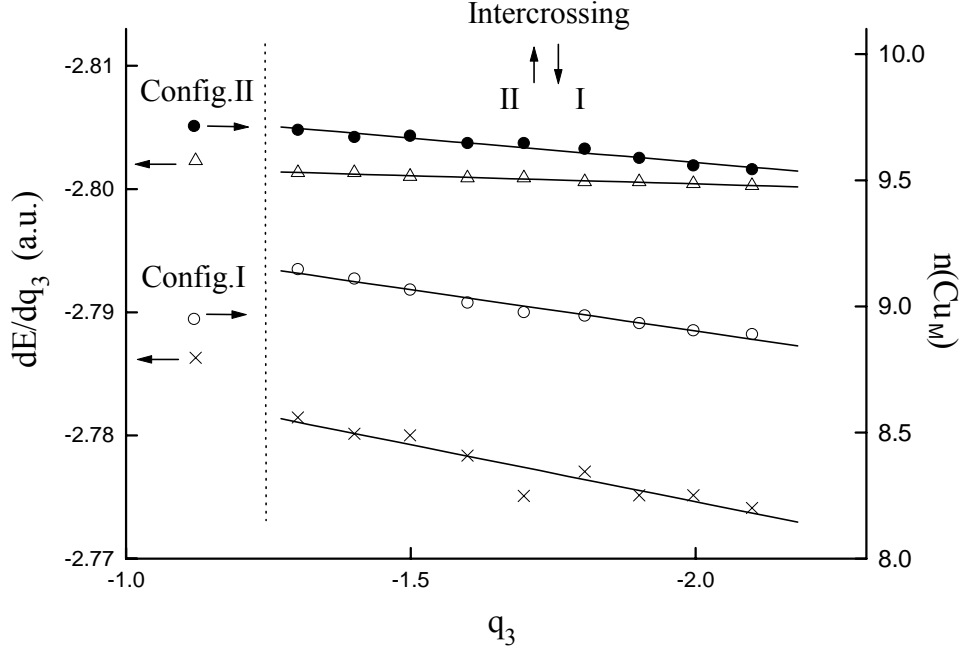

FIG. 10. Dependence of the first derivative d$E$/d$q_3$ of the total electronic energy and the electronic occupation number $n$ at the central copper atom $Cu_M$ on the charge $q_3$ (Fig. 7), (PA). Two configurations have been obtained corresponding to two different CBF sequences, $(-,+,-)$ and $(+,-,+)$. The arrows indicate the turn over with respect to the ground state configuration. An absolute difference of the first derivative occurs between the two different configurations I and II. The calculations were performed for a cluster with two negative charges.

The derivation of the remaining interaction rules in Fig. 6 is straightforward. In Fig. 6(a),(3) a practically equivalent situation to the cyclic case of Fig. 5(a) exists, i.e. $E_c = 0$. That means that the middle copper atom $Cu_M$ is electronically not influenced by the polarization states at the corner copper atoms. The same is true for the case in Fig. 6(b),(1), where the induced asymmetric polarizations at the middle copper atom are cancelled out. In Fig. 6(b),(2) the central copper atom will be differently polarized, because an attractive and repelling polarization occurs simultaneously thereby leading to an annihilation effect. Consequently, $Cu_M$ will not be promoted into a definite polarized state. Besides that, it has to be assumed that the hole self-energy is also reduced which is based on additional equivalently polarized states on next neighbour copper atoms (signed by the dotted ellipse in Fig. 6(b),(2)). Definitely, an undisturbed hole-hole coupling state can only occur for an one-dimensional linear hole-hole arrangement with a coupling copper atom ($Cu_M$) being included which is not polarized otherwise.

## V. THE ELECTRONIC STATE OF THE COPPER OXIDE PLANE

The most essential results obtained until now can be summarized in three points.

1. Within the electronic state of the $CuO_2$ plane a charge and bonding fluctuation state (CBF) exists which creates two different copper sites and which is accompanied by a division of the Brioullin zone by half.
2. In the case of hole doping a quadrupolar polarization induced localization of hole density occurs at particular oxygen atoms with a nominal hole density of $q_h = 1$ per oxygen atom.
3. There is the possibility of an attractive hole-hole interaction (of non-magnetic origin) of the localized topological holes (Figs. 6(a),(1) and (2)) which depends on the hole topology and the phase of the CBF state (Fig. 6).

At this point, it is necessary to give an idea of the total electronic state within the $CuO_2$ plane and its dependence on the number of doped holes. In the following, the problem is primarily considered for hole concentrations in the range between $n_h = 0.125$ holes/copper and $n_h = 0.25$ holes/copper. These concentrations describe the predominant region where superconductivity occurs. Besides that, it covers the so-called underdoped and overdoped regions which differ often distinctly as observed in various experiments.

## A. The influence of the Coulomb monopole interaction

For the moment it is assumed that all hole-hole couplings in Fig. 6 (point 3. above) are switched off. In this case the problem is reduced to an energetic consideration of doped localized hole states within a doubly negatively charged $CuO_2^{2-}$ plane where the charge carrier interaction is reduced to long-range Coulomb monopole interactions. With respect to the long-range Coulomb interactions the doubly negatively charged $CuO_2^{2-}$ plane can be considered to consist of singly charged electronic subunits $CuO^{1-}$ along the $x$ and $y$ axes where the negative charge is predominantly localized at the oxygen atoms. The location of hole density at particular oxygen atoms means that hole doping occurs with respect to one of these subunits directed either in $x$ or $y$ direction. Hence, at a certain distance a hole carrying oxygen place occurs to be neutral ($q_h = 1$). That means, every hole doping leads approximately to a complete loss of the repulsive Coulomb energy with respect to the hole carrying oxygen place at a certain distance from this oxygen place. Therefore, the total lowering of the electronic energy during the transition from the undoped to the hole doped state is in a first place proportional to the number of doped holes for small degrees of hole doping. Merely, if the Coulomb interaction between two hole sites are considered this strong proportionality has to be corrected, because the loss of the repulsive Coulomb interaction is doubly counted. Hence, the only thing one has to do during the transition from the hole undoped to the hole doped state is to calculate or estimate the loss of the total Coulomb repulsive energy between the hole carrying oxygen atoms for different topological hole configurations at a given degree of hole doping. This method will be applied in the following discussions.

The Coulomb forces will not be any longer in the equilibrium state if a localized neutral hole state is formed. The isotropic repulsive Coulomb forces are attempted to fill up the local hole state with electronic density. However, a localized topological hole state is energetically favoured as it has been concluded before. As a result, the isotropic repulsive Coulomb interaction will create a localized hole state with a coupled negatively charged environment. Each hole state causes the same tendency. Hence, the isotropic Coulomb forces will drive the hole system into a state of equally distributed holes. Corresponding equilibrium distributions are depicted in Fig. 11(a) for hole concentrations of nominal $n_h = 0.125$ holes/copper and $n_h = 0.25$ holes/copper, respectively. However, such equilibrium hole distributions are not necessarily the energetically most stable configurations.

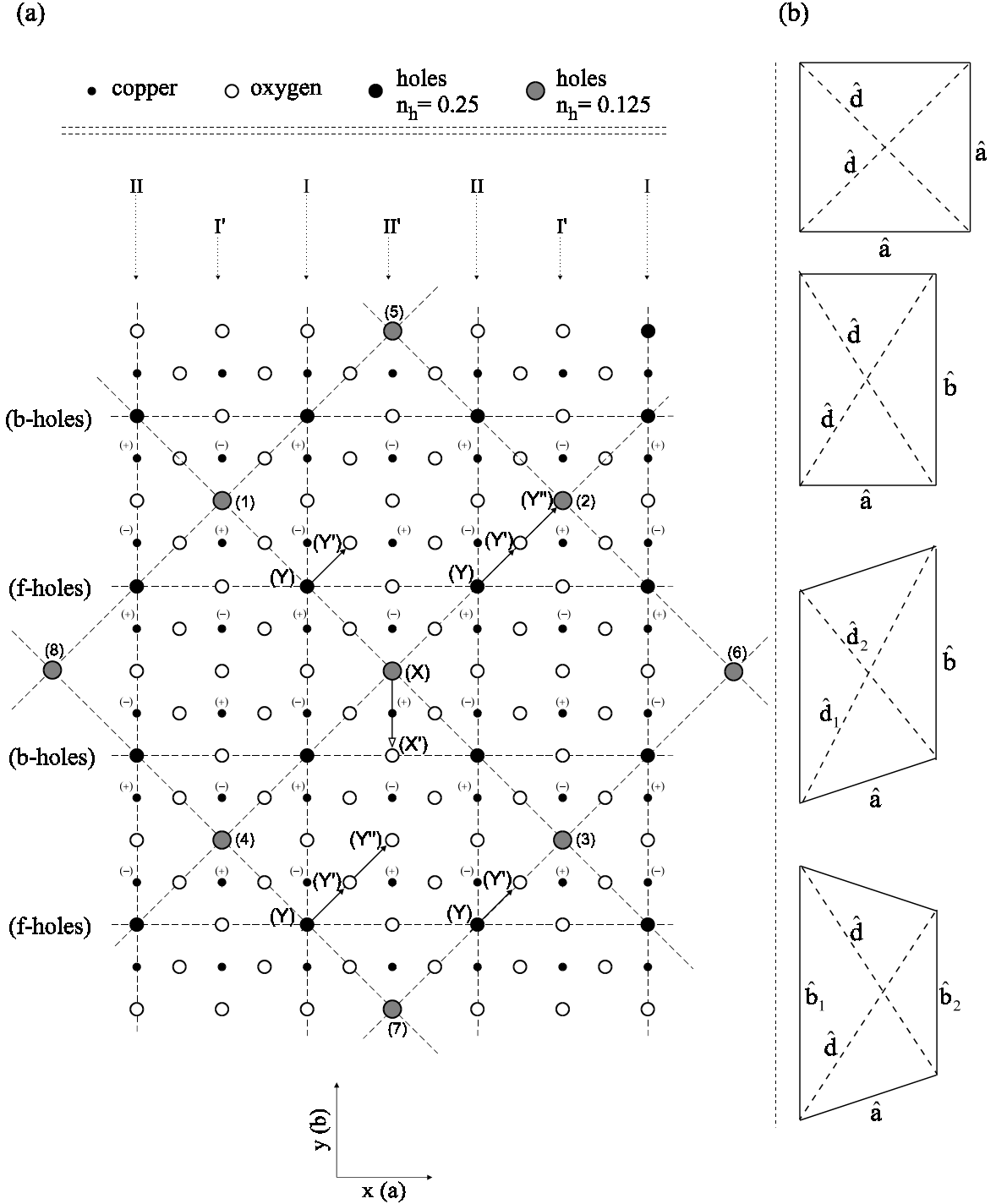

FIG. 11 (a) isotropic hole-hole arrangements for $n_h = 0.125$ holes/copper and $n_h = 0.25$ holes/copper within the CuO$_2$ plane. The $X \rightarrow X'$ transition describes a single hole displacement for $n_h = 0.125$ holes/copper. The transitions $Y, Y \rightarrow Y', Y''$ as well as $Y, Y \rightarrow Y'', Y'''$ describe complex displacements of holes within each second hole string along the $x$ coordinate (f-holes) for $n_h = 0.25$ holes/copper. (b) a square hole structure with nominally $n_h = 0.125$ holes/copper as given in (a) can be transferred into a rectangular (in (a) given by the b-holes or f-holes) parallelogram or trapezium next neighbour hole-hole arrangement. The parallelogram and trapezium hole structure result from appropriate hole shifts of one lattice spacing in $y$ direction within the rectangular hole structure.

First, the Coulomb monopole interaction of a topological hole state with other holes in the first coordination sphere will be considered. This coordination sphere is shown in Fig. 11(a) by the points (1) to (8) around $X$ in the case of $n_h = 0.125$ holes/copper. For $n_h = 0.25$ holes/copper equivalent points would result within the respective topological hole grid as depicted in Fig. 11(a). Nominally, four hole-hole distances belong to one hole place within this first coordination sphere. In square hole arrangements as given in Fig. 11(a) two hole distances with $r_{X,(i)} = \hat{a} = \sqrt{8} \cdot a$ and two hole distances with $r_{X,(i)} = \sqrt{2} \cdot \hat{a} = 4a$ result for $n_h = 0.125$ holes/copper and in the case of $n_h = 0.25$ holes/copper two hole distances with $r_{X,(i)} = \hat{a} = 2a$ and $r_{X,(i)} = \sqrt{2} \cdot \hat{a} = \sqrt{8} \cdot a$ are obtained, respectively. If only the first coordination sphere $s = 1$ is included the repulsive hole-hole Coulomb interaction per hole is given by $E(1) = 1.2071 \cdot \left(q_h^{'}\right)^2 / a$ for $n_h = 0.125$ holes/copper where $q_h^{'}$ stands for an effective charge which corrects for a point charge approximation with $q_h = 1$. If one hole is displaced for example from its equilibrium position as given by the $(X) \rightarrow (X')$ transition in Fig. 11(a) the Coulomb repulsive energy is totally changed by $\Delta E\left((1, X \rightarrow X')\right) = E(1, X') - E(1, X) = 0.0533 \cdot \left(q_h^{'}\right)^2 / a$. The new state $E(1, X')$ which includes the $(X) \rightarrow (X')$ transition has obviously a higher Coulomb repulsive energy as the highly symmetrical initial state and is therefore not favoured.

A different situation may occur if a complex reorganization of the entire hole structure is included. Hence, the original square hole structure as depicted in Fig. 11(a) is considered to be changed subsequently in a rectangular, parallelogram and trapezium structure of the next neighbour hole arrangement (pictograms in Fig. 11(b)).

First, the case with $n_h = 0.125$ holes/copper is considered. If the original square hole structure in Fig. 11(a) is rectangularly deformed resulting in hole distances $\hat{a} = 2a, \hat{b} = 4a$ and subsequently rotated by 45° a regular rectangular structure along the coordinate axes $x, y$ is formed. That means, this new hole structure is identical with the hole structure of $n_h = 0.25$ holes/copper but consists only of one sort of holes, so-called b-holes or f-holes in Fig. 11(a). The surface of the structure units is the same with respect to the square and rectangular structure. In a next step, this rectangular structure is changed into a structure where the hole units are forming parallelograms. That can be reached if the holes of the rectangular structure are particularly shifted in $y$ direction, i.e. holes are shifted which belong either to the hole strings I or to the hole strings II in Fig. 11(a). The case in which the relative shift between the strings I and II corresponds to one lattice spacing $b$ will be considered in particular. The surface of the structure unit is again not changed. Lastly, the parallelogram structure is transferred to a trapezium structure, i.e the hole units are forming trapezia. That can be reached, if each second side of the parallelogram structure is mirrored with respect to the small side of the corresponding rectangular unit (see pictograms in Fig. 11(b)). In this case a trapezium hole structure as given by the b-holes in Fig. 12 is created. The surface of the structure unit is again conserved. Under these conditions one obtains for the hole-hole Coulomb interaction per hole with respect to the first coordination sphere:

$$E_Q(1)\left[ = 1.2071 \cdot \frac{\left(q_h^{'}\right)^2}{a} \right] >$$

$$E_R(1)\left[ = 1.1972 \cdot \frac{\left(q_h^{'}\right)^2}{a} \right] > \qquad \text{(for } n_h = 0.125 \text{ holes/copper)}, \qquad (16)$$

$$E_P(1)\left[ = 1.1603 \cdot \frac{\left(q_h^{'}\right)^2}{a} \right] \cong E_T(1)\left[ = 1.1611 \cdot \frac{\left(q_h^{'}\right)^2}{a} \right]$$

where $Q, R, P$ and $T$ represent the quadratic, rectangular, parallelogram and trapezium structure, respectively. The number (1) represents again the first coordination sphere as defined above. These calculations show that the parallelogram and trapezium structure being energetically nearly degenerate would be energetically favoured when only Coulomb monopole interactions are included. Nevertheless, it can not be expected that such strongly ordered structures are already stable. The isotropic repulsive Coulomb monopole interactions would always force the hole system to achieve an isotropic equilibrium state. Hence, these results suggest that an ordered equilibrium hole structure can not be supposed that is stable in time. Instead, pronounced quantum fluctuations have to be taken into account, i.e. local fluctuations of the hole topology between nearly degenerate configurations.

Similar results can be obtained for the case of $n_h = 0.25$ holes/copper. The quadratic structure in Fig. 11(a) can be changed into a parallelogram structure by shifting one of the strings I or II by one lattice spacing $b$ along the $y$ coordinate. For the Coulomb energy with respect to the first coordination sphere one obtains:

$$E_Q(1)\left[=1.7071\cdot\frac{(q_h')^2}{a}\right] > E_P(1)\left[=1.6718\cdot\frac{(q_h')^2}{a}\right] \qquad \text{(for } n_h = 0.25 \text{ holes/copper).} \qquad (17)$$

Also in this case, the isotropic equilibrium structure is energetically not the most stable structure. In the following, the total number of holes will be subdivided into $n_h' = 0.125$ holes/copper termed as b-holes and $n_h'' = 0.125$ holes/copper termed as f-holes. The meaning of this differentiation will be clearly recognizable later on. In doing so, lines of holes along the $x$ direction in Fig. 11(a) can be differentiated in b-holes and f-holes. The subsequent terms $R_b$, $P_b$, $T_b$ are now representing the rectangular-, parallelogram- and trapezium structure of the b-hole subspace. Then the Coulomb repulsive interaction is calculated with all f-holes shifted into positions $Y'$. It results:

$$\overline{E}_{R_b}\left(1,Y',Y'\right)\left[=1.7325\cdot\frac{(q_h')^2}{a}\right]$$

$$> \overline{E}_{P_b}\left(1,Y',Y'\right)\left[=1.7146\cdot\frac{(q_h')^2}{a}\right] \qquad \text{(for } n_h = 0.25 \text{ holes/copper).} \qquad (18)$$

$$> \overline{E}_{T_b}\left(1,Y',Y'\right)\left[=1.7061\cdot\frac{(q_h')^2}{a}\right]$$

where $\overline{E}$ describes the averaged energy over all different hole positions. The Coulomb energy is obviously rising in comparison to structures with a higher symmetry as given in Eq. (17). But the trapezium structure is relatively the lowest one. In a further step, half of the f-holes are shifted in $Y'$ positions and the other half in $Y''$ positions as depicted in Fig. 11(a). Then one obtains:

$$\overline{E}_{R_b}\left(1,Y',Y''\right)\left[=1.7721\cdot\frac{(q_h')^2}{a}\right]$$

$$> \overline{E}_{P_b}\left(1,Y',Y''\right)\left[=1.7097\cdot\frac{(q_h')^2}{a}\right] \qquad \text{(for } n_h = 0.25 \text{ holes/copper).} \qquad (19)$$

$$> \overline{E}_{T_b}\left(1,Y',Y''\right)\left[=1.6867\cdot\frac{(q_h')^2}{a}\right]$$

In this case, the energy is distinctly higher for a rectangular b-hole configuration and relatively lowered for the parallelogram and the trapezium structure of the b-holes. The trapezium structure creates again the smallest energy which is close to the value obtained for the highly symmetric parallelogram structure as given in Eq. (17).

Of course, the above results were obtained by the inclusion of the first coordination sphere only. In order to get a general validity one has to analyze the influence of the higher coordination spheres, in particular, because the coordination spheres are mixing if the f-holes are shifted in $Y'$ and $Y''$ positions.

First, the case of $n_h = 0.125$ holes/copper is considered. It can be shown, that a rectangular hole structure is basically energetically favoured with respect to every coordination sphere s (see Appendix B) compared to a square hole structure. Hence, one has to compare in detail merely the regular rectangular hole structure with their related parallelogram and trapezium hole structures. Under the condition of regularly ordered topological hole states hole-hole lattice constants can be defined, with $a_h$ being the hole-hole lattice constant in $x$ direction and $b_h$ the corresponding constant in $y$ direction. If the reference hole is placed at the zero point $(0,0)$ the hole-hole distances are given by

$$\left| r_{o,ij} \right| = \sqrt{\left( i \cdot a_h \right)^2 + \left( j \cdot b_h \right)^2} \tag{20}$$

with $i,j$ representing the hole coordinates in $x$ and $y$ direction, respectively. For the rectangular hole state with $n_h = 0.125$ holes/copper the hole lattice constants are given by $a_h = 2a$ and $b_h = 4a$. In the parallelogram or trapezium case the hole distances are either equivalent to Eq. (20) or deviate by a shift of one lattice spacing $b \, (= a)$ in $y$ direction (Fig. 12). The term $j \cdot b_h$ in Eq. (20) has therefore to be replaced by $(j \cdot b_h - a)$ or $(j \cdot b_h + a)$ where an absolute increase and shortening of the $y$ coordinate with one lattice spacing $b \, (= a)$ occurs always in pairs (see Appendix B). This kind of deviation from a regular rectangular structure occurs for odd $i$ and all $j$ values in the case of the parallelogram hole structure and for odd $i$ and even $j$ values as well as even $i$ and odd $j$ values in the case of the trapezium hole structure (Fig. 12). Consequently, for the parallelogram and trapezium hole structure only half of the hole distances deviate from a regular rectangular hole structure. However, even if it is assumed that all hole positions deviate in the above sense the energetic differences to the corresponding regular rectangular hole structure can already be neglected for $s \geq 3$ (see Appendix B). What remains is the second coordination sphere. It can be shown that the square hole structure is again the energetically highest one. The energy increase of the trapezium hole structure in comparison to the parallelogram hole structure should not be much higher than around 10 meV. Therefore, one can basically assume the qualitative relations of Eq. (16) in summing over all coordination spheres. That means, the parallelogram and trapezium hole structure can be considered as nearly energetically degenerate if hole-hole couplings as given in Fig. 6 are not yet included.

Now, the case of $n_h = 0.25$ holes/copper is considered with the inclusion of the second coordination sphere $s = 2$. It can be shown, that a parallelogram hole structure is further energetically favoured compared to a square hole structure

$$E_Q(1+2)\left[ = 3.4551 \cdot \frac{\left( q_h' \right)^2}{a} \right] > E_P(1+2)\left[ = 3.4356 \cdot \frac{\left( q_h' \right)^2}{a} \right] \quad \text{(for } n_h = 0.25 \text{ holes/copper)} \tag{21}$$

with 1+2 as the sum of the first and second coordination sphere. Then, the hole-hole Coulomb energies will be considered with additional shifts of the f-holes into $Y'$ and $Y''$ positions (Fig. 11(a)). It results:

$$\overline{E}_{R_b}\left(1+2, Y^{'}, Y^{"}\right)\left[= 3.5294 \cdot \frac{(q_h^{'})^2}{a}\right]$$

$$> \overline{E}_{P_b}\left(1+2, Y^{'}, Y^{"}\right)\left[= 3.4758 \cdot \frac{(q_h^{'})^2}{a}\right] \quad \text{(for } n_h = 0.25 \text{ holes/copper)}. \quad (22)$$

$$> \overline{E}_{T_b}\left(1+2, Y^{'}, Y^{"}\right)\left[= 3.4560 \cdot \frac{(q_h^{'})^2}{a}\right]$$

The qualitative relations are the same as given for the first coordination sphere (Eq. (19)). In particular, it is obvious that the trapezium b-hole structure is the relatively lowest structure in energy further on. The inclusion of higher coordination spheres $s > 2$ does not change the qualitative relations of Eq. (22) where the energetic differences between the configurations diminish very fast and become zero similar as given for the difference between the regular rectangular $R$ and disturbed rectangular $R'$ structure in Fig. 15 (Appendix B).

The Coulomb hole-hole repulsive energies of the regular square and parallelogram structure in Eq. (21) are somewhat lower than the energies for the particular structures in Eq. (22). Assuming a point charge approximation $\left(q_h^{'} = 1\right)$ an energetic difference of e.g. $\overline{E}_{T_b}\left(1+2, Y^{'}, Y^{"}\right) - E_P\left(1+2\right) = 80$ meV results. At least, this magnitude of quantitative alterations has to be taken into account in further discussions.

In the hitherto discussions no individual electronic rearrangement of the particular oxygen atoms has been included. The highly negatively charged oxygen orbitals are very soft, i.e. they possess the ability for an effective breathing behaviour. In general, individual quantities $q_h^{'}$ have to be used for every particular hole-hole Coulomb interaction. In this way the energetic differences between particular topological hole configurations as discussed above may be further diminished. Hence, it seems to be very realistic to assume that a large variety of nearly degenerate topological hole configurations exists if the electronic system is restricted to exclusively Coulomb monopole interactions. It means, under inclusion of only Coulomb monopole interactions quantum fluctuations has to be taken into account, i.e. local fluctuations between different hole topologies which are energetically nearly degenerate.

## B. The influence of the induced hole-hole interactions

In section A the electronic ground state of particular topological hole configurations has been discussed including only long-range Coulomb interactions. It will be specifically changed if the hole-hole couplings given in Fig. 6 are switched on.

First, the situation for $n_h = 0.125$ holes/copper is considered, again. A regular square hole topology as given in Fig. 11(a) does not create attractive couplings as depicted in Figs. 6(a),(1) and (2), and gives not any possibility to form non-interacting states similar as depicted in Fig. 6(a),(3). Therefore, the ground state energy is additionally increased. For a regular rectangular hole structure as given e.g. by the b-holes in Fig. 11(a) the situation appears completely different. Here, the topological hole arrangement creates equal numbers of attractive and repelling hole-hole couplings in $x$ direction (Figs. 6(a),(1) and (4)). If the parallelogram hole structure is formed the repelling hole-hole configurations according to Fig. 6(a),(4) are resolved but non-interacting states like Fig. 6(a),(3) or additional interactions like Figs. 6(a),(1) and (2) are not created. Therefore, the electronic system will occupy a not well-defined electronic state with a total hole-hole coupling energy which should not be essentially different from the energy of the rectangular hole structure. The situation is dramatically changed if a trapezium topological hole structure is formed as shown by the b-hole grid in Fig. 12. In this case all hole-hole couplings are proven to be attractive according to the coupling rules in Figs. 6(a),(1) and (2). Therefore, this state is clearly the lowest in energy bearing in mind the qualitative relations of Eq. (16). It can be deduced from Fig. 9 that a total attractive hole-hole interaction of $|E_{C(tot)}| > 160$ meV per b-hole can be assumed in minimum for the b-hole structure of Fig. 12. In reality one has to assume an energy that is even further lowered if chains of attractive hole states are continuously formed like the b-holes in Fig. 12. The reason for that is that an attractive hole-hole pair state creates an electronic polarization at all involved atoms. These polarizations are accompanied with a promotion of the electronic system into higher excited states. If an infinite chain of hole-hole pair states is considered, a pair state localized immediately before and after the considered pair state already creates the electronic excitations which are necessary for forming the polarizations of the considered pair state. In consequence, the net hole-hole interaction energy of continuously coupled pair states should be lower than the sum of the pair interaction energies of separated hole-hole coupling states. Hence, one has to look for hole configurations forming continuous and complete attractively coupled hole-hole states. This is given for the trapezium case in Fig. 12 where nominally 3/2 attractive topological hole-hole configurations occur per hole without any repelling state. This is the maximum number of two-dimensionally undisturbed coupled attractive hole-hole configurations which are possible within a $CuO_2$ plane. All these aspects give rise to the important conclusion that the b-hole grid in Fig. 12 represents a ground state which is energetically distinctly separated from other topological hole configurations. It means that the electronic state will be self-consistently frozen out into a particular topological b-hole grid as given in Fig. 12 in which hexagons of topological b-holes are formed within a honeycomb structure. The deduced minimum of the absolute hole – hole coupling energy of 160 meV may be even higher. This is based on the fact that $E_S$ in Fig. 9 is lower than the undisturbed value resulting from Fig. 3, as well as the general fact that the polarizability of the charged $CuO_2^{2-}$ plane may be higher in comparison to the doubly charged cluster used in Fig. 9. However, within this context one has to remark that the conclusion of a frozen electronic state is only valid if exclusively electronic interactions within an absolutely rigid lattice are considered. I will show elsewhere that a particular topological hole state as given in Fig. 12 occurs only as a snapshot within the total electronic state.

Now, the case of $n_h = 0.25$ holes/copper is considered. Only when a trapezium hole configuration (b-holes) is formed the maximum number of undisturbed attractive interactions can be further reached. The remaining holes (f-holes) within the particular configurations are driven to form non-interacting states like Fig. 6(a),(3). That can be reached by shifting the topological hole states according to $Y', Y''$ in Fig. 11(a). Any other hole configuration which is not equivalent (degenerate) to Fig. 12 creates either additional repelling hole-hole interactions and/or destroys some attractive interactions. Therefore, the qualitative relations of Eq. (22) are additionally supported favouring the trapezium b-hole structure as the structure with the lowest energy. At this point a comparison with the regular parallelogram hole structure of Eq. (21) has to be carried out. It can be assumed that the more favourable Coulomb monopole interaction ($\approx$-80 meV) of the regular parallelogram hole structure is already compensated for by the lower absolute self-energy $|E_S|$ of the holes. A regular parallelogram hole structure additionally creates identical hole polarization states at next-heighboured copper atoms thereby lowering the self-energy as discussed in Sec. IV. If the quantitative alterations of the hole self-energy deduced from Fig. 3 and Fig. 9 are taken as a basis it can be concluded that an increase of the self-energy of more than 80 meV occurs. In addition, the same number of attractive and repelling couplings (Figs. 6(a),(1) and (4)) occur within a parallelogram hole structure. This causes a further increase of the electronic energy in comparison to an exclusively attractively coupled hole configuration as given for the trapezium case. Hence, a trapezium attractive b-hole structure which includes non-interacting holes (f-holes) as depicted in Fig. 12 has the lowest electronic energy for $n_h = 0.25$ holes/copper. That means that the electronic state is also self-consistently frozen out into a particular hole configuration as given in Fig. 12 for $n_h = 0.25$ holes/copper.

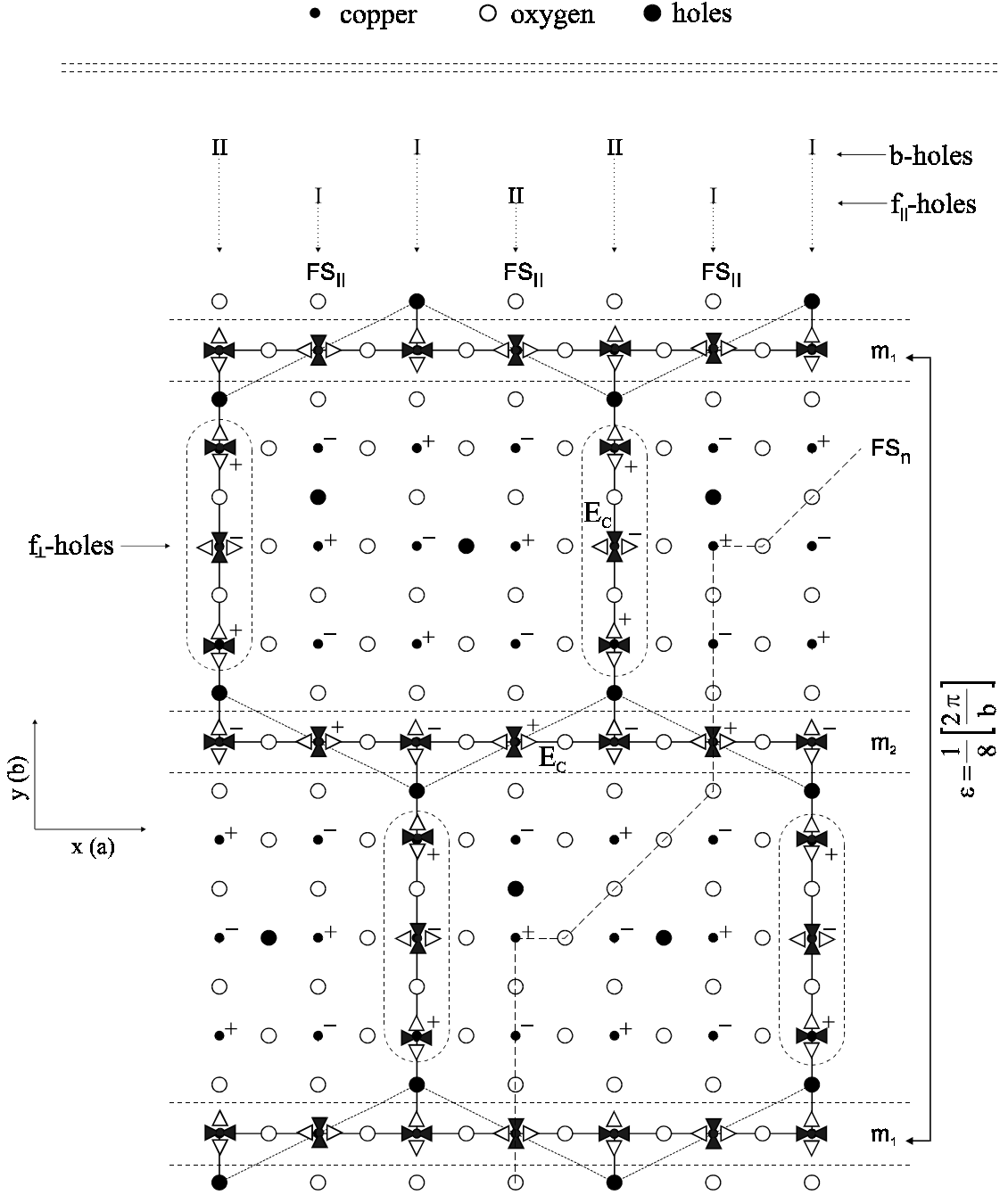

FIG. 12. Topological hole confuguration for $n_h = 0.25$ holes/copper representing an electronic ground state configuration. +,– indicate the different copper states corresponding to the CBF state. The pictogram representation indicate attractive hole-hole interactions according to Figs. 6(a),(1) and (2) where the hole-hole correlation energy $E_C$ occurs. These holes ($n_h/2$) which are involved in the formation of the attractive couplings (pictogram lattice, full dark lines) are termed as b-holes (bonded holes). The remaining holes ($n_h/2$) are non-bonding according to Fig. 6(a),(3) and termed as f-holes (free-holes). The b-hole system forms hexagons of holes within a honeycomb hole structure (dotted lines). It exists undisturbed in the range of 0.125 holes/copper $\leq n_h \leq 0.25$ holes/copper. The b-holes and $f_{II}$-holes are aligned along the y coordinate forming one-dimensional collective electronic states along the stripes I, II. The $f_\perp$-holes form hole stripes along the x coordinate. In the case of b-holes the particular stripes are coupled by the correlation energy $E_C$. Lines signed by $FS_{II}$ include charge carriers which are placed at the Fermi surface along the y coordinate, $FS_n$ indicates a possible part of the Fermi surface with contributions along the nodal direction [1,1].

All the above aspects lead to the conclusion that the dominant electronic state is a two-dimensionally ordered, attractively coupled topological b-hole grid with a trapezium hole structure. This ordered b-hole structure occurs in the concentration range of $n_h = 0.125$ holes/copper to $n_h = 0.25$ holes/copper. In the intermediary concentration range between $n_h = 0.125$ holes/copper and $n_h = 0.25$ holes/copper the non b-hole parts $\Delta n_h = n_h - 0.125$ holes/copper will be distributed over the b-hole grid as non-interacting topological double hole states (f-holes). The distribution of these topological f-hole pairs over the b-hole grid happens in a way that the total electronic energy is minimized. In the concentration range of $n_h < 0.125$ holes/copper the b-hole grid will be locally dissolved with the result that the two-dimensional ordering of the b-hole state contains b-hole vacancies or the b-hole ordering is locally destroyed (b-hole clustering). In the concentration range of $n_h > 0.125$ holes/copper it is necessary that two additional f-holes are locally placed within the b-hole grid. From Fig. 12 it is easy to recognize that such a topological distortion leads inevitably either to the destruction of some attractive topological hole-hole couplings according to the rules given in Fig. 6 or/and to strong distortions of the undisturbed polarisability of the coupling copper atoms $Cu_M$ or/and to an unfavourable rise of the hole self-energy as a result of identical hole polarized states at next-neighboured copper atoms. Hence, it can be concluded that either the regular two-dimensional topological ordering of the b-hole system is locally destroyed or the two-dimensionally coupled electronic state within the b-hole system is locally strongly distorted for $n_h > 0.25$ holes/copper. All in all it means that for $n_h < 0.125$ holes/copper as well as for $n_h > 0.25$ holes/copper the undisturbed two-dimensionally coupled electronic b-hole system is more and more locally destroyed for lower $n_h$ values than 0.125 holes/copper or for higher $n_h$ values than 0.25 holes/copper.

If the two-dimensionally coupled electronic subsystems (b-hole stripes I, II in Fig. 12) which form a b-hole structure as given in Fig. 12 proves to be the precondition for superconductivity one has to expect that the conditions to form superconducting states are continuously diminished the lower the $n_h$ values than $n_h = 0.125$ holes/copper and the higher the $n_h$ values than $n_h = 0.25$ holes/copper are. All experimental investigations which had the aim to analyze the range of superconductivity in dependence on $n_h$ have revealed this fact.[74-81] In a forthcoming article I will give a prove of this conclusion from a microscopic idea of the formation of the superconducting state.

## VI. CONCLUSIONS AND A FIRST COMPARISON WITH EXCEPTIONAL EXPERIMENTAL RESULTS

The deductive proof in this work of the electronic state within the $CuO_2$ planes of the HTC materials reveal the general conclusion that electronic rearrangements occur giving rise to various symmetry breakings. In the hole undoped case, the electronic state differs from the crystal symmetry by the existence of a commensurate superstructure with a period of a double lattice spacing ($2a, 2b$). Two different copper sites are formed differing in the electronic population as well as the Cu-O bonding state. That leads basically to a state with fluctuating charges and fluctuating bonding and antibonding parts of the Cu-O bonds (CBF state). This state is doubly degenerate with respect to its mirror symmetric state. If electrons are removed from the doubly negatively charged $CuO_2^{2-}$ plane the formation of topological hole states centered at particular oxygen atoms will be favoured. The formation of topological hole states is causally related to the quadrupolar polarization state of the copper atoms which creates an in-plane anisotropy of the EFG ($V_{xx} \neq V_{yy}$) and of the quadrupole moment ($Q_{xx} \neq Q_{yy}$) on the copper atoms. In that way topological hole states are formed which are aligned along the coordinate axes of the $CuO_2$ plane, i.e. along particular ..Cu-O-Cu... bonding lines. The topological hole states are characterized in a way that hole density corresponding to nominally one hole is localized at a particular oxygen atom. All in all it means that stripes of such hole states will be formed along particular ..Cu-O-Cu... bonding lines.

The effect of the localization of hole density leads first and foremost to a local symmetry breaking of the electronic states. However, it has been evidently proven that the possibility of an attractive coupling between such topological hole states can exist. Basically two attractive topological hole-hole configurations are possible (Figs. 6(a),(1) and (2)). These additional hole-hole couplings remove the degeneracy of a multitude of topological hole structures thereby favouring a strongly ordered topological hole structure. This electronic state which is given by a topological hole structure in which one part of the holes (b-holes) forms hexagons within a honeycomb structure turns out to be the most stable state. Therefore, an additional global symmetry breaking results. For a hole concentration of $n_h = 0.125$ holes/copper 32 equivalent and strongly ordered topological hole structures can exist which are energetically degenerate (these include two polarization directions $x, y$ and two CBF states). In the intermediary hole concentration range of 0.125 holes/copper $< n_h \leq 0.25$ holes/copper the number of energetically equivalent hole configurations is further increased, but the degeneracy with respect to the b-hole system has further a value of 32. The most deciding result is therefore that an absolutely stable topological hole configuration (b-holes) exists within the entire concentration range of 0.125 holes/copper $\leq n_h \leq 0.25$ holes/copper.

The electronic structure as given in Fig. 12 has been derived on the assumption of the Born-Oppenheimer approximation. It neglects any lattice degrees of freedom and hence does not yet represent the complete quantum state. However, the deduced topological b-hole structure is a universal stable structure which confirms the assumption that it forms the basis structure for the electronic states within the $CuO_2$ planes. Therefore, the b-hole structure should be the determining quantity for the formation of a superstructure being of an assumed static nature (static stripes) which has been found in Neutron Scatterings in $La_{2-x-y}Nd_ySr_x)CuO_4$[47] as well as for superstructure peaks measured in $La_{2-x}Sr_xCuO_4$[49] which are considered to be dynamic (dynamic stripes). Recently performed ARPES (angle resolved photoelectron spectroscopy) experiments distinctly point out such a fact. It has been shown that a duality between the ARPES spectra in [1,0] and [0,1] direction exists between the materials (($La_{2-x-y}Nd_ySr_x)CuO_4$) and (($La_{1.85}Sr_{0.15})CuO_4$).[82] In a forthcoming paper I will show that the difference between a dynamically constrained electronic structure and the formation of a fully dynamic superstructure lies in the fact that in the constrained case the electronic b-hole subsystem shown in Fig. 12 is topologically pinned with the result that the degeneracy of the electronic system with respect to the possible topological hole states is reduced. Commonly it means that the b-hole system is dynamically polarized only into one direction ($x$ or $y$) and a superposition of degenerate states occurs which result predominantly from phase shifts of the b-hole system along of one coordinate axis $x$ or $y$. In the case of fully dynamic superstructures the topological hole structure will fluctuate between all possible topological hole states with a general structure as depicted in Fig. 12 (including different f-hole concentrations and possible b-hole vacancies) where the b-hole system is temporarily polarized in $x$ and $y$ direction. This kind of electron dynamics forms a new quantum state which I will term as *"Topological Resonance"* (TR) state. In some cases, it is sufficient for an experimental comparison to discuss the considered problems by regarding the particular structures as given in Fig. 12. The interpretation of the fully dynamic behaviour commonly requires extended considerations, but some basic information can be related to the static picture, too, which will be particularly done in the following. At this stage, I will not carry out a comparative discussion of other stripe models in order to limit the size of this article.

The hitherto performed calculations do not permit a direct determination of the spin state. However, from theoretical arguments (Sec. III) it was concluded that the antiferromagnetic spin structure and the CBF state are caused by the same electronic renormalization behaviour resulting in a collinear structure between spin state and CBF state. It was concluded (Sec. IV, A) that the doping of holes occurs within the highest occupied bands with pronounced $s, d_{x^2-y^2}$ hybridizations. This hybridization should prevent any well-ordered antiferromagnetic structure in two dimensions (see Sec. III). Therefore, the antiferromagnetic spin state will not be directly affected by the hole doping. Only renormalization and polarization influences by the local hole state have to be considered. Hence, hole doping leads essentially to a polarization of the local copper spin states but does not annihilate the local copper spin itself. The antiferromagnetic spin coupling is causally related to the Cu-O overlap population as it was concluded in Sec. III. On the other hand, the Cu-O bonding strengths in $x$ and $y$ direction will be anisotropically changed, if a local topological hole state is created (Fig. 2(b)). This means that the antiferromagnetic spin state will be specifically stressed by the topological hole structure caused by the local spin anisotropies, but the local magnetic moments at the copper atoms are nearly unchanged. Consequently, a given topological hole structure creates a corresponding spin structure which is directly related to the given hole topology but on the basis of a continuously existing antiferromagnetic coupling.

First, the f-holes will be neglected (i.e. $n_h \leq 0.125$ holes/copper). In this case, the two kinds of pictograms in Fig. 12 can also be identified with two spin states being oppositely directed. In this way, an alternating spin structure with the period $2a$ should result in $x$ direction along the lines $m_1$, $m_2$ in Fig. 12. The basically highly periodic ordering of the b-holes in $x$ direction with the period $4a$ out of the lines $m_1$, $m_2$ preserves large parts of antiferromagnetic orderings of at least one antiferromagnetic subgrid. Therefore, the continuously existence of large antiferromagnetic orderings in $x$ direction can be presupposed. However, the situation in y-direction is completely different. The alternating b-hole distances 3b and 5b in $y$ direction create only a hole and spin periodicity of $8b$ for a complete b-hole structure (i.e. for $n_h = 0.125$ holes/copper). If the hole concentration is lowered, some b-hole bonds will be broken. Consequently, the ideal spin periodicity of $8b$ along the $y$ direction in Fig. 12 will be modulated in relation to the number of the b-hole vacancies $\Delta n_h^{vac} = \left(0.125 - n_h\right)$ holes/copper.[83] At least for not to large vacancy concentrations, large parts with an antiferromagnetic ordering in $x$ direction should be preserved. This is especially caused by the fact that maintaining double-stripes (neighboured stripes I and II in Fig. 12) which are periodically arranged in $x$ direction is basically favoured (see Ref. [88]).

There is distinct experimental support for the above conclusions. The fact that the local spin moments will be practically unchanged by hole doping may be reflected in $\mu$SR experiments. $\mu$SR experiments probe the magnetic state very locally. B. Nachumi $et$ $al$.[84] concluded that in various systems $La_{1.875}Ba_{0.125-y}Sr_yCuO_4$ (with y=0,0.025,0.065), $La_{1.6-x}Nd_{0.4}Sr_xCuO_4$ (with x=0.125,0.15) and $La_{1.6-x}Nd_{0.4}Ba_xCuO_4$ (with x=0.125) the ordered copper moment is nearly unchanged, $\approx 0.3$ $\mu_B$. Earlier investigations of $La_2CuO_{4-y}$ (with varying y) leads to the same conclusions with a magnetic moment of $\approx 0.5$ $\mu_B$.[85] Furthermore, they showed[84] that in $La_{1.45}Nd_{0.4}Sr_{0.15}CuO_4$ magnetism and superconductivity exist microscopically in the same sample volume which supports the above deduced collinear coupling between hole structure and magnetic state.[86] It is further concluded, that on the time scale of $\mu$SR ($<10^{-6}$ s) the spin order may be an incommensurate static modulation of the antiferromagnetic order. The width of the internal magnetic field distribution $\Delta B$ is a measure of the degree of disorder of the magnetic state. It is shown by $\mu$SR , that in the range of about $n_h = 0.09$ to 0.11 holes/copper $\Delta B$ approaches rapidly zero.[87] This can be interpreted as the formation of strongly ordered and largely extended topological b-hole structures like Fig. 12 with all the degenerate b-hole structures (32 in minimum) being superimposing within a fast fluctuating fully dynamical TR state below the μSR time scale. The life time of particular states as given in Fig. 12 is in the range of some $10^{-14}$ seconds as I will show elsewhere. Hence, the expectation values for the spin and charge states are averaged quantities which are the same in this case for all copper and oxygen sites, respectively.

A strong support of the deduced electronic state arises from neutron scattering experiments. First, one has to give a general remark with respect to elastic (NS) and inelastic (INS) neutron scatterings. It seems that the NS and INS experiments reveal to the same state, however, there is a general difference with respect to the electronic state as given in Fig. 12. As deduced before, a strong coupling between the local magnetic moment and the topological hole state exists. In general that means that excited spin fluctuations will simultaneously excite charge density fluctuations within the strongly coupled b-hole system. The latter is definitely quantized through its strong topological order. That means, for inelastic spin excitations the quantum conditions for the electronic states of the b-hole system will determine the excitation spectrum. If the b-hole system is complete the modulation length of $8b$ in Fig. 12 happens to be a characteristic length within the one-dimensional electronic states (stripes) along I and II in Fig. 12. I will show elsewhere that in the superconducting state inelastic excitations with respect to the b-hole system will only be possible for scattering vectors $\kappa$ having a component in the polarization direction of the b-hole system which is related to the periodicity of the entire b-hole state . Or in other words, the phase of the scattering amplitude must be the same for equivalent lattice points of the b-hole system. Therefore, an equivalent condition as given by Eq. (23) for the NS case of the spin Bragg scattering (see below)  results for INS but now in straight relation to the topological b-hole structure. Hence, for a complete b-hole system the scattering wavelength in Fig. 12 is restricted to the incommensurate modulation vector

$$Q_y = \frac{1}{8}\left(\frac{2\pi}{b}\right) = \varepsilon\left(\frac{2\pi}{b}\right)$$ with respect to the antiferromagnetic reciprocal lattice vector $\tau = (1/2, 1/2)$. If  b-hole vacancies exist (i.e. for $n_h < 0.125$ holes/copper) one has to assume that either certain lines $m_1$, $m_2$ are shifted in $y$ direction in Fig. 12 or that ordered topological b-hole vacancies may occur within the superconducting state. Consequently, particularly increased b-hole distances in $y$ direction in Fig. 12 or ordered b-hole vacancies may lead to changed periodicities of the entire b-hole state with the result that excitation vectors with

$$Q_y = \frac{1}{8+n}\left(\frac{2\pi}{b}\right) = \varepsilon\left(\frac{2\pi}{b}\right)$$ ($n$: positive integer) will be possible. All in all it means that for $n_h \geq 0.125$ holes/copper an INS excitation spectrum will exist which is strongly centered at $n = 0$ and for $n_h < 0.125$ holes/copper one has to expect an excitation spectrum which results from a folding of all possible excitation vectors being related to the particular b-hole periodicities which on their part depend on the b-hole vacancy concentration $\Delta n_h^{\text{vac}}$. Contrary to INS, NS experiments test only the spin state. This means, the elastic spin scattering spectrum will detect exclusively the spin topology which is, however, also decisively determined by the hole topology.

The INS experiments of K. Yamada *et al.*[49] reveal the above conclusions. They found superstructure peaks corresponding to an incommensurability of $\varepsilon = 1/8$ along the particular coordinate axis x or y within the entire area of  0.125 holes/copper $\leq n_h \leq 0.25$ holes/copper. In addition, it is shown that $\varepsilon$ decreases with decreasing $n_h$ for $n_h < 0.125$ holes/copper where simultaneously a strong line broadening is observed if  the hole concentration deviates from $n_h = 0.125$ holes/copper. Both findings can be related to the folding character of the spectra in dependence on the vacancy concentration $\Delta n_h^{\text{vac}}$. For  $\Delta n_h^{\text{vac}} = 0.0625$ holes/copper nominally half of the b-hole pairs in Fig. 12 have been annihilated. For this hole concentration it is possible that two-dimensionally ordered b-hole structures may be maintained further on.[88] These structures result from the structure in Fig. 12 if every second pair of stripes (I,II) is annihilated (parallel configuration)[88] . If additionally b-hole pair states along of one of the existing stripes (I or II) in $y$ direction in Fig. 12 are alternating left and right positioned with respect to the unchanged stripe an ordered b-hole state occurs (zigzag configuration) with a periodicity of $16b$ in $y$ direction. This configuration should be most probably.[88] The periodicity of $16b$ corresponds to a changed modulation wave vector of the original b-hole state corresponding to $n = 8$, i.e. the INS spectrum should be centered near $Q_y = \dfrac{1}{8+(n=8)}\left(\dfrac{2\pi}{b}\right) = \dfrac{1}{16}\left(\dfrac{2\pi}{b}\right)$. That has been indeed found experimentally.[49] However, one has to assume that the parallel and zigzag configurations are energetically not very far from each other so that also parts of the parallel configuration could be found. The elementary cell of the b-hole structure with a parallel configuration of the stripes is $(8a,8b)$[88] so that the ordinary antiferromagnetic scattering vector $\kappa$=(1/2,1/2) may contribute to inelastic scatterings of the b-hole state. Pronounced spectral parts at $\kappa$=(1/2,1/2) are visible in the above experiments (Fig. 4(a) in Ref. [49]). If the hole concentration is further lowered ($n_h < 0.0625$ holes/copper) more and more b-hole couplings will be annihilated. A pronounced two-dimensionally ordered b-hole structure is then not very probable. If the existence of a two-dimensionally ordered b-hole structure is the precondition for the occurrence of superconductivity it means further that for hole concentrations around (above) $\Delta n_h = 0.0625$ holes/copper the conditions for the formation of superconducting states should be distinctly improved. It is experimentally found that the superconducting transition temperature $T_c$ is basically strongly increasing with rising $n_h$ between 0.0625 and 0.125 holes/copper.[49] Moreover, it is important to note that the absolutely highest correlation lengths of the incommensurate structures have been observed for hole concentrations very near to $n_h = 0.125$ holes/copper, i.e. if practically no b-hole vacancies exist.[49,54,89]

The above discussions were related to inelastic excitations of the spin and charge state. Now, the elastic neutron spin Bragg scatterings will be considered which directly reflect only the spin state. It was concluded that the local magnetic moment is nearly unchanged under hole doping. Therefore, the effect of the topological hole state is merely a deformation/modulation of the antiferromagnetic spin state. The stress of the b-hole system on the antiferromagnetic state is strongly different with respect to the $x$ and $y$ coordinate as already discussed above. It was concluded (Sec. III) that the antiferromagnetic spin interactions are directly related to the electronic renormalizations. These renormalizations will be additionally influenced by the b-hole state due to quadrupolar polarizations and directly by the hole density distribution. The latter changes the electronic states predominantly along the $y$ coordinate in Fig. 12 where the holes of the topological b-hole state are placed. It means, the antiferromagnetic couplings will be comparatively small influenced along the $x$ coordinate but strongly asymmetric stressed by the alternating periodicity of the b-hole topology along the $y$ coordinate. Therefore, an antiferromagnetic spin ordering along the x coordinate will be widely conserved, but distinct incommensurate spin modulations have to be expected along the $y$ coordinate in Fig. 12. It is very naturally to assume that the periodicity of these modulations will be directly related to the periodicity of $8b$ of the b-hole system in $y$ direction as long as an undisturbed b-hole state exists (for $n_h$=0.125 holes/copper). If an undisturbed b-hole system exists, the strong symmetry of the b-hole state guarantees a strong symmetry of the spin modulations along the [1,1] and [-1,1] directions of the $CuO_2$ plane, i.e. the modulation lengths along this directions will be identical.

The physical situation turns out to be much more complex if the f-holes have to be included (i.e. for 0.125 holes/copper < $n_h \le 0.25$ holes/copper). The topological f-hole system will additionally stress the antiferromagnetic spin state. In the intermediary concentration range of 0.125 holes/copper < $n_h < 0.25$ holes/copper the f-hole system possess various degrees of freedom to form topological hole structures because there are no definite couplings between the f-holes of a given polarization direction, i.e. within the $f_{||}$-holes or within the $f_\perp$-holes. However, two basically different topological f-hole structures can be distinguished.[83] First of all there are *collinear* f-hole structures, defining here an unit cell of a topologically ordered f-hole system ($f_\parallel$-holes, $f_\perp$-holes) that is collinearly oriented to the b-hole system, i.e. the f-hole system is also strongly oriented along the crystallographic axis *a* or *b*. The second possible structure is a *non-collinear* f-hole structure which is given when the orientation of the ordered f-hole structure deviates from the crystal axes, i.e. the unit cell of an ordered f-hole structure is rotated with respect to the crystallographic axes *a*, *b*. In the collinear case, the spin polarizations caused by the b-hole and f-hole structures interfere constructively whereas in the non-collinear case they may destructively superimpose. The latter means, a topological f-hole structure which is not oriented along the crystallographic axes rotates in the simplest case the axes oriented modulations of the antiferromagnetic state (caused by the b-hole system) out of the crystallographic axes. In consequence, the spin modulation lengths along the [1,1] and [-1,1] directions of the $CuO_2$ plane will not be identical. If the f-hole system is collinear to the b-hole system the influence is merely restricted to possible changes of the modulation length, i.e. some deviations from the incommensurability of $\varepsilon = 1/8$ may occur.

For a theoretical interpretation of the outlined spin modulations one can employ a theoretical description as given by S.W. Lovesey.[90] The essential result is that Bragg scattering occurs for

$$\boldsymbol{\kappa} = \boldsymbol{\tau} + \lambda \cdot \mathbf{Q} \qquad (23)$$

with $\kappa$ the Bragg scattering vector, $\tau$ the vector of the reciprocal lattice and $\mathbf{Q}$ the vector of the spin modulations ($\lambda$ integer numbers). From the here outlined ideas it follows that a strongly ordered topological hole state creates a correspondingly ordered spin state which has an absolute modulation vector corresponding to

$$Q_0 = \varepsilon \left( \frac{2\pi}{a} \right) \quad \text{or} \quad Q_0 = \varepsilon \left( \frac{2\pi}{b} \right) \qquad (24)$$

if only b-holes exist or if the f-holes are collinearly directed to the b-holes. A modulation vector of $\mathbf{Q} = (0, Q_0)$ results if the b-holes are polarized in *y* direction (Fig. 12) and of $\mathbf{Q} = (Q_0, 0)$ if the b-holes are polarized in *x* direction. In the pure b-hole case or if the f-hole structure possesses the same periodicity of 8*b* (8*a*) as the b-hole system it is very natural that the spin modulations will possess the same modulation length, i.e. $\varepsilon = 1/8$ should be observed. This has been experimentally proven many times.[89] The fact that the measurements do not always reveal the absolutely exact value of $\varepsilon = 1/8$ depends probably on the comparatively small coherence lengths of the incommensurate modulation state. Coherence lengths which correspond to 3 - 14 unit cells of the b-hole system have been observed at best.[89] In addition, for $n_h$ values slightly below 0.125 holes/copper the occurrence of b-hole vacancies may reduce the $\varepsilon$ values (see below). Nevertheless, there is also the possibility that the spin modulation length is incommensurately influenced by the f-hole system, especially if the f-hole period is different to the b-hole period. J.M. Tranquada *et al.*[46] found increased values of $\varepsilon$ ($\varepsilon > 1/8$) for larger hole concentrations in $La_{1.6-x}Nd_{0.4}Sr_xCuO_4$ (for x = 0.15, 0.20), i.e. for larger f-hole concentrations.

For hole concentrations of $n_h = 0.15625$ holes/copper, there is already the possibility that various highly symmetrical and largely extended topological f-hole structures can be formed having collinear or non-collinear character. If the symmetry of the material is tetragonal ($a = b$) it is not very likely that the f-holes form non-collinear structures which include asymmetric modulation structures with respect to the axes [1,1] and [-1,1], but they can not definitely be excluded. It seems more probable, however, that such non-collinear structures will be formed if the crystal is orthorhombically deformed ($a \neq b$). If the b-hole system in Fig. 12 represents an internal coordinate system the modulation vector defined by the axes [1,1],[-1,1] is given by

$$\mathbf{Q} = \begin{pmatrix} Q_{[1,1]} \\ Q_{[-1,1]} \end{pmatrix} = Q_0 \begin{pmatrix} \sin(45^0 - \theta) \\ \cos(45^0 - \theta) \end{pmatrix} \tag{25}$$

with $\theta$ the rotation angle of the polarization direction of the spin modulations from the given polarization axis of the b-hole system. $\theta$ should be in proportion to the difference of the angle of the polarization direction between the f-hole and b-hole system. The collinear case corresponds to $\theta = 0$. In consequence, if the f-hole system causes a rotation of the spin modulations by $\theta \neq 0$ the wave vector components along the axes [1,1],[-1,1] will be different. This phenomenon has indeed been observed by Y.S. Lee et al.[54] within the orthorhombic La$_2$CuO$_{4+x}$ (stage 4) material. They found a rotation angle of $\theta = 3.3°$ which corresponds to different incommensurate modulations along the two orthorhombic axes ([1,1],[-1,1]) corresponding to $\varepsilon = 0.114$ and $\varepsilon = 0.128$ (in orthorhombic units). The absolute modulation vector $Q_0$ is correspondingly $\varepsilon = 0.121$. Further indications for such rotations ($Y$ shift) are found in La$_{2-x}$Sr$_x$CuO$_4$.[91] A recently found $Y$ shift in La$_{1.88}$Sr$_{0.12}$CuO$_4$ [92] can not be traced back to the f-hole system, because there are no f-holes for x = 0.12. If x = 0.12 reflects exactly $n_h$ = 0.12 holes/copper which is lower than the exact value of 0.125 holes/copper of the undisturbed b-hole state b-hole vacancies occur or the b-hole structure has to be modified. If non-collinear ordered b-hole vacancies are formed the influence is principally the same as already discussed for the f-holes resulting in $\theta \neq 0$. If one assumes that complete b-hole strings in $x$ direction in Fig. 12 have been annihilated than a rearranged b-hole structure may be formed where nominally every 6[th] b-hole string in $x$ direction ($m_1, m_2$) in Fig. 12 can be considered to be shifted by one lattice spacing in $y$ direction. It means, the b-hole – b-hole coupling in $y$ direction is widely lost at this places with the result that the two neighboured (well-ordered) b-hole structures may almost freely shift against each other in $x$ direction to occupy an appropriate CBF position. This may explain the frequently found incommensurability of only $\varepsilon \approx 0.120$ as well as the $Y$ shift of about $\theta = 3°$.[92] An alternative explanation for the $Y$ shift was given under the contrary assumption of strong spin density modulations where kings (steps) within stripes are assumed to be responsible for these rotations.[58] This model suggests a continuously increasing $Y$ shift with increasing hole density (for $n_h > 0.125$ holes/copper). From the here proposed topological hole state it results that for $n_h$ = 0.25 holes/copper all f-hole places are occupied (Fig. 12) which means that this collinear f-hole structure can only create a collinear f-hole spin structure, i.e. in this case no $Y$ shift should be found at least for a tetragonal crystal structure.

Much more problematic than the Bragg scatterings of the spin states proves to be the Bragg scatterings with respect to the charge density modulations. The dynamic TR state consists of a superposition of degenerate states like this one in Fig. 12. Especially it means that a superposition of equivalent b-hole states occurs which are phase shifted to each other and which commonly can exist in two polarization states ( polarized in $x$ and $y$ direction). The life time of particular hole configuartions as given in Fig. 12 is in the range of some $10^{-14}$ seconds as already mentioned above. Therefore, it can be assumed that the scattered neutron feels an averaged b-hole structure, because the interaction time of the neutron with the target is much longer. Under these conditions the Bragg scattering amplitude concerning the b-hole system can be represented by

$$F(\text{b} - \text{holes}) = \sum_p^{x,y} c_p \left[ \sum_j c_j \left( \sum_{u=1}^{4} \sum_{\tau_{b-holes}} \int dV \, n_{\tau_{b-holes}} \, e^{i(\tau_{b-holes} - \kappa)\tau + \tau_{b-holes} \cdot \Delta\tau_{u,j}} \right) \right]_p \tag{26}$$

with $\tau_{b-holes}$ a reciprocal lattice vector and $n_{\tau_{b-hole}}$ the corresponding Fourier-coefficients. The summations occur with respect to four identical but phase shifted b-hole subgrids (see Fig. 12), the particular b-hole configurations $j$ of a given b-hole polarization with the weighting coefficients $c_j$ and the two possible polarization directions $x$, $y$ of the b-hole system weighted by $c_p$. The scattering amplitude is decisively changed by the additional relative phase shifts $\tau_{b-hole}\Delta\mathbf{r}_{u,j}$ of the b-hole states. In the fully dynamically TR state all possible b-hole configurations (32 in minimum) contribute to the scattering amplitude. In this case commonly no superstructure will be detected, because all oxygen atoms possess the same b-hole density on average if the 32 b-hole states are energetically degenerate. A constrained dynamical TR state, however, may result in a continuously existing superstructure also in the time averaged state. If one assumes that the b-hole system occupies basically only one polarisation state ($x$ or $y$ polarized) and the superimposed b-hole configurations differ e.g. by a phase factor which is predominantly restricted to shifts of $\Delta r_{u,j}$ being perpendicular to the given reciprocal lattice vector $\tau_{b-holes}$ then $\tau_{b-holes}\Delta\mathbf{r}_{u,j}$ is zero and a superstructure of the b-hole state may be detected. Of special interest is the case if $\tau_{\text{b-holes}}$ lies along the polarisation direction of the b-hole state ($y$ direction in Fig. 12) and the superimposed b-hole structures are restricted to states where the b-hole system is exclusively phase shifted along of the perpendicular coordinate axis ($x$ coordinate in Fig. 12). It results a time averaged b-hole superstructure with a period of 4 lattice spacing in y-direction corresponding to a scattering vector of $\kappa = 2 \cdot \tau_{ob-holes}$ where $\tau_{ob-holes} = 2\pi/8b$ is the shortest reciprocal lattice vector of the b-hole system in $y$ direction in Fig. 12 .

Very recently, the first direct proof of the existence of incommensurate charge density modulations was given by NS experiments. H. A. Mook et al.[93] have shown that incommensurate crystal Bragg peaks occur for 8 lattice spacing within the material $YBa_2Cu_3O_{6.35}$ ($T_c$=39 K). The crystal used in these experiments were twinned so that the crystal axes $a$ and $b$ could not be distinguished. For this oxygen content one can assume $n_h \gtrsim 0.0625$ holes/copper, i.e. a b-hole structure with ordered vacancies may be formed.[88] If one assumes that the b-hole structure is topologically pinned one has to start from a b-hole system which exists only within of one polarization state. If one further assumes that such b-hole states superimpose which result from a phase shift of the b-hole system exclusively along of the polarization of the b-hole state a superstructure with a period of 8 lattice spacing can be observed. The reason is that for $n_h = 0.0625$ holes/copper every second pair of stripes (I,II) in Fig. 12 can be assumed to be vacant.[88] In this case a time averaged superstructure with a period of 8 lattice spacing occurs in $x$ direction in Fig. 12 what is in line with the experiments of H. Mook et al. . Additionally, inelastic spin excitations with $\epsilon$=1/16 were observed.[93] This is in agreement with the experiments of K. Yamada et al.[49] for a concentration of hole vacancies of $\Delta n_h = 0.0625$ holes/copper which were related to the superstructure formation of $16b$ (zigzag configuration) in polarization direction of the b-hole state (see discussion in relation to the experiments of K. Yamada et al.) .[88] Besides that, larger areas of hole free regions exist for the above concentration of hole vacancies giving larger extended and widely undisturbed antiferromagnetically ordered substructures. Additionally, parallel configuration parts may exist.[88] Both effects give rise to inelastic spin scatterings with the antiferromagnetic scattering vector $\kappa$=(1/2,1/2) (see above). This was also reflected in the INS experiments by a broad spectral distribution around the antiferromagnetic position (1/2,1/2,c).[93] The time dependence of the b-hole state within the dynamic TR state is not deciding for the inelastic spin excitations, because only the general quantum conditions of the b-hole state are the determining quantities.

The general ideas mentioned above are more impressively reflected by the very recently experiments of M. Fujita et al. .[94] Simultaneously charge density (CDW) and spin density (SDW) orders were measured. They investigated by NS experiments the relationship between crystal structure, CDW/SDW orders and superconductivity in $La_{1.875}Ba_{0.125-x}Sr_xCuO_4$ with x = 0.05, 0.06, 0.075, and 0.085. It can be supposed that $n_h \cong 0.125$ holes/copper is basically given and the crystal structure is modified from the LTT (low-temperature tetragonal) to the LTO (low-temperature orthorhombic) phase with rising x. Incommensurate superlattice peaks are found for the CDW orders corresponding to 4 lattice spacing and for SDW orders corresponding to $\varepsilon \cong 1/8$. The most remarkable result is, however, that CDW and SDW peaks are simultaneously found in the LTT phase whereas only a SDW peak but no CDW peak is found in the LTO phase. These surprising findings support the proposal introduced in this paper that the spin moments are unchanged during hole doping and that a dynamic stressing of the spin state by ordered b-hole structures occurs within the TR state. If one identifies the LTT phase (low $T_c$ values) as the phase where the b-hole structure is pinned with the result that the b-hole state is polarized only in one direction and the superposition of the b-hole structures result predominantly from phase shifts of the b-hole state in $x$ direction in Fig. 12 then a CDW and SDW order is seen by elastic neutron scatterings. Neutron scattering vectors of $\kappa = 2 \cdot \tau_{\text{b-holes}} = n \cdot (1/4) \cdot 2\pi/b$ ($\tau_{\text{b-holes}} = n \cdot (1/8) 2\pi/b$ in Fig. 12, see above) result for the time averaged b-hole state (CDW order). Contrary to the b-hole state, the copper spins are locally fixed, continuously existing and the entire spin structure within the $CuO_2$ plane is dynamically but coherently stressed by a fluctuating b-hole structure. It means, there occurs a continuously existing superstructure of the entire spin state with the modulation vector $Q_{\text{b-holes}} = 2\pi/8b, (=2\pi/8a)$ giving rice to quasi-stationary Bragg reflections. Therefore, $\kappa = \tau \pm Q = \tau \pm (\varepsilon = 1/8) \cdot 2\pi/b$ with the antiferromagnetic lattice vector $\tau = (1/2, 1/2)$ is expected for the SDW order. If the LTO phase (high $T_c$ values) represents the fully dynamic TR state with phase shifts of the b-hole structures also in $y$ direction and two existing polarization states of the b-hole system then not any superstructure of the b-hole system can be observed (see above). In consequence, no incommensurate CDW peak will be measured. The measured incommensurability of the magnetic superstructure is again not influenced by the b-hole dynamics based on the quasi-stationary Bragg reflections. Therefore, in the LTO phase an incommensurate spin modulation with $\varepsilon = 1/8$ is also measured. In summary, the experiments of M. Fujita et al. reflect impressively the here deduced b-hole state as the basis structure within the TR state. That the LTT phase is connected with a pinned b-hole state may be explained from the fact, that a b-hole system which exists only within of one polarization state may principally reduce orthorhombic distortions by appropriate choosing the polarization direction (along the $a$ or $b$ axis).

The new experiments of M.Fujita et al.[95] may give further evidence that ordered b-hole structures begin to form near $n_h = 0.0625$ holes/cooper which are simultaneously the precondition for the occurrence of superconductivity. They found a strong correlation between the occurrence of a parallel incommensurability $\varepsilon$ ($\theta \cong 0$) and the onset of superconductivity at a concentration of $n_h$ near 0.0625 holes/cooper. If there are widely ordered b-hole structures the spin modulation vector is basically oriented near a coordinate axis, i.e. near the b-hole polarization axis with $\theta$ being zero or not fare from zero (see above). For hole configurations where widely ordered b-hole structures are lost (for $n_h < 0.0625$ holes/cooper) the antiferromagnetic spin state dominates with possible modulations along one orthorhombic axis ($\theta = 45°$, diagonal modulation).

Very recent experiments on the system $La_{1.875}Ba_{0.125-x}Sr_xCuO_4$ published by M.Fujita et al.[96] verify an $Y$ shift also for the CDW order within the low-temperature less-orthorhombic phase [LTLO] with x=0.075. The most remarkable result of this paper is that there is a close relation between $Y$ shifts of the CDW order ($\theta_{\text{ch}}$) and of the SDW order ($\theta_{\text{m}}$) which are both directly related to the orthorhombic distortions of the crystal. If a tetragonal structure (LTT) exists there is no reason why the ordered b-hole structure in Fig. 12 should be strongly disturbed by structural influences. In consequence, charge density modulations along the $b$ axis in Fig. 12 are observed, $\theta_{\text{ch}}=0$. Orthorhombic distortions, however, may break up a large extended b-hole structure in particular b-hole structures which are phase shifted to each other. If the b-hole structure is cut along the $y$ coordinate in Fig. 12 the two particular b-hole structures may be phase shifted to each other e.g. by the lattice vector ($1a, 1b$) in order to occupy equivalent CBF positions. The expended energy for such a process is not very high because the necessary hole - hole pair interactions according to Figs. 6(a), (1) and (2) can be continuously realized. If the extent of the particular b-hole subsystems is equivalent then the effective scattering amplitudes with respect to a reciprocal lattice vector of the b-hole system in $y$ direction e.g. $\kappa = 2 \cdot \tau_{ob-holes}$ ( $\tau_{ob-holes} = 2\pi/8b$ ) (see above) must be multiplied by $\cos\left(\left(2 \cdot \tau_{ob-holes}\right) \cdot \left(b/2\right)\right)$ which is responsible for the relative phase shift of the b-hole subsystems in $y$ direction leading to a reduction of the total scattering amplitudes. If a phase shift corresponding to the lattices vector $(0,2b)$ or $(2a,2b)$ occurs the factor is $\cos\left(\left(2 \cdot \tau_{ob-holes}\right) \cdot b\right)$. This leads even to an annihilation of the scattering amplitudes since $\cos\left(\left(2 \cdot \tau_{ob-holes}\right) \cdot b\right) = 0$ . Therefore, it cannot be expected that the overall measured CDW superstructure is exactly aligned along the $y$ direction, $\theta_{ch} \neq 0$. Even this is reflected in the experiments.[96]

One of the most important characteristics of the deduced electronic state is the existence of energetically deep lying hole states. They are given by the b-hole system which is energetically shifted to lower energies by the magnitude of $|E_{C(tot)}|$ (>160 meV). The b-hole system consists of stripes along particular ..Cu-O-Cu.. bonding lines which are intrinsically coupled by the correlation energy $E_C$ (in $x$ direction in Fig. 12). In the non-superconducting state, the electronic states which create a particular b-hole stripe form an independent one-dimensional electronic subspace. The coupling between the stripes by $E_C$ occurs merely as a mean field influence, i.e. no *phase stiffness* is reached between the stripes. Therefore, every electronic movement perpendicular to the direction of the stripes will feel $2 \cdot |E_{C(tot)}|$ (> 320 meV) as a gap,[97] because the charge carriers have to be promoted from the given one-dimensional electronic subspace. In consequence, the effective mass $m_\perp$ for every electronic movement perpendicular to the direction of the stripes is infinite ($m_\perp \rightarrow \infty$), i.e. away from the polarization direction ($y$ direction in Fig. 12) of the b-hole system. However, these conclusions are only strongly valid in the case of pinned b-hole states, i.e. dynamically constrained TR states where b-hole configurations exist being polarized only into one direction $x$ or $y$. The effective mass $m_{\parallel}$ along the stripes is finite based on the strong collective nature of the electronic states within the stripes. Contrary to that, the effective mass of the f-hole system is always finite along and perpendicular to the polarization direction of the stripes of the b-hole system, because $f_{\parallel}$-holes and $f_\perp$-holes exist always simultaneously. Therefore, the f-hole system exhibits basically a two-dimensional electronic behaviour.

A decisive proof can be given by Hall effect measurements, because the b-holes cannot contribute to the Hall current in the pinned case whereas the Hall current of the f-hole system will not significantly be changed if the topological hole state is pinned. Even this is found by Hall effect measurements of T. Noda *et al.* on $La_{2-x-y}Nd_ySr_xCuO_4$ with x = 0.10, 0.12, 0.13, 0.14 and y = 0.4, 0.6.[98] Below a characteristic temperature $T_o$ the Hall coefficient $R_H = \dfrac{\rho_{xy}}{B_z}$ changes significantly in dependence on x. For x = 0.1, 0.12 $R_H$ drops to zero when lowering the temperature whereas for x = 0.13, 0.15 $R_H$ remains finite for T→0. This reflects the different behaviour of the b-hole and f-hole system where below a pinning temperature ($T_o$) the Hall current is zero for a pinned b-hole state ($k_BT_o$ represents an activation barrier for de-pinning) but basically finite for the f-hole system. For x = 0.1, 0.12 ($n_h$ = 0.1, 0.12 holes/copper) only b-holes exist whereas for x = 0.13, 0.15 ($n_h$ = 0.13, 0.15 holes/copper) b-holes and f-holes exist. The doping dependence of the in-plane resistivity ($\rho_{xx}$, $\rho_{yy}$) does not show any distinct behaviour at x = 1/8. This may indicate the gapless behaviour of the electronic states along the stripes, i.e. a finite effective mass $m_{\parallel}$ of the b-hole system along the stripes as a consequence of a collective electronic behaviour. If the pinning of the b-hole system is suppressed then the two-dimensionally fluctuating b-hole structures (fully dynamic TR state) can contribute to the Hall current. This is impressively reflected very recently by the experiments of S. Arumugan *et al.* .[99] It was shown that a pressure induced de-pinning in $La_{1.48}Nd_{0.4}Sr_{0.12}CuO_4$ which is a consequence of a structural transition from the LTT to the LTO phase (see the above discussions of the NS experiments of M. Fujita *et al.*) leads to an additional Hall conductivity.

Very recently, from scanning tunneling microscopy (STM) on the nanoscale region the existence of two different domain structures within the $CuO_2$ planes has been proposed.[100] There are so-called α-domains with a mean energy gap of $\overline{\Delta} = 35.6 - 50$ meV in the superconducting state and so-called β-domains which represent lower energetic states. The authors investigated two different hole concentrations with $n_h = 0.14$ and $0.18$ holes/copper and found that the number of α-domains are much smaller for $n_h = 0.14$ holes/copper in comparison to $n_h = 0.18$ holes/copper. The α-domains can be identified by an existing electronic f-hole subsystem which tends to a stronger clustering of the topological f-hole states if the f-hole concentration

$$n_h \text{ [f-holes]} = (n_h - n_h \text{ [b-holes]}) \text{ holes/copper} = (n_h - 0.125) \text{ holes/copper} \qquad (27)$$

is stronger lowered in comparison with a completed f-hole structure in the case of $n_h = 0.25$ holes/copper (Fig. 12). The existence of a collective electronic state (f-holes) near the Fermi-surface along the $FS_{||}$-lines in Fig. 12 allows an electron tunneling from or under inclusion of this extended collective electronic state, also in the case of a local nanoscale tunneling. That means, the total hole self-energy $\Sigma E_S$ along a FS-line will not dramatically be changed if the additionally hole state which is created during the tunneling process is largely distributed along the extended collective f-hole state ($E_S \sim q_h^2$). If no collective electronic f-hole state exists (β-domains), the tunneling occurs from a local one electron state, i.e. the additionally created hole is strongly localized, too. In this case an additional localization energy corresponding to a singular hole results during the electron tunneling from the Fermi surface which is not very fare from $E_S$.

With respect to the above STM experiments J. Zaanen[101] has metaphorically characterized the electronic state as an electronic two fluid state comparable to a "mixture of oil and vinegar". The here deduced electronic state can be characterized more strikingly by a two fluid state of strongly interacting but non-mixing fluids. Considering a static picture as given in Fig. 12 one can say that the one fluid state (b-holes) is an one-dimensional fluid which flows within a valley and the other fluid is a two-dimensional fluid ($f_{||}$-holes, $f_{\perp}$-holes) on a raised ground plateau with strong couplings between the two fluids by topological quantum constraints.

## APPENDIX A: OXYGEN ENERGIES IN DEPENDENCE ON THE HOLE DOPING

In the following changes of the oxygen atomic energies in dependence on the oxygen valence state and the surrounding charge distribution are investigated. Under the precondition of an isotropic outer charge distribution ($O_h$-symme-try) one can state that the absolute value of the electronic energy increa-ses nonlinearly with increasing number of valence electrons (Fig.13). However, in the range between 8 and 7 valence electrons the nonlinearity is very small. If the symmetry of the outer charge distribution is lowered ($C_{4v}$-symmetry), two essential results can be outlined. Firstly, an energy lowering occurs basically in proportion to the value of the outer EFG (Fig. 14). Beyond that, the energy is stronger de-creased when the number of valence electrons is lowered. The order of magnitude of these energy shifts is enough to exceed some nonlinear rising while reducing the number of valence electrons in the isotropic case at least from 8 up to about 7 valence electrons. The positively charged copper atoms have to be included in the next neighbour charges, too, but that does not basically change the conditions. It was found that the increase in energy, caused by a reduced absolute ligand field strength through these positive charges is largely compensated for by a simultaneous energy decrease, which is the result of a further reduction of the local symmetry. Of course, it cannot exactly be proven that these considerations also hold to be true for the oxygen valence states within the $CuO_2^{2-}$ plane but these results give well-founded arguments that a nearly linear relation between valence population and total atomic energy exists in a certain valence range. The oxygen atoms tend to the -2 charge state within the undoped $CuO_2^{2-}$ plane, i.e. 8 valence electrons, and the doping of holes leads to hole states with predominant oxygen character. Hence, it seems justified to assume that a nearly linear relation between hole density $q_h$ and onsite atomic energy exists up to nominally one hole per oxygen site ($q_h = 1$).

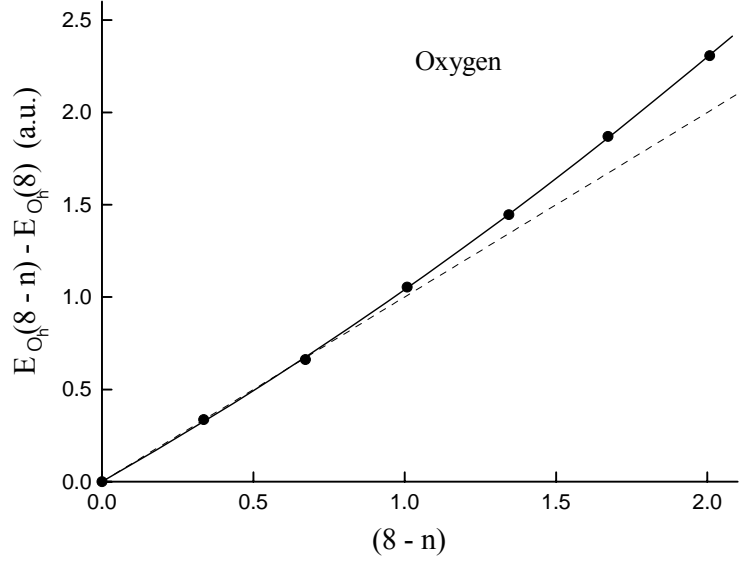

FIG. 13. Dependence of the electronic energy $E$ of the oxygen atom on the number of holes $(8 - n)$ with $n$ being the number of valence electrons (PA). $O_h$-symmetry of the surrounding charges is maintained.

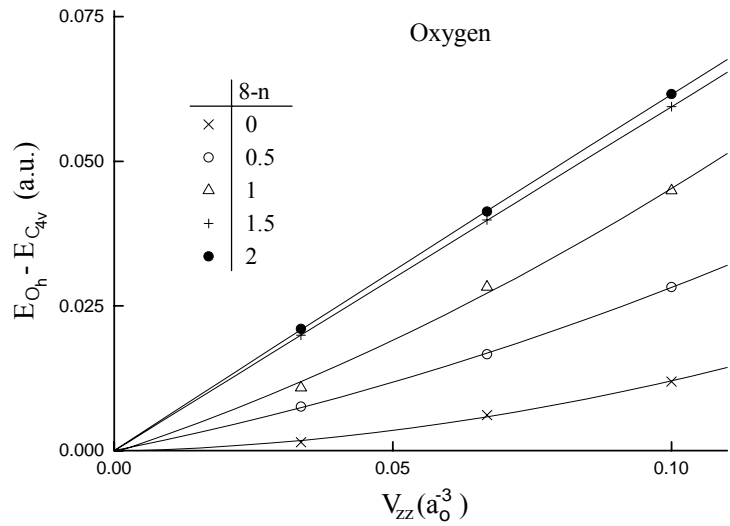

FIG. 14. Shift of the electronic energies of the oxygen atom when reducing the symmetry of the surrounding charges. It is depicted for different numbers of holes in dependence on degree of the reduced symmetry given by the electric field gradient ($V_{zz}$) at the oxygen atom (PA).

## APPENDIX B: COULOMB HOLE-HOLE MONOPOLE INTERACTIONS OF HIGHER COORDINATION SPHERES

In the following the total Coulomb monopole interaction of a hole with holes in higher coordination spheres $s$ $(s \geq 2)$ is considered. When doing so, one has to calculate the interactions with holes placed in $x$ direction with $i \in [i,s]$, $i \in [i,-s]$ and $i$ ranging from $-s$ to $s$ and holes placed in $y$ direction with $j \in [s,j]$, $j \in [-s,j]$ and $j$ going from $-s$ to $s$ . If one intends to compare the nominal Coulomb interaction per hole and per coordination sphere between the regularly square and rectangular hole lattices the sums

$$2 \cdot \sum_{i=1}^{s-1} E(i,s) = 2 \cdot \frac{\left(q_h^{'}\right)^2}{\sqrt{i^2 a_h^2 + s^2 b_h^2}} \text{ and } 2 \cdot \sum_{j=1}^{s-1} E(s,j) = 2 \cdot \frac{\left(q_h^{'}\right)^2}{\sqrt{s^2 a_h^2 + j^2 b_h^2}} \tag{B1}$$

have to be calculated with $a_h$ and $b_h$ being the hole lattice constants in $x$ and $y$ direction, respectively. Here, the axes and diagonal values $(s,0)$, $(-s,0)$, $(0,s)$, $(0,-s)$, $(s,s)$, $(s,-s)$, $(-s,s)$, $(-s,-s)$ have been omitted. The axes and diagonal values belong to similar hole-polygons in every coordination sphere and, therefore, they obey an equivalently qualitative relation as already given by Eq. (16). Hence one has to compare only the sums according to Eqs. (B1) for the square and rectangular case of the topological hole grids. For the hole concentration of $n_h = 0.125$ holes/copper the hole lattice constants are $a_h = b_h = \sqrt{8}a$ in the square case and $a_h = 2a$ , $b_h = 4a$ in the rectangular case. Then, the difference of the Coulomb interactions according to Eqs. (B1) between the square and rectangular hole lattices is given by:

$$\Delta E_{Q,R}(s) = E_Q(s) - E_R(s)$$
$$= \frac{\left(q_h^{'}\right)^2}{a} \left( \sum_{i=1}^{s-1} \left( \frac{1}{\sqrt{2}\sqrt{i^2 + s^2}} - \frac{1}{\sqrt{i^2 + 4s^2}} \right) + \sum_{j=1}^{s-1} \left( \frac{1}{\sqrt{2}\sqrt{s^2 + j^2}} - \frac{1}{\sqrt{s^2 + 4j^2}} \right) \right) . \tag{B2}$$

In Fig. 15 the normalized energy difference $\Delta E_{Q,R}^{'}(s) = \Delta E_{Q,R}(s) \cdot a \big/ \left(q_h^{'}\right)^2$ is depicted. For high $s$ values the sums in Eq. (B2) can be replaced by integrals and the limiting quantity of $\Delta E_{Q,R}^{'}(s \to \infty)$ can be calculated. From the $\Delta E_{Q,R}^{'}(s)$ curve in Fig. 15 two important results are obtained. First, the Coulomb hole-hole monopole interaction is basically higher for a square hole lattice in comparison to the rectangular hole lattice. Therefore, a rectangular hole lattice is energetically always favoured. Secondly, the difference in the Coulomb repulsive energy between the two hole lattices is nearly constant for each coordination sphere with $s > 2$.

In a next step the difference of the Coulomb interaction between a regular rectangular hole structure and a disturbed rectangular hole structure is considered. The parallelogram and trapezium hole structures differ from the rectangular hole structure by the fact that some hole positions are shifted by one lattice spacing $b(= a)$ along the $y$ direction (Fig. 12). A pair of hole positions $(i,j)$ and $(i,-j)$ are considered. It can be stated that if the hole shift of one lattice spacing $b$ results in an absolute increase of the $y$ coordinate for one of the two hole positions an equivalent absolute de-crease of the $y$ coordinate by one lattice spacing occurs in the other hole position (Fig. 12). This is true within the trapezium hole structure as well as the parallelogram hole structure. Therefore, for these positions the Coulomb interactions in Eq. (B1) can be replaced by a mean Coulomb interaction according to

$$E(s,j) = \left( \frac{\left(q_h^{'}\right)^2}{\sqrt{s^2 a_h^2 + (j \cdot b_h - b)^2}} + \frac{\left(q_h^{'}\right)^2}{\sqrt{s^2 a_h^2 + (j \cdot b_h + b)^2}} \right) \bigg/ 2 \tag{B3}$$

$$E(i,s) = \left( \frac{\left(q_h^{'}\right)^2}{\sqrt{i^2 a_h^2 + \left(s \cdot b_h - b\right)^2}} + \frac{\left(q_h^{'}\right)^2}{\sqrt{i^2 a_h^2 + \left(s \cdot b_h + b\right)^2}} \right) \Big/ 2 \quad . \tag{B4}$$

In the following the difference in the Coulomb interaction between a regular rectangular hole structure ($R$) and a disturbed rectangular structure ($R'$) is considered where in the latter all Coulomb interactions have been replaced by Eqs. (B3),(B4). It results:

$$\Delta E_{R,R'}(s) = E_{R'}(s) - E_R(s)$$

$$= \frac{\left(q_h^{'}\right)^2}{a} \left( \begin{array}{c} \displaystyle\sum_{i=1}^{s-1} \left( \frac{1}{\sqrt{i^2 + 4s^2}} - \left( \frac{1}{\sqrt{i^2 + \left(2s + \frac{1}{2}\right)^2}} + \frac{1}{\sqrt{i^2 + \left(2s - \frac{1}{2}\right)^2}} \right) \Big/ 2 \right) \\[20pt] + \displaystyle\sum_{j=1}^{s-1} \left( \frac{1}{\sqrt{s^2 + 4j^2}} - \left( \frac{1}{\sqrt{s^2 + \left(2j + \frac{1}{2}\right)^2}} + \frac{1}{\sqrt{s^2 + \left(2j - \frac{1}{2}\right)^2}} \right) \Big/ 2 \right) \end{array} \right) . \tag{B5}$$

In Fig. 15 the normalized energy difference

$$\Delta E_{R,R'}^{'}(s) = \Delta E_{R,R'}(s) \cdot \frac{a}{\left(q_h^{'}\right)^2}$$

is depicted. It can be seen that the absolute energy differences are much smaller in comparison with the corresponding differences between a squared and rectangular hole structure. The most important result is, however, that the energy difference approaches very rapidly zero with rising coordination sphere. Already for $s = 3$ an energetic difference in the order of only 10 meV has to be expected which is negligibly small, especially if one includes an additional coulomb shielding being realistic which may effect the total coulomb interaction at least in this order.

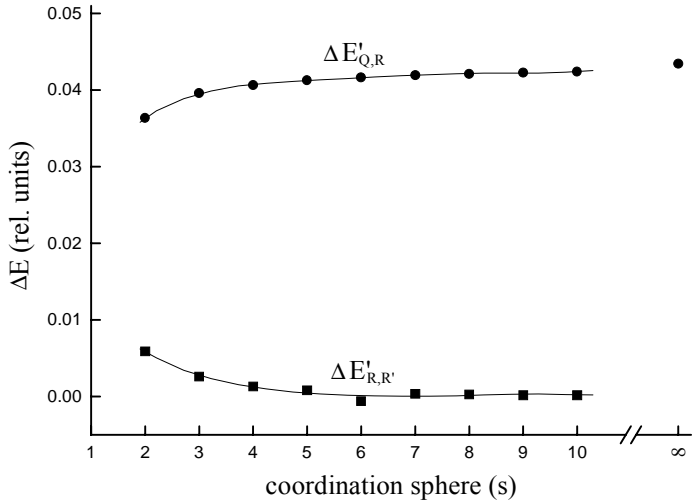

FIG. 15. Normalized energy difference $\Delta E_{Q,R}^{'}(s) = \Delta E_{Q,R}(s) \cdot a / \left(q_h^{'}\right)^2$ per hole of the hole-hole Coulomb monopole interaction between quadratic and rectangular hole structure in dependence on the coordination sphere $s$ (for $n_h = 0.125$ holes/copper). Normalized energy difference per hole $\Delta E_{R,R'}^{'}(s) = \Delta E_{R,R'}(s) \cdot a / \left(q_h^{'}\right)^2$ of the hole-hole Coulomb monopole interaction between a regular rectangular and a disturbed rectangular hole structure in dependence on the coordination sphere $s$ (for $n_h = 0.125$ holes/copper).

# REFERENCES

[1] P. Prelovsek, T.M. Rice and F.C. Zhang, J. Phys. C 20, L229 (1987).

[2] W. Weber, Phys. Rev. Lett. 58, 1371 (1987).

[3] A.S. Alexandrov, P.P. Edwards, Physica C 331, 97 (2000).

[4] P.W. Anderson, Science 235, 1196 (1987); P.W. Anderson, G. Baskaran, Z. Zou, and T. Hsu, Phys. Rev. Lett. 58, 2790 (1987); G. Baskaran, Z. Zou, and P.W. Anderson, Solid State Commun. 63, 973 (1987); S.A. Kivelson, D.S. Rokhsar, and J.P. Sethana, Phys. Rev. B 35, 8865 (1987); P.W. Anderson, *Theory of Superconductivity in the High-$T_c$ Cuprates* (Princeton University Press, Princeton, 1997).

[5] J.R. Schrieffer, X.G. Wen, and S.C. Zhang, Phys. Rev. B 39, 11663 (1989); A. Kampf and J.R. Schrieffer, *ibid*. 41, 6399 (1990).

[6] V.J. Emery, Phys. Rev. Lett. 58, 2794 (1987); V.J. Emery and G. Reiter Phys. Rev. B 38, 4547 (1988); V.J. Emery and G. Reiter, *ibid*. 38, 11938 (1988).

[7] A.J. Millis, H. Monien, and D. Pines, Phys. Rev. B 42, 167 (1990); P. Monthoux, A. Balatsky, and D.Pines, Phys. Rev. Lett. 67, 3448 (1991).

[8] E. Dagotto, A. Nazarenko, and A. Moreo, Phys. Rev. Lett. 74, 310 (1995).

[9] N.E. Bickers, D.J. Scalapino, and S.R. White, Phys. Rev. Lett. 62, 961 (1989).

[10] E. Dagotto, Rev. Mod. Phys., 66, 763 (1994).

[11] J. Zaanen, M.L. Horbach, and W. van Saarloos, Phys. Rev. B 53, 8671 (1996).

[12] V.J. Emery, S.A. Kivelson, and O. Zachar, Phys. Rev. B 56, 6120 (1997).

[13] Ch. Nayak and F. Wilczek, Phys. Rev. lett. 78, 2465 (1997).

[14] C.M. Varma, S. Schmitt-Rink, and E. Abrahams, Solid State Commun. 62, 681 (1987).

[15] H.-B. Schüttler, M. Jarrell, and D.J. Scalapino, Phys. Rev. Lett. 58, 1147 (1987).

[16] W. Weber, Z. Phys. B 70, 323 (1988); W. Weber, A.L. Shelankov, and X. Zotos, Physica C 162-164, 307 (1989).

[17] R. Friedberg and T.D. Lee, Phys. Rev. B 40, 6745 (1989); Phys. Lett. A 152, 417 (1991); *ibid*. 152, 423 (1991).

[18] J. Frank, in *Physical Properties of High Temperature Superconductors IV*, edited by P.M. Ginsberg (World Scientific, Singapore, 1994).

[19] A. Nazarenko and E. Dagotto, Phys. Rev. B 53, R 2987 (1996); N. Bulut and D.J. Scalapino, *ibid*. 54, 14971 (1996). T. Sakai, D. Poilblanc, and D.J. Scalapino, *ibid*. 55, 8445 (1997).

[20] X.-G. Wen, F. Wilczek, and A. Zee, Phys. Rev. B 39, 11413 (1989).

[21] R.B. Laughlin, Physica C 234, 280 (1994).

[22] B. Batlogg, Phys. Today 44, 44 (1991).

[23] W. E. Pickett, Rev. Mod. Physics 61, 433 (1989).

[24] Y. Guo, J.-M. Langlois, and W.A. Goddard, Science 239, 896 (1988).

[25] R.L. Martin, J. Chem. Phys. 98, 8691 (1993).

[26] J. Casanovas, J. Rubio, and F. Illas, Phys. Rev. 53, 945 (1996); D. Muñoz, F. Illas, and I. de P.R. Moreira, Phys. Rev. Lett. 84, 1579 (2000).

[27] L.F. Mattheiss, Phys. Rev. Lett. 58 1028 (1987).

[28] J. Yu., A.J. Freeman, and J.-H. Xu, Phys. Rev. Lett. 58, 1035 (1987).

[29] R. Micnas, J. Ranninger, and S. Robaszkiewicz, Rev. Mod. Phys. 62, 113 (1990).

[30] P. Prelovsek and X. Zotos, Phys. Rev. B 47, 5948 (1993). P.Prelovsek and I. Sega, *ibid*. 49, 15241 (1994).

[31] V.J. Emery, S. Kivelson, and H-Q Lin, Phys. Rev. Lett. 64, 475 (1990); V.J. Emery and S. Kivelson, Physica C 209, 597 (1993).

[32] J. Zaanen and O. Gunnarson, Phys. Rev. B 40, 7391 (1989).

[33] H.J. Schulz, Phys. Rev. Lett. 64, 1445 (1990).

[34] M. Kato, K. Machida, H. Nakanishi, and M. Fujita, J. Phys. Soc. Jpn. 59, 1047 (1990).

[35] M. Inui and P.B. Littlewood, Phys. Rev. B 44, 4415 (1991).

[36] J. Vergés, F. Guinea, and E. Louis, Phys. Rev. 46, 3562 (1992).

[37] D. Poilblanc and T.M. Rice, Phys. Rev. B 39, 9749 (1989).

[38] T. Giamarchi and C. Lhuillier, Phys. Rev. B 43, 12943 (1991).

[39] G. An and J.M.J. van Leeuwen, Phys. Rev. B 44, 9410 (1991).

[40] H.J.M. van Bemmel, D.F. B. ten Haaf, W. van Saarlos, J.M.J. van Leeuwen, and G. An, Phys. Rev. Lett. 72, 2442 (1994).

[41] H. Yoshizawa, S. Mitsuda, H. Kitazawa, and K. Katsumata, J. Phys. Soc. Jpn. 57, 3686 (1988).

[42] R.J. Birgeneau, Y. Endoh, Y. Hidaka, K. Kakurai, M.A. Kastner, T. Murakami, G. Shirane, T.R. Thurston, and K. Yamada, Phys. Rev. B 39, 2868 (1989).

[43] S.-W. Cheong, G. Aeppli, T.E. Mason, H. Mook, S.M. Hayden, P.C. Canfield, Z. Fisk, K.N. Clausen, and J.L. Martinez, Phys. Rev. Lett. 67, 1791 (1991).

[44] T.E. Mason, G. Aeppli, S.M. Hayden, A.P. Ramires, and H.A. Mook, Phys. Rev. Lett. 71, 919 (1993).

[45] M. Matsuda, K. Yamada, Y. Endoh, T.R. Thurston, G. Shirane, R.J. Birgeneau, M.A. Kastner, I. Tanaka, and H. Kojima, Phys. Rev. B. 49, 6958 (1994).

[46] J.M. Tranquada, B.J. Sternlieb, J.D. Axe, Y. Nakamura, and S. Uchida, Nature (London) 375, 561 (1995); J.M. Tranquada, J.D. Axe, N. Ichikawa, Y. Nakamura, S. Uchida, and B. Nachumi, Phys. Rev. B 54, 7489 (1996); J.M. Tranquada, J.D. Axse, N. Ichikawa, A.R. Moodenbaugh, Y. Nakamura, and S. Uchida, Phys. Rev. Lett. 78, 338 (1997).

[47] J.M. Tranquada, Physica B 241-243, 745 (1998); J. M. Tranquada, N. Ichikawa, and S. Uchida, Phys. Rev. B 59, 14712 (1999).

[48] K. Yamada, C.H. Lee, Y. Endoh, G. Shirane, R.J. Birgeneau, and M.A. Kastner, Physica C 282-287, 85 (1997).

[49] K. Yamada, C.H. Lee, K. Kurahashi, J. Wada, S. Wakimoto, S. Ueki, H. Kimura, Y. Endoh, S. Hosoya, G. Shirane, R.J. Birgeneau, M. Greven, M.A. Kastner, and Y.J. Kim, Phys. Rev. B 57, 6165 (1998).

[50] P. Dai, H.A. Mook, and F. Doğan, Phys. Rev. Lett. 80, 1738 (1998).

[51] H.A. Mook, P.Dai, S.M. Hayden, G. Aeppli, T.G. Perring, and F. Dogan, Nature 395, 580 (1998).

[52] H.A. Mook and F. Doğan, Nature 401, 145 (1999).

[53] H.A. Mook, P. Dai, F. Doğan, and R.D. Hunt, Nature 404, 729 (2000).

[54] Y.S. Lee, R.J. Birgenau, M.A. Kastner, Y. Endoh, S. Wakimoto, K.Yamada, R.W. Erwin, S.-H- Lee, and G. Shirane, Phys. Rev. B 60, 3643 (1999); Y. Endoh, R.J. Birgeneau, M.A. Kastner,G. Shirane, and K. Yamada, Physica B 280, 201 (2000); H. Kimura, H. Matsushita, K. Hirota, Y. Endoh, K. Yamada, G. Shirane, Y.S. Lee, M.A. Kastner, and R.J. Birgeneau, Phys. Rev. B 61, 14366 (2000).

[55] J. Zaanen, O.Y. Osman, and W. van Saarloos, Phys. Rev. B 58, R11868 (1998).

[56] H. Eskes, O. Y. Osman, R. Grimberg, W. van Saarloos, and J. Zaanen, Phys. Rev. B 58, 6963 (1998).

[57] J. Zaanen, Phys. Rev. Lett. 84, 753 (2000).

[58] M. Bosch, W. van Saarlos, and J. Zaanen, Phys. Rev. B 63, 92501 (2001).

[59] J. Bonča, J.E. Gubernatis, M. Guerrero, E. Jeckelmann, and S.R. White, Phys. Rev. B 61, 3251 (2000).

[60] A. Sadori and M. Grilli, Phys. Rev. Lett. 84, 5375 (2000).

[61] A.L. Chernyshev, A.H. Castro Neto, and A.R. Bishop, Phys. Rev. Lett. 84, 4922 (2000).

[62] S.R. White and D.J. Scalapino, Phys. Rev. Lett. 80, 1272 (1998); 81, 3227 (1998); S.R. White and D.J. Scalapino, Phys. Rev. B 61, 6320 (2000).

[63] M. Votja and S. Sachdev, Phys. Rev. Lett. 83, 3916 (1999); E. Arrigoni, A.P. Harju, W. Hanke, B. Brendel, and S.A. Kivelson, Phys. Rev. B 65, 134503 (2002).

[64] G.B. Martins, C. Gazza, J.C. Xavier, A . Feiguin, and E. Dagotto, Phys. Rev. Lett. 84, 5844 (2000).

[65] P.O. Löwdin, Rev. Mod. Phys. 35, 496 (1963) [and discussions to the HF problem in this volume]; Advan. Chem. Phys. 14, 283 (1969).

[66] R. Poirier, R. Kari, and G. Csizmadia, *Physical Science Data 24, Handbook of Gaussian Basis Sets* (Elsevier 1985).

[67] P.J. Hay and W.R. Wadt, J. Chem. Phys. 82, 270 (1985); ibid. 82, 285 (1985); ibid. 82, 299 (1985).

[68] C. J. Calzado and J.P. Malrieu, Phys. Rev. B 63, 214520 (2001).

[69] J.M. Tranquada, S.M. Heald, A.R. Moodenbaugh, and M.Suenaga, Phys. Rev. B 35, 7187 (1987).

[70] A. Biaconi, A.C. Castellano, M.D. Santis, P. Rudolf; P. Lagarde, A.M. Flank, and A. Marcelli, Solid State Communications 63, 1009 (1987).

[71] N. Nücker, J. Fink, B. Renker, D. Ewert, C. Politis, P.J. Weijs, and J.C. Fuggle, Z. Phys. B 67, 9 (1987).

[72] D. van der Marel, J. van Elp, G.M. Sawatzky, and D. Heitmann, Phys. Rev. B 37, 5136 (1988).

[73] N. Nücker, J. Fink, J.C. Fuggle, P.J. Durham, and W.M. Temmerman, Phys. Rev. B 37, 5158 (1988).

[74] M.W. Shafer, T. Penney, and B.L. Olson, Phys. Rev. B 36, 4047 (1987); Solid State Ionics 39, 63 (1990).

[75] J.B. Torrance, Y. Tokura, A.I. Nassal, A. Bezinge, T.C.Huang, and S.S.P. Parkin, Phys. Rev.Lett. 62, 2317 (1988).

[76] M.A. Subramanian, A.R. Strzelecki, J. Gopalakrishnan, and A.W. Sleight, J. Solid State Chem. 77, 196 (1988).

[77] A. Manthiram and J.B. Goodenough, Appl. Phys. Lett. 53, 420 (1988).

[78] J.M. Tarascon, P. Bardoux, G.W. Hull, R. Ramesh, L.H. Greene, M. Giroud, M.S. Hedge, and W.R. McKinnon, Phys. Rev. B39, 4316 (1988).

[79] W.A. Groen, D.M. de Leeuw, and L.F. Feiner, Physica C 165, 55 (1990).

[80] R.J. Birgeneau and G. Shirane in *Physical Properties of High Temperature Superconductors I*, edited by D.M. Ginsberg (World Scientific, Singapore, 1989).

[81] H. Takagi, B. Batlogg, H.L. Kao, J. Kwo, R.J. Cava, J.J. Krajewski, and W.F. Peck, Jr., Phys. Rev. Lett. 69, 2975 (1992).

[82] X.J. Zhou, T. Yoshida, S.A. Kellar, P.V. Bogdanov, E.D. Lu, A. Lanzara, M. Nakamura, T. Noda, T. Kakeshita, H. Eisaki, S. Uchida, A. Fujimori, Z. Hussain, and Z.-X. Shen, Phys. Rev. Lett. 86, 5578 (2001).

[83] At least for small changes from the undisturbed b-hole state ($n_h = 0.125$ holes/copper) the distribution of the b-hole vacancies can be supposed to be periodically ordered. That is founded on the basic fact, that the electronic system tends to form topologically ordered hole states. It applies also to the f-hole subspace. I will treat this subject in detail elsewhere.

[84] B. Nachumi, Y. Fudamoto, A. Keren, K.M. Kojima, M. Larkin, G.M. Luke, J. Merrin, O. Tschernyshyov, Y.J. Uemura, N. Ichikawa M. Goto, H. Takagi, S. Uchida, M.K. Crawford, E.M. McCarron, D.E. MacLaughlin, and R.H. Heffner, Phys. Rev. B 53, 8760 (1998).

[85] Y.J. Uemura, W.J. Kossler, X.H. Yu, J. R. Kempton, H.E. Schone, and D. Opie, C.E. Stronach, D.C. Johnston, M.S. Alvarez, and D.P. Goshorn, Phys. Rev. Lett. 59, 1045 (1987).

[86] It is assumed that the existence of the b-hole state is the precondition for the occurrence of superconductivity.

[87] Ch. Niedermayer, C. Bernhard, T. Blasius, A. Golnik, A. Moodenbaugh, and J.I. Budnick, Phys. Rev. Lett. 80, 3843 (1998).

[88] For $n_h = 0.0625$ holes/copper nominally half of the b-holes in Fig. 12 are vacant. In order to maximize the number of attractive couplings as given in Figs. 6(a),(1) and (2) it is favourable if pairs of neighboured stripes (I and II) continuously exist. It means, every second pair of stripes (I,II) in Fig. 12 has to be considered to be vacant for $n_h = 0.0625$ holes/copper. In the non-superconducting case one has to assume that the particular pairs of b-hole stripes are not or only slightly coupled to each other. Therefore, they may nearly freely shift in y direction against each other in order to minimize the coulomb monopole interactions. In the supeconducting case, however, one has to assume that these stripes are well ordered in a two dimensional array. Two particular highly symmetrically b-hole arrays may be formed. One of this consists in a simple formation of parallel stripes by annihilation of every second pair of stripes (I,II) in Fig. 12. A b-hole unit cell of ($8a$,$8b$) results. If b-hole pair states which belong to one of the existing stripes (I or II) in Fig. 12 are changed in a way that they are alternately left and right positioned with respect to the unchanged stripe an ordered b-

hole state occurs with a zigzag configuration of the particular b-hole trapezia along the y direction in Fig. 12 . A b-hole unit cell of ($8a$,$16b$) results. The last configuration should possess a somewhat lower coulomb repulsive energy and therefore has to be considered as being most stable. All these symmetry problems will, however, really comprehensible only when I will discuss the microscopic formation of the superconducting state.

[89] K. Hirota, Physica C 357-360, 61 (2001).

[90] S.W. Lovesey, *Theory of Neutron Scattering from Condensed Matter* (Oxford Univ. Press 1984), Vol. 2, Chap. 12.6.

[91] H. Matsushita, H. Kimura, M. Fujita, K. Yamada, K. Hiroda, and Y. Endoh, J. Phys. Chem. Solids 60, 1071 (1999).

[92] H. Kimura, H. Matsushita, K. Hirota, and Y. Endoh, K. Yamada, G. Shirane, Y.S. Lee, M.A. Kastner, and R.J. Birgeneau, Phys. Rev. B 61, 14366 (2001).

[93] H.A. Mook, P. Dai, and F. Doğan, Phys. Rev. Lett. 88, 97004 (2002).

[94] M. Fujita, H. Goka, K. Yamada, and M. Matsuda, Phys. Rev. Lett. 88, 167008 (2002).

[95] M. Fujita, K. Yamada, H. Hiraka, P.M. Gehring, S.H. Lee, S. Wakimoto, and G. Shirane, Phys. Rev. B 65, 64505 (2002).

[96] M. Fujita, H. Goka, K. Yamada, and M. Matsuda, Phys. Rev. B 66, 184503 (2002).

[97] The electronic states within the b-hole system are basically pair-states as I will show elsewhere. Therefore, electronic excitations out of the b-hole subspace are pair-breaking excitations which explains the factor 2.

[98] T. Noda, H. Eisaki, S. Uchida, Science 286, 265 (1999).

[99] S. Arumugam, N. Môri, N. Takeshita, H. Takashima, T. Noda, H. Eisaki, and S. Uchida, Phys. Rev. Lett. 24, 247001 (2002).

[100] K.M. Lang, V. Madhavan, J.E. Hoffman, E.W. Hudson, H. Eisaki, S. Uchida, and J.C. Davis, Nature 415, 412 (2002).

[101] J. Zaanen, Nature 415, 379 (2002).